\documentclass[12pt]{iopart}

\usepackage[X2,T1]{fontenc}
\usepackage[utf8]{inputenc}
\usepackage[russian,english]{babel}
\newcommand{\ru}[1]{\foreignlanguage{russian}{#1}}

\usepackage{iopams}  
\usepackage{graphicx}  
\usepackage{overpic}  
\usepackage{bm}  
\def\be{\begin{equation}}
\def\ee{\end{equation}}
\def\ber{\begin{eqnarray}}
\def\eer{\end{eqnarray}}
\def\grad{{\mbox{\boldmath ${\nabla}$}}}

\def\cd{\!\cdot\!}

\def\a0{\gamma}
\def\b0{\tilde \gamma}

\def\bsp{{{\mbox{\boldmath${\sigma}$}}}}

\def\bvpfsc{{\hat{\mbox{\scriptsize \boldmath$p$}}_{\!f}}}
\def\bvpfscu{{\hat{\mbox{\scriptsize \boldmath$p$}}_{\!f\uparrow}}}
\def\bvpfscd{{\hat{\mbox{\scriptsize \boldmath$p$}}_{\!f\downarrow}}}
\def\bvpf{{\hat{\mbox{\boldmath$p$}}_{\!f}}}
\def\bvpfu{{\hat{\mbox{\boldmath$p$}}_{\!f\uparrow}}}
\def\bvpfd{{\hat{\mbox{\boldmath$p$}}_{\!f\downarrow}}}
\def\bvp{{{\mbox{\boldmath$p$}}}}

\def\bvR{{\mbox{\boldmath \mbox{$\scriptscriptstyle R$}}}}

\def\bvvf{{\mbox{\boldmath$v$}_{\!f}}}
\def\bvvfu{{\mbox{\boldmath$v$}_{\!f\uparrow}}}
\def\bvvfd{{\mbox{\boldmath$v$}_{\!f\downarrow}}}

\def\qcgrad{{i\hbar\, \bvvf\!\cdot\!\grad_{\!\!\bvR}\, }}
\def\qcgradbr{{i\hbar\, (\bvvf\!\cdot\!\grad_{\!\!\bvR})\, }}
\def\iPhi{\mathit{\Phi}}
\def\Av{{\mbox{\boldmath$A$}}}
\def\Bv{{\mbox{\boldmath$B$}}}

\def\Mv{{\mbox{\boldmath$M$}}}
\def\Rv{{\mbox{\boldmath$R$}}}
\def\Sv{{\mbox{\boldmath$S$}}}
\def\Jv{{\mbox{\boldmath$J$}}}
\def\Jvsc{{\mbox{\scriptsize\boldmath$J$}}}
\def\ev{{\mbox{\boldmath$e$}}}
\def\evsc{{\mbox{\scriptsize\boldmath$e$}}}
\def\fv{{\mbox{\boldmath$f$}}}
\def\ftv{{\mbox{\boldmath$\tilde f$}}}
\def\gv{{\mbox{\boldmath$g$}}}
\def\gtv{{\mbox{\boldmath$\tilde g$}}}
\def\jv{{\mbox{\boldmath$j$}}}
\def\rv{{\mbox{\boldmath$r$}}}

\def\Ds{\Delta}
\def\Dt{{\mbox{\boldmath ${\Delta}$}}}
\def\Ts{\tilde \Delta}
\def\Tt{{\mbox{\boldmath ${\tilde \Delta}$}}}

\def\nut{{\mbox{\boldmath ${\nu}$}}}

\def\tut{{\mbox{\boldmath ${\tilde \nu}$}}}

\def\tAa{\stackrel{\leftrightarrow}{A^a}}
\def\tAan{\stackrel{\leftrightarrow}{A^a_0}}
\def\tVt{\stackrel{\!\!\leftrightarrow}{V^t}}
\def\tWa{\stackrel{\!\!\leftrightarrow}{W^a}}
\def\tmu{\stackrel{\leftrightarrow}{\mu}}
\def\tchi{\stackrel{\leftrightarrow}{\chi}}
\def\tV{\stackrel{\leftrightarrow}{V}}
\def\tG{\stackrel{\leftrightarrow}{G}}

\newcommand{\bea}{\begin{eqnarray}}
\newcommand{\eea}{\end{eqnarray}}

\def\grad{{\bf \nabla }}
\newcommand{\vsigma}{\mbox{$\bm{\sigma}$}}

\renewcommand{\phi}{\varphi}
\renewcommand{\epsilon}{\varepsilon}

\def\vk{{\mbox{\boldmath$k$}}}
\def\vp{{\mbox{\boldmath$p$}}}
\def\vv{{\mbox{\boldmath$v$}}}
\def\vR{{\mbox{\boldmath$R$}}}
\def\vRs{{\mbox{\scriptsize\boldmath$R$}}}
\def\vQ{{\mbox{\boldmath$Q$}}}
\def\vQs{{\mbox{\scriptsize\boldmath$Q$}}}

\def\vRs{{\mbox{\scriptsize\boldmath$R$}}}

\def\vJ{{\mbox{\boldmath$J$}}}
\def\vB{{\mbox{\boldmath$B$}}}

\newcommand{\vepsilon}{\varepsilon}

\newcommand{\mbfsc}[1]{\mbox{\scriptsize\boldmath $#1$}}

\renewcommand{\vec}[1]{{\mathbf{#1}}}

\usepackage{fancyhdr}
\begin{document}

\review[Spin-polarized supercurrents for spintronics]{Spin-polarized supercurrents for spintronics: a review of current progress\footnote{
This is an author-created, un-copyedited version of an article 
published in {\it Reports on Progress in Physics} {\bf 78}, 104501 (50 pp) (2015).
IOP Publishing Ltd is not responsible for any errors or omissions in this version of the manuscript or any version derived from it. 
The Version of Record is available online at doi:10.1088/0034-4885/78/10/104501.
}}

\author{Matthias Eschrig}

\address{
Department of Physics, Royal Holloway, University of London, Egham Hill, Egham, Surrey TW20 0EX, United Kingdom}
\ead{matthias.eschrig@rhul.ac.uk}
\begin{abstract}
During the past 15 years a new field has emerged, which combines
superconductivity and spintronics, with the goal to pave a way for
new types of devices for applications combining the virtues of both
by offering the possibility of long-range spin-polarized supercurrents.
Such supercurrents constitute a fruitful basis for the study of fundamental physics as they combine macroscopic quantum coherence with microscopic exchange interactions, spin selectivity, and spin transport. 
This report follows recent developments in the controlled creation of long-range equal-spin triplet supercurrents in ferromagnets and its contribution to spintronics. The 
mutual proximity-induced modification of order in superconductor-ferromagnet hybrid structures introduces in a natural way such evasive phenomena as triplet superconductivity, odd-frequency pairing, Fulde-Ferrell-Larkin-Ovchinnikov pairing, long-range equal-spin supercurrents, $\pi$-Josephson junctions, as well as long-range magnetic proximity effects. All these effects were rather exotic before 2000, when improvements in nanofabrication and materials control allowed for a new quality of hybrid structures.
Guided by pioneering theoretical studies, experimental progress evolved rapidly, and  
since 2010 triplet supercurrents are routinely produced and observed. We have entered a new stage of studying new phases of matter previously out of our reach, and of
merging the hitherto disparate fields of superconductivity and spintronics to a new research direction: super-spintronics.
\end{abstract}

\newpage
$\;$
\newpage
\tableofcontents
\maketitle

\section{Introduction: historical background}

\subsection{Superconductivity interacting with magnetism}
Quantum phenomena fascinate us and challenge our imagination for over 100 years since the theoretical foundations of quantum physics were laid.
The two prime examples for macroscopic quantum phenomena are magnetism and superconductivity. 
Soon after the discovery of superconductivity in 1911 in Leiden by Kamerlingh-Onnes
\cite{Delft10,KO11} it became clear that a magnetic field acts in peculiar ways on superconductors. 
Silsbee unified in 1927 the observations of Kamerlingh-Onnes of a critical transport current density and a critical magnetic field \cite{Silsbee}, and in 1933, following a suggestion by 
von Laue, Mei{\ss}ner and Ochsenfeld performed experiments
showing that the new state of matter is a true thermodynamic phase and that it expels magnetic fields from its interior \cite{Meissner}. 
Following this pivotal discovery, Shubnikov pioneered in the 1930's
the field of interplay between superconductivity and magnetic field, culminating in the discovery of what is now called the intermediate state, or the Shubnikov phase 
\cite{Shubnikov35},\footnote{Unfortunately, Shubnikov's work ended abruptly when he was arrested on the 6th of August and executed on the 10th of November 1937 during the Stalin-Yezhov terror on the basis of falsified charges of antirevolutionary activities; he was officially rehabilitated in 1957.} 
where magnetic flux partially penetrates the superconductor in a well defined region of the temperature-field phase diagram.
After the notion of a macroscopic quantum wave function was introduced for the superconducting state by Ginzburg and Landau in 1950 \cite{Ginzburg50}, Abrikosov theoretically explained in the 1950's type II superconductivity, where a magnetic field penetrates the superconductor in a regular array of flux lines carrying quantized flux \cite{Abrikosov52}. 
Abrikosov's vortex phase exists between the upper critical field $H_{\rm c2}$ and the lower critical field $H_{\rm c1}$ in type II superconductors. 

Soon the microscopic theories of superconductivity
by Bardeen, Cooper, and Schrieffer (BCS) in 1957 \cite{Bardeen57}, by 
Bogoliubov in 1958 \cite{Bogolyubov58}, and by 
Gor'kov in 1958-59 \cite{Gorkov59} brought all previous studies on a firm ground.

Ginzburg, in an attempt to formulate a theory for ferromagnetic superconductors (he considered the possibility of superconductivity in gadolinium),
concentrated in 1956 on the electromagnetic (or so-called orbital) mechanism, where a suppression of superconductivity occurs via the interaction of the Cooper pairs with the vector potential of the magnetic field due to their charge \cite{Ginzburg56}. 
In the orbital mechanism the magnetic field leads to an additional kinetic energy of the condensate, and when this energy exceeds the condensation energy, the superconducting state is destroyed.
As shown in 1966 by Werthamer, Helfand, and Hohenberg, the orbital critical field 
for conventional isotropic, diffusive superconductors in the absence of spin-orbit and spin paramagnetic effects is $H_{\rm orb}\equiv H_{\rm c2}|_{T=0} \approx 0.7 T_{\rm c}[{-d} H_{\rm c2}/{d} T]_{T=T_{\rm c}}$ \cite{Werthamer66}, where $T_{\rm c}$ is the superconducting transition temperature.

An alternative mechanism, the exchange mechanism, was suggested in 1958 by Matthias, Suhl, and 
Corenzwit \cite{Matthias58} in order to explain the variation of $T_{\rm c}$ in lanthanum with rare earth impurities and the correlation between the appearance of ferromagnetism and superconductivity in ruthenides. They observed that it is not the dipole field of the effective moments of the rare earth elements that causes decrease of superconductivity in these systems, but the spin of the solute atoms; the $T_{\rm c}$ does not correlate with van Vleck's famous curve of $\mu_{\rm eff}$ \cite{vanVleck32}, but rather with spin.
In these cases an exchange interaction mediated by conduction electrons, responsible for ferromagnetism, occurs in a material that by itself is superconducting.
The exchange interaction via conduction electrons tries to align spins in a ferromagnet, whereas the spins in a Cooper pair are opposite for the usual case of singlet superconductors. 
These antagonistic tendencies led to the so-called paramagnetic effect of pair breaking. 

In 1959, Anderson and Suhl
showed in a seminal paper \cite{Anderson59} that in a system with coexisting superconductivity and ferromagnetism the magnetism is modified due to the suppression of the spin susceptibility at small wavevectors.  
As a consequence, magnetism becomes under certain conditions inhomogeneous with a new, finite, wavevector. 
This new state was termed the cryptoferromagnetic state. 
The wavelength of the magnetic  inhomogeneity is much smaller than the superconducting coherence length (the typical extend of a Cooper pair) and much larger than the lattice spacing, thus keeping neighboring spins almost parallel, however reducing the exchange field averaged over a typical Cooper pair volume.  
In real (anisotropic) materials the corresponding magnetic structure is often a one-dimensional magnetic domain structure.  

\subsection{Pair breaking by paramagnetic impurities and external fields}
If one introduces paramagnetic impurities into a singlet superconductor, there is
the effect of pair breaking due to scattering from paramagnetic impurities, which reduces $T_{\rm c}$ with increasing impurity concentration. 
Following previous works by 
Herring \cite{Herring58} and by Suhl and Matthias \cite{Suhl59} for the limit of small concentrations, 
Abrikosov and Gor'kov developed in 1960 a theory covering the full range of concentrations up to the critical value, when superconductivity is destroyed; the $T_{\rm c} $ dependence vs. concentration is described in terms of a single ``pair breaking parameter'' $\rho $ by
the Abrikosov-Gor'kov formula \cite{AbrikosovGorkov60}. In particular, these authors showed that the possibility of gapless superconductivity exists in metals with paramagnetic impurities (this peculiar state was further studied in 1964 by Skalski, Betbeder-Matibet, and Weiss \cite{Skalski64}, by 
Maki \cite{Maki64}, and by de Gennes \cite{deGennes64}).
The Abrikosov-Gor'kov model, which employs the Born approximation, works well, e.g. for rare earth (except cerium) impurities. For transition metal impurities in superconductors 
Yu \cite{Luh65}, Shiba \cite{Shiba68},
and Rusinov \cite{Rusinov69} discovered within the framework of a full $t$-matrix treatment of the problem that local states (now called the Yu-Shiba-Rusinov states) are present within the BCS energy gap due to multiple scattering between conduction electrons and paramagnetic impurities.

De Gennes and 
Tinkham noted in 1964 \cite{deGennesTinkham64} that for a time-reversal invariant superconducting order parameter, pair breaking can be related to
the asymptotic long-time behavior of the time-reversal correlation function, $\eta(t)=\lim_{t\to\infty} \langle \hat{\Theta}^\dagger (0) \hat{\Theta}(t) \rangle $, with the time reversal operator $\hat{\Theta}$. 
An exponential decay, $\eta(t)= \exp^{-2t/\tau_\Theta}$, with $\tau_\Theta$ some correlation time, corresponds to pair breaking with pair breaking parameter $\rho=(2\pi T\tau_\Theta )^{-1}$.
If, on the other hand, a nonzero limiting value $0<\eta<1$ appears, then this leads to an effective weakening of the pairing interaction.

When applying a magnetic field to a superconductor,
apart from the orbital effect of the penetrating field, there is also an appreciable Zeeman coupling between the electronic spins and the magnetic field.  
It leads to a splitting between the electronic spin bands similarly as in a weak ferromagnet the exchange energy leads to a band splitting. 
This effect has been experimentally studied in the early 1970's by Meservey and Tedrow 
in a series of classical papers
(see \cite{Meservey94} for a review).
If the magnetic field exceeds a certain value, superconductivity becomes energetically unfavorable due to this Zeeman coupling. 
This limiting field is the Pauli paramagnetic limiting field or the so-called Chandrasekhar-Clogston limiting field, and was predicted independently in 1962 by  
Chandrasekhar \cite{Chandrasekhar62} and  by 
Clogston \cite{Clogston62}. It amounts to $H_{\rm p}=\sqrt{2}\Delta(T=0)/g\mu_{\rm B}$, where $\Delta $ is the 
excitation gap in the superconductor at zero magnetic field, $g\approx 2$ the electron $g$-factor, and $\mu_{\rm B}$ the Bohr magneton. The ratio 
$\alpha=\sqrt{2}\tilde H_{\rm orb}/H_{\rm p}$, where $\tilde H_{\rm orb}$ is the upper critical field in the absence of the Pauli term,
was discussed in 1966 by Maki, and is known as the Maki parameter \cite{Maki66}.

\subsection{Ferromagnetic superconductors}
Coexisting singlet superconductivity and ferromagnetism is rare, and can be achieved either by finding suitable crystalline materials (classical examples are the rare-earth ternary compounds ErRh$_4$B$_4$ \cite{ErRh4B4} and HoMo$_6$S$_8$ \cite{HoMo6S8} in a narrow temperature region below the Curie temperature), 
or by introducing magnetic ions in a superconducting material that order ferromagnetically. 
In the latter case, at high impurity concen\-tra\-tions, the possibility to obtain coexistence between ferromagnetism and superconductivity was theoretically predicted in 1964 by Gor'kov and Rusinov \cite{GorkovRusinov64}. Furthermore,
for the case that ferromagnetism and superconductivity coexist or an external magnetic field is applied, 
Fulde and 
Maki \cite{FuldeMaki66} showed that the Abrikosov-Gor'kov formula holds with a modified pair breaking parameter.
For an early review on magnetic superconductors see \cite{Buzdin84}.

\begin{figure}
\begin{center}
\includegraphics[width=0.5\linewidth]{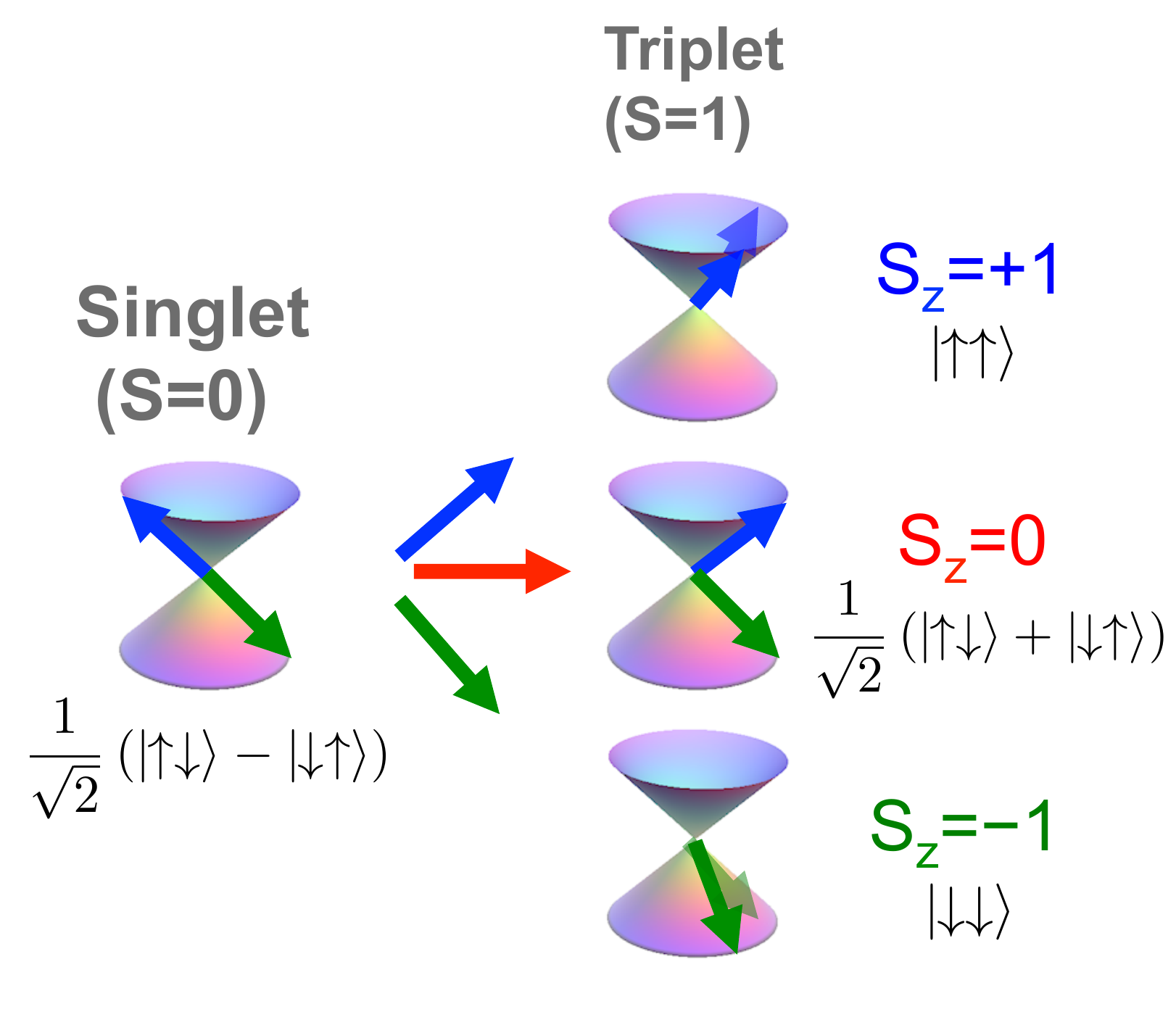}
\end{center}
\caption{
Two electrons, each of which has spin $s=\frac{1}{2}$, can combine their spin angular momenta $\vec{s}_1$ and $\vec{s}_2$ to build a pair with total spin $S=0$ or a pair with total spin $S=1$. 
In the first case, the pair is in a spin-singlet state (shown on the left). In the second case, the pair is in one of the three possible spin-triplet states: with spin projection $S_z=0, \pm 1$ (shown on the right). 
The pairs with $S_z=\pm 1$ are called ``equal-spin'' pairs with respect to the spin quantization axis (here $z$-axis). The spin state $|S,S_z\rangle $ is fully characterized by the two quantum numbers $S$ and $S_z$. 
We use a spin-vector representation to visualize pair angular momenta, where the expectation value of the cosine of the relative angle between the two spin vectors within a pair, 
$\langle S,S_z|\vec{s}_1\cdot\vec{s}_2|S,S_z\rangle /\sqrt{\vec{s}_1^2\vec{s}_2^2}$,
is $-1$ for singlet states, and $\frac{1}{3}$ for triplet states. Thus,
angular momenta can be visualized as antiparallel for singlet pairs, whereas for triplet pairs they
enclose in average a relative angle of $\approx 70.53^o$ (even for ``equal spin'' pairs).
}
\label{SingletTriplet}
\end{figure}

In addition to the above cases of conventional singlet superconductors, there has been discovered a large number of unconventional superconductors, which are distinguished from the conventional ones by the fact that they break additional symmetries, e.g. the lattice symmetry of the normal state, or the spin rotational symmetry. 
The latter is the case for superconductors exhibiting spin-triplet pairing (see figure \ref{SingletTriplet}).
In particular ferromagnetic superconductors are under suspicion of such a pairing state, as equal-spin triplet pairing is not sensitive to the exchange mechanism in the way singlet pairing is. 
This refers to some heavy fermion compounds, like UGe$_2$ \cite{UGe2}, URhGe \cite{URhGe}, UCoGe \cite{UCoGe}, and UIr \cite{UIr}, in which superconductivity can coexist with ferromagnetism, and which have been studied in the past 15 years \cite{Aoki11}.

The main problem for coexistence between superconductivity and ferromagnetism is that a strong exchange effect destroys superconductivity unless the pairing is of the triplet kind. However, not many superconductors support triplet pairing. Furthermore, such systems are typically $p$-wave superconductors 
\cite{Balian63}, which are very sensitive to pair breaking by normal impurities 
(conventional superconductors are insensitive to scattering from normal impurities, which is the content of a theorem by Abrikosov, Gor'kov, and Anderson \cite{Abrikosov59,Anderson59a}).
For this reason new avenues have been chosen to study such systems for $s$-wave and $d$-wave superconductors, which are much more common in nature. These avenues combine the phenomena of proximity induced superconductivity, Cooper pairs with finite center of mass momentum, and odd-frequency $s$-wave spin-triplet pairs.

\subsection{Cooper pairs with finite center of mass momentum}

When the paramagnetic limiting field is smaller than the orbital critical field, then the possibility of an inhomogeneous superconducting state with finite pair momentum arises near $H_{\rm p}$. Such a state was theoretically predicted in 1964
independently by Fulde and Ferrell \cite{Fulde64} and by Larkin and Ovchinnikov \cite{Larkin64}.
It is known in the western literature as FFLO state and in the eastern as LOFF state.

\begin{figure}
\begin{center}
\includegraphics[width=0.6\linewidth]{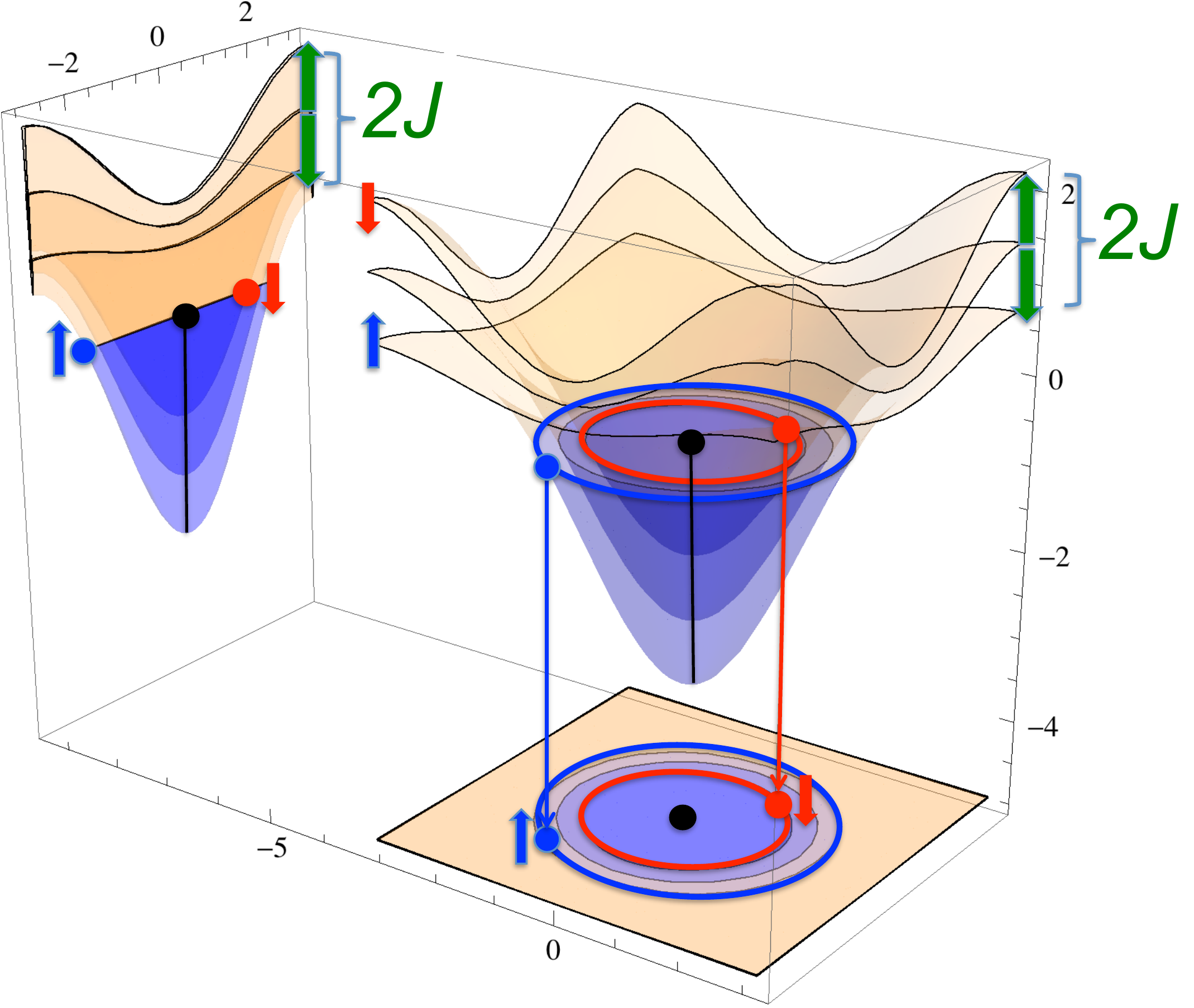}
\end{center}
\caption{The FFLO effect: due to the exchange splitting of the energy bands (filled states are shaded blue in the picture) by $\pm J$ the Fermi surfaces split as well. A Cooper pair with opposite spins can be accommodated at the Fermi surface only on the cost of a finite center of mass momentum. 
Note that spin is quantized here in direction of the exchange field $\vJ$, entering the Hamiltonian as ${\cal H}_{\rm exch}=-\vJ\cdot \vsigma$.
}
\label{FFLO}
\end{figure}

Consider superconductivity in a weakly spin-polarized ferromagnetic material (or in an external magnetic field). 
Electronic spin bands are shifted in energy with respect to each other by an amount $2J$ (which in general can be anisotropic): $\varepsilon(\vp) \to \varepsilon(\vp) - \vJ(\vp) \cdot \vsigma $. 
The exchange field can be related to an effective magnetic field via $ \mu\vB_{\rm eff}=
\vJ$ (for free electrons the magnetic moment is $\mu=\mu_{\rm e}<0$).
We use the convention that spin is quantized in direction of the exchange field.
The energy
shift translates into a splitting of the Fermi surface for the two spin species (see Figure \ref{FFLO}), which results e.g. for small $J$ from the relation 
\be
\hbar \bvvf \cdot \vQ=\bvvf \cdot (\vp_{f\uparrow}-\vp_{f\downarrow})= 2J(\vp_{f})
\ee
obtained by linearizing the dispersion relation around the Fermi energy
 [here the Fermi velocity $\bvvf $ is assumed to be approximately equal for the two spin bands due to the small splitting, and $\vp_f=(\vp_{f\uparrow}+\vp_{f\downarrow})/2$].
Thus, the system only allows for opposite-spin Cooper pairs 
$\left\{\vp_{f\uparrow}, -\vp_{f\downarrow}\right\}$ 
and 
$\left\{\vp_{f\downarrow}, -\vp_{f\uparrow}\right\}$
built from electrons 
at the Fermi energy if they carry a finite center of mass momentum $\pm \hbar \vQ/2=
\pm(\vp_{f\uparrow}-\vp_{f\downarrow})/2$ ($\hbar \vQ $ is the momentum of the Cooper pair).
This can lead to an order parameter characterized by a linear combination of terms with different Cooper pair momenta
\be
\Delta_{\rm FFLO} (\vR )= \sum_{\nu} \Delta_{\nu} e^{i\vQs_\nu \cdot \vRs} .
\ee
The preferred directions are chosen by the 
system (spontaneously or due to crystal anisotropy and boundary conditions). 
Fulde and Ferrell suggested an order parameter characterized by only one wavevector $\vQ $, leading to a spatially homogeneous modulus and a spatially varying phase.
If the order parameter is characterized by more than one wavevector $\vQ_\nu $, then the FFLO state exhibits an inhomogeneous order showing a periodic structure.
Larkin and Ovchinnikov proposed that the ground state may show a one-dimensional modulation (neglecting orbital effects).
However, more complex structures are possible in the general case of anisotropic metals.
In general, the pair amplitudes develop an $|S=1, S_z=0\rangle $ spin-triplet component, and the inhomogeneous modulus of the order parameter is accompanied by an oscillation of the magnetization around its average, determined by the product of the singlet and triplet pair amplitudes.

Originally the state had been studied 
for a superconductor with ferromagnetically aligned impurities. 
However, it is known today that the FFLO state is very sensitive to disorder \cite{Aslamazov69,Takada70}, and thus most likely is found in clean materials in an external field. 
Gruenberg and 
Gunther studied in 1966 the influence of orbital effects in type II superconductors on the possibility to observe an FFLO state and found that although orbital effects are detrimental to the formation of the FFLO state, such a state can exist at finite temperatures in the mixed state of a pure superconductor provided the Maki parameter $\alpha $ exceeds 1.8 \cite{Gruenberg66}.
Interesting candidates are quasi-two-dimensional superconductors in
a magnetic field applied nearly parallel to the conducting layers. The theory for such systems was developed by Bulaevski\u{i} \cite{Bulaevskii73} in 1973, 
by Burkhardt and Rainer \cite{Burkhardt94} and by Shimahara \cite{Shimahara94} in 1994,
and by Buzdin and Brison \cite{Buzdin96} in 1996.
For quasi-one-dimensional systems theories in terms of soliton lattice solutions were developed by 
Buzdin and Tugushev in 1983 \cite{Buzdin83}, by Buzdin and Polonskii in 1987 \cite{Buzdin87}, and
by Dupuis in 1995 \cite{Dupuis95}.
In compounds with large spin susceptibility and high $H_{\rm c2}$, as in heavy fermion and intermediate-valence systems, a generalized FFLO state was discussed by 
Tachiki and co-workers, where the order parameter is spatially modulated, and planar nodes of the order parameter are periodically aligned perpendicular to the vortices \cite{Tachiki96}.

The FFLO state only appears below a certain temperature $T^\ast$, which determines a tri-critical point $(T^\ast, H^\ast)$ in the $(T,H)$-diagram at which the normal, BCS, and the FFLO phases meet.
A generalized Ginzburg-Landau theory for the vicinity of this tri-critical point was derived by 
Buzdin and Kachkachi \cite{Buzdin97}. 
A self-consistent calculation of the field versus temperature phase diagram and order parameter structures for the FFLO states of quasi-two-dimensional $d$-wave superconductors is presented in Ref.~\cite{Vorontsov05}.
Recently, a symmetry classification of the pairing amplitudes in FFLO states coexisting with vortices was suggested \cite{Yokoyama10}.
For reviews on this topic see Refs. \cite{Casulbuoni04,Matsuda07,Zwicknagl10,Beyer13}.

The FFLO state
has not been unambiguously found in bulk materials to date, although there are currently candidates under debate, as the heavy fermion material CeCoIn$_5$ \cite{Radovan03,Bianchi03,Kenzelmann08,Koutroulakis10}, or 
some organic superconductors such as $\kappa$-(BEDT-TTF)$_2$Cu(NCS)$_2$ \cite{Singleton00,Lortz07,Bergk11,Mayaffre14}.
However, a similar state can be realized in superconductor-ferromagnet hybrid structures, where 
superconducting pairs leak into a ferromagnet due to the superconducting 
proximity effect, as we discuss in detail further below (for earlier reviews see \cite{Izyumov02,Eschrig04,Bergeret05a,Buzdin05,Lyuksyutov07}).

\subsection{Proximity effect}

The pivotal role for realizing the FFLO state in such hybrid structures
is played by the proximity effect between a superconducting and a normal conducting material. 
The theory for the proximity effect was developed largely by de Gennes \cite{deGennes63,deGennes64a,deGennes66,Deutscher69}, and by Werthamer \cite{Werthamer63,Hauser64} in the 1960's. 
It describes the effect of penetration of Cooper pairs in a normal metal where the two conduction electrons (or holes) of the pair stay correlated with each other over a distance that depends on the amount of disorder in the normal material and on temperature.  
In a clean material, the penetration depth is determined by the thermal coherence length 
$\xi_{\rm N,c}=\hbar v_f/2\pi k_{\rm B}T$. It gives the distance, a quasiparticle with velocity $v_f$ travels during time $\hbar/2\pi k_{\rm B}T$, before thermal decoherence sets in.
For ballistic motion and atomically smooth interfaces, the penetration depth depends on the angle of impact of the quasiparticles, i.e. the Fermi velocity should be replaced by the
Fermi velocity component in the normal metal perpendicular to the interface, $v_{f,x}$.
The notion of a coherence length determined by the Fermi velocity and a characteristic superconducting energy scale was introduced by 
Pippard in 1953 \cite{Pippard53}. 

In the presence of impurities, scattering decreases the penetration of Cooper pairs into the normal metal. 
If the mean free path is $\ell $, and the motion is diffusive, then 
the quasiparticle travels in average between two scattering events a time $\tau=\ell/v_f$.
For a random walk, the variance for the $x$-component is given by $\sigma_x^2=\frac{1}{3} \frac{t}{\tau} \ell^2=tv_f \ell /3$, where $t$ is the elapsed time. 
Coherence is lost when the total path traversed 
during the random walk is equal to $\xi_{\rm N,c}$, i.e. $v_ft=\xi_{\rm N,c}$.
If we define the diffusive coherence length by $\xi^2_{\rm N,d}=\sigma_x^2$, we obtain
$\xi_{\rm N,d}^2=\xi_{\rm N,c}\ell/3$. Hence, the diffusive coherence length is given by  $\xi_{\rm N,d}=\sqrt{\xi_{\rm N,c} \ell /3}$, or in other words,
$\xi_{\rm N,d}=\sqrt{\hbar D/2\pi k_{\rm B}T}$ with the diffusion constant 
$D=v_f\ell /3$.\footnote{In $d$ dimensions the factor of 3 is replaced by a factor of $d$.}

The proximity amplitude is proportional to the pair potential in the superconductor,
$\Delta $, and the product of the transmission amplitudes through the interface of the two particles comprising a pair. 
Inversely, the loss of Cooper pairs in the superconductor leads to the so-called inverse proximity effect of a weakening of the superconducting pair potential near the interface in the superconductor. 

\subsection{Andreev reflection}

A very closely related phenomenon is the so-called Andreev reflection, discovered by Andreev in 1964 
\cite{Andreev64} and by de Gennes and Saint-Jaimes in 1963/64 \cite{deGennesJaimes63,Jaimes64} (see Ref. \cite{Deutscher05} for a review). 
It describes the correlations between particles and holes at the normal side of the interface due to penetration of pairs. In fact, Andreev reflection and proximity effect are two sides of one and the same coin.

Let us assume that
a spin $\uparrow $ conduction electron 
in the normal metal with energy $\mu+\epsilon $ near the chemical potential $\mu $
moves in positive $x$-direction towards a superconductor. 
The excitation energy $\epsilon $ is assumed to
be within the energy gap $\Delta $ of the superconductor. 
The electron can be
transmitted into the superconductor only together with another electron with spin $\downarrow $, energy $\mu-\epsilon $, and momentum approximately opposite, 
to build a Cooper pair that enters the condensate
at the pair chemical potential $2\mu$ (after some momentum relaxation has taken place).
For $\Delta \ll \mu$
it is possible to approximate the electronic dispersion 
$E(\vp)\approx \mu + \bvvf\cdot (\vp -\bvpf)
$, 
with the  Fermi momentum $\bvpf $ and
Fermi velocity $\bvvf \equiv \vv(\bvpf) $.
Within this approximation, the group velocity of the incoming electron is
equal to $\bvvf $, and its excitation energy $\epsilon = \bvvf \cdot (\vp -\bvpf)$.
\begin{figure}
\begin{center}
\includegraphics[width=0.7\linewidth]{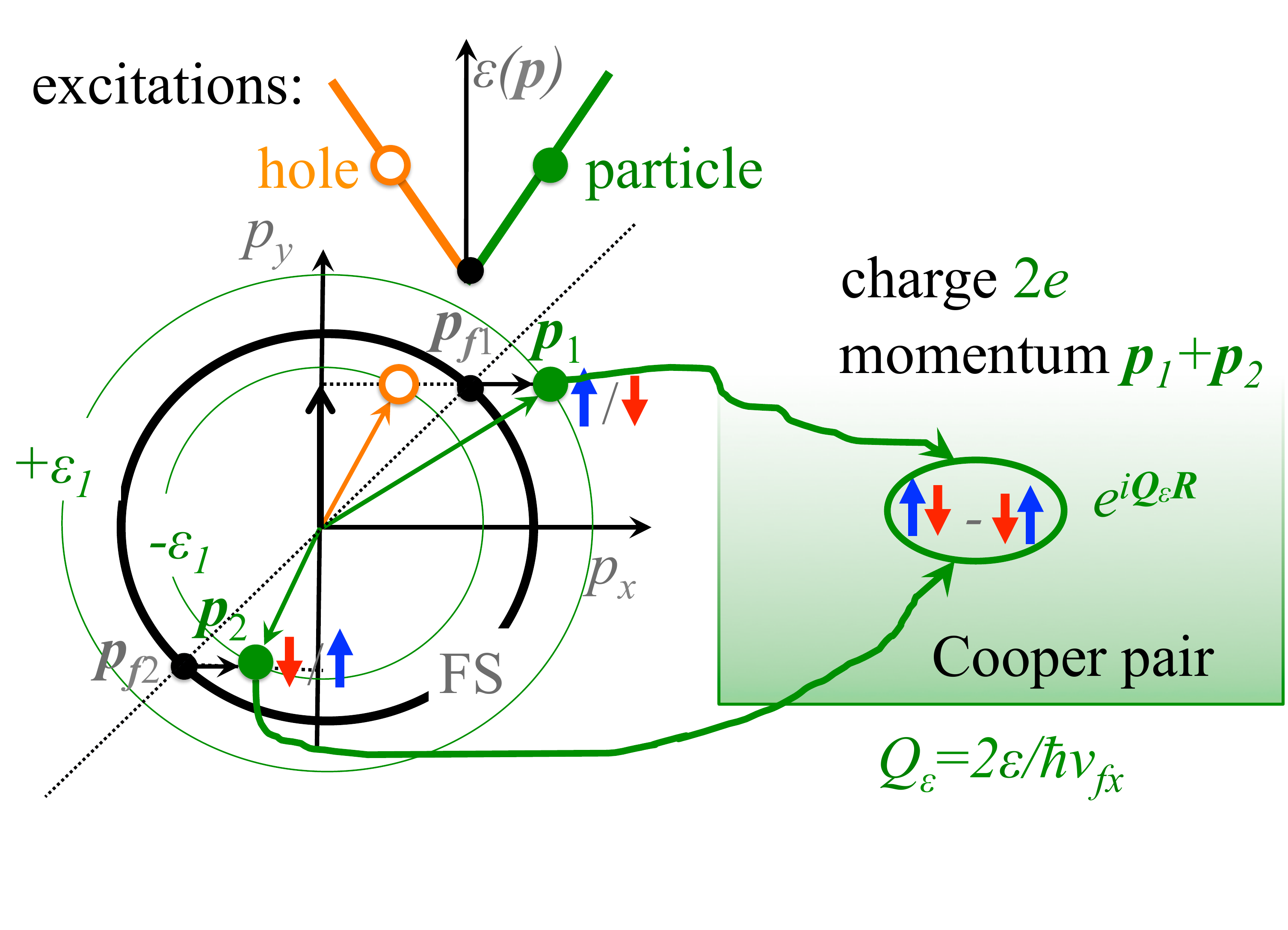}
\end{center}
\caption{Andreev reflection: An electron at crystal momentum $\vp_1$ pairs with another one at $\vp_2$ under the conditions that their energies and $p_y$-component of momentum are opposite to each other. This leads to a finite $p_x$ component and a resulting center of mass momentum $\hbar \vQ_\epsilon/2 $ of the Cooper pair. A hole is reflected, which is drawn in this plot at the negative momentum and energy of the missing electron (in the excitation picture all excitation states inside the Fermi surface have reversed momentum and energy and are considered as hole excitation with positive excitation energy). The resulting Cooper pair transfers its momentum partially to the interface between the normal metal and the superconductor (for the case that there is a Fermi surface or Fermi velocity mismatch), and 
partially to the condensate in the superconductor when entering it. The amount transferred to the condensate results in a supercurrent consistent with current conservation in the scattering process.
}
\label{Andreev}
\end{figure}

Let us consider a Landau quasiparticle with momentum $\vp_{1}=\vp_{f1}+\delta \vp_1$, spin $\uparrow $, and excitation energy $\epsilon_1= \vv_{f1} \cdot \delta \vp_1$.
We will consider pairing with another quasiparticle with momentum 
$\vp_{2}=-\vp_{f1} + \delta \vp_2$, spin $\downarrow $, and excitation energy $\epsilon_2=-\vv_{f1}\cdot \delta \vp_{2}$ (see Figure \ref{Andreev}). 
We require that $\epsilon_2=-\epsilon_1$. This leads to the relation $\vv_{f1}\cdot (\delta \vp_{2}-\delta \vp_1)=0$,
which allows us to choose $\vp_{f1}$ such that $\delta \vp_1=\delta \vp_2=\delta \vp$ holds\footnote{Choose $\tilde{\vp}_{f1}=\vp_{f1}+(\delta \vp_1-\delta \vp_2)/2$, then $\vp_1-\tilde{\vp}_{f1}=\vp_2+\tilde{\vp}_{f1}=\delta \vp=(\delta \vp_1+\delta \vp_2)/2$. This is again (in leading order) a Fermi surface point due to $\vv_{f1}\cdot (\tilde{\vp}_{f1}-\vp_{f1})=0$, i.e. the shift is tangential to the Fermi surface.}.
The missing electron in the spin-$\downarrow $ band is equivalent to a hole excitation characterized by energy $-\epsilon_2=+\epsilon_1 $, momentum $-\vp_2=\vp_{f1} - \delta \vp$, group velocity $\vv(\vp_2)\approx -\vv_{f1}$, and spin $\uparrow $ (a missing spin $\downarrow $). 
The process can be considered as a scattering of a particle with charge $e$, momentum $\bvp_f+\delta \vp$, and velocity $\bvvf$ into a hole with charge $-e$, momentum 
$\bvp_f-\delta \vp$ and velocity $-\bvvf$.
The electron-hole pair gives rise to a current density of $e\bvvf+(-e)(-\bvvf)=
2e\bvvf$, twice the current density of the incoming particle.
The incoming electron (velocity $\vv_{f1}$) is approximately ``retro-reflected'' (velocity $\approx -\vv_{f 1}$) as a hole \cite{Andreev64}. 
This process is called Andreev reflection, contrasting 
the familiar specular reflection. 

A possible small angle between the reflected hole
and the incoming particle is due to variation of the Fermi velocity between the momenta $\bvp_f$ and $\bvp_f\pm \delta \vp$, and is of order of 
$\epsilon/E_{f}$, with Fermi energy $E_{f}$ \cite{Stone96}. 
Retroreflection is perfect for perpendicular impact if no supercurrent flows along the surface.
In the Andreev particle-hole conversion process, energy and spin are conserved,
however charge is not, and momentum is only approximately conserved. 
The charge $2e$ and (partially) the momentum $\hbar \vQ=2\delta \vp$
are transferred to the superconducting condensate:
two electrons with opposite spin enter the superconductor to create a Cooper pair
with non-zero pair momentum,
which joins the condensate, leading to a (small) supercurrent for
finite excitation energies. The momentum $\hbar \vQ $ is thereby partially transferred to the interface (provided a Fermi surface or Fermi velocity mismatch exists), and partially to the condensate resulting into a supercurrent consistent with total charge conservation.
In equilibrium, and in the absence of a macroscopic supercurrent, this current is exactly canceled by the inverse process when a Cooper pair enters the normal metal and converts a hole into an electron.

For an atomically clean interface, 
if there is a macroscopic supercurrent with Cooper pair momentum $\vp_S$ parallel to the interface present in the superconductor, then momentum conservation requires that the created Cooper pair enters the condensate with the required Cooper pair momentum parallel to the interface, such that
$2\delta \vp_{||}=\vp_S$. 
One obtains 
(we assume $v_{fx}>0$)
$2\delta p_{x}=
(2\epsilon-\bvvf\cdot \vp_S)/v_{fx}$.
In particular, for $\vp_S=0$ only $p_x$ can change in the Andreev scattering process. 
In this case, as $\delta \vp_{||}$ must be zero, the Cooper pair momentum is given by 
$Q_\epsilon=2\epsilon /v_{fx}$ (see Figure \ref{Andreev}).

The Andreev reflection process is phase coherent, i.e.  the phases of the incoming particle and the Andreev-reflected hole are correlated. For a superconducting order parameter $\Delta =|\Delta |e^{i\chi }$,
the phase difference is $\phi(\epsilon)-\chi $,
between an incoming electron and retroreflected hole,
and is $\phi(\epsilon)+\chi $ between an incoming hole and retroreflected electron, where $\phi(\epsilon)=-\arccos (\epsilon /|\Delta |)$ describes the phase delay due to the fact that
during Andreev reflection the particle and hole amplitudes penetrate into the superconductor
over a length scale $\xi^{(\rm S)}_\epsilon =\hbar v^{(\rm S)}_{f}/\sqrt{|\Delta|^2-\epsilon^2}$.\footnote{For a supercurrent along the interface, $\chi(\vR) = \chi(\vR_0)+\frac{1}{\hbar}\vp_{\rm S} \cdot (\vR-\vR_0)$, with $\vR$ the spatial coordinate on the interface.
The time the Cooper pair needs to enter the condensate is $\tau^{(\rm S)}_\epsilon=\xi^{(\rm S)}_\epsilon/v^{(\rm S)}_{f}$, which also determines the time delay between the maximum of an incoming electron wave packet
and the maximum of the retroreflected hole wave packet,
$\tau^{(\rm S)}_\epsilon= \hbar \partial_\epsilon \phi (\epsilon)=
\hbar/\sqrt{|\Delta|^2-\epsilon^2}$ \cite{kummel}.  
}

During their motion in the normal metal, the phase coherence is lost 
due to the slight difference in momentum of the electron and the hole, 
given by $2\delta p_x=2\epsilon / v_{fx}$. When averaged over all momentum directions, the pair correlation function decays away from the interface algebraically $\sim \xi^{(\rm N)}_{\rm c}/x$ as function of $x$, where we define $\xi^{(\rm N)}_{\rm c}=\hbar v_f/2\epsilon $.  For observables in equilibrium, the decay length can be
obtained by replacing 
$\epsilon$ in the expression for the momentum $\delta p_x$ by the lowest Matsubara energy \cite{Matsubara55} $i\pi k_{\rm B}T$. This leads to an exponential decay on the length scale
$\hbar v_{f}/2\pi k_{\rm B}T$ along each ballistic trajectory, which is precisely the coherence length from the proximity effect. 

For diffusive motion, non-magnetic scattering events conserve the time reversal
symmetry, and thus scatter time-reversed states in a similar way. In particular
a particle-hole pair with momenta both close to $\vp_f$ will scatter in
a particle hole pair with momenta both close to $\vp'_f$ (as scattering amplitude for $\vp_f\to \vp_f'$ is the same as for $-\vp_f'\to -\vp_f$ due to time reversal symmetry). Thus,
scattering events change the momentum of particle and hole by the same amount, 
however do not destroy their coherence. This means, the dephasing between the particle
and hole takes place on a semiclassical trajectory 
as in the clean case, however the trajectory changes direction multiple times in a random way. 
The diffusive process away from the interface is characterized by a diffusion equation for the pair amplitude $f$, given by $\partial_t f = D\partial^2_x f$. With $f\sim e^{i(k_x x-2\epsilon t/\hbar)}$ this implies $-2i\epsilon = -\hbar Dk_x^2 $, resulting in a complex wave vector given
for decay in positive $x$ direction by $k_x=(i\pm 1)\sqrt{|\epsilon|/\hbar D}$. 
This gives rise to an exponential decay on the length scale $\sqrt{\hbar D/|\epsilon |}$, accompanied by an oscillation on the same length scale.
For observables in equilibrium we can replace $\epsilon \to i\pi k_{\rm B}T$, leading to
$k_x^2=-2\pi k_{\rm B}T/\hbar D$ and
an exponential decay on the
length scale $\xi^{(\rm N)}_{\rm d}=\sqrt{\hbar D/2\pi k_{\rm B}T}$, 
again matching the length scale for the proximity effect in diffusive metals.
The two phenomena, proximity effect and Andreev reflection are intertwined and cannot be discussed separately from each other.
The above discussion assumes that the phase coherence is not weakened by
additional effects.
If phase coherence is weakened by inelastic processes, as for example magnetic scattering, this puts limitations on the proximity effect. In this case, one has to replace 
$\epsilon $ by 
$\bar \epsilon =\epsilon+i\alpha_{\rm s}$ with the spin-flip scattering rate $\alpha_{\rm s}=\hbar/\tau_{\rm s}$, where $\tau_{\rm s}$ is the corresponding life time.
This means that in the ballistic limit exponential decay of pair correlations sets in on the length scale $\hbar v_f/2\alpha_{\rm s}=v_f\tau_{\rm s}/2$ (the factor 2 takes into account that both particles and holes are affected).
The corresponding coherence lengths are obtained by replacing $\pi k_{\rm B} T $ by $\pi k_{\rm B}T+\alpha_{\rm s}$.

\begin{figure}
\begin{center}
\includegraphics[width=0.7\linewidth]{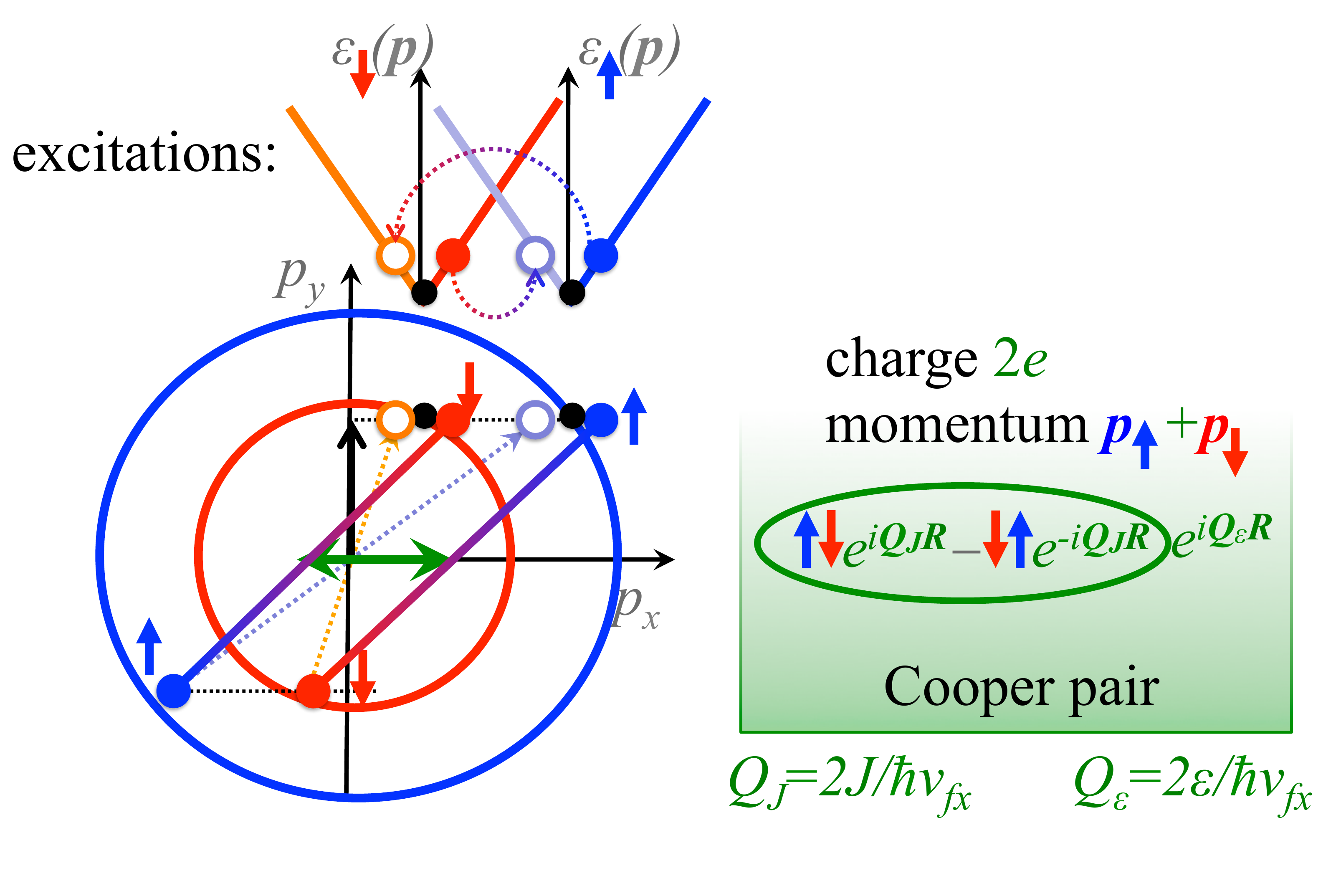}
\end{center}
\caption{Andreev reflection in a ferromagnet: an electron with energy $\epsilon $ and spin $\uparrow$ pairs with an electron with energy $-\epsilon$ and spin $\downarrow$, resulting into a total momentum of $\hbar( \vQ_\epsilon \pm \vQ_J)$. For sufficiently small spin band splitting a Cooper pair of the form shown to the right results, which contains singlet and triplet components with amplitudes $\cos (Q_J x)$ and $i\sin (Q_J x)$, respectively. The Cooper pair enters the condensate in the superconductor, transferring its momentum partially to the interface and partially to the singlet condensate, ensuring current conservation. The triplet component decays on the superconducting coherence length scale into the superconductor.
}
\label{Andreev2}
\end{figure}
\begin{figure}
\begin{center}
\includegraphics[width=0.7\linewidth]{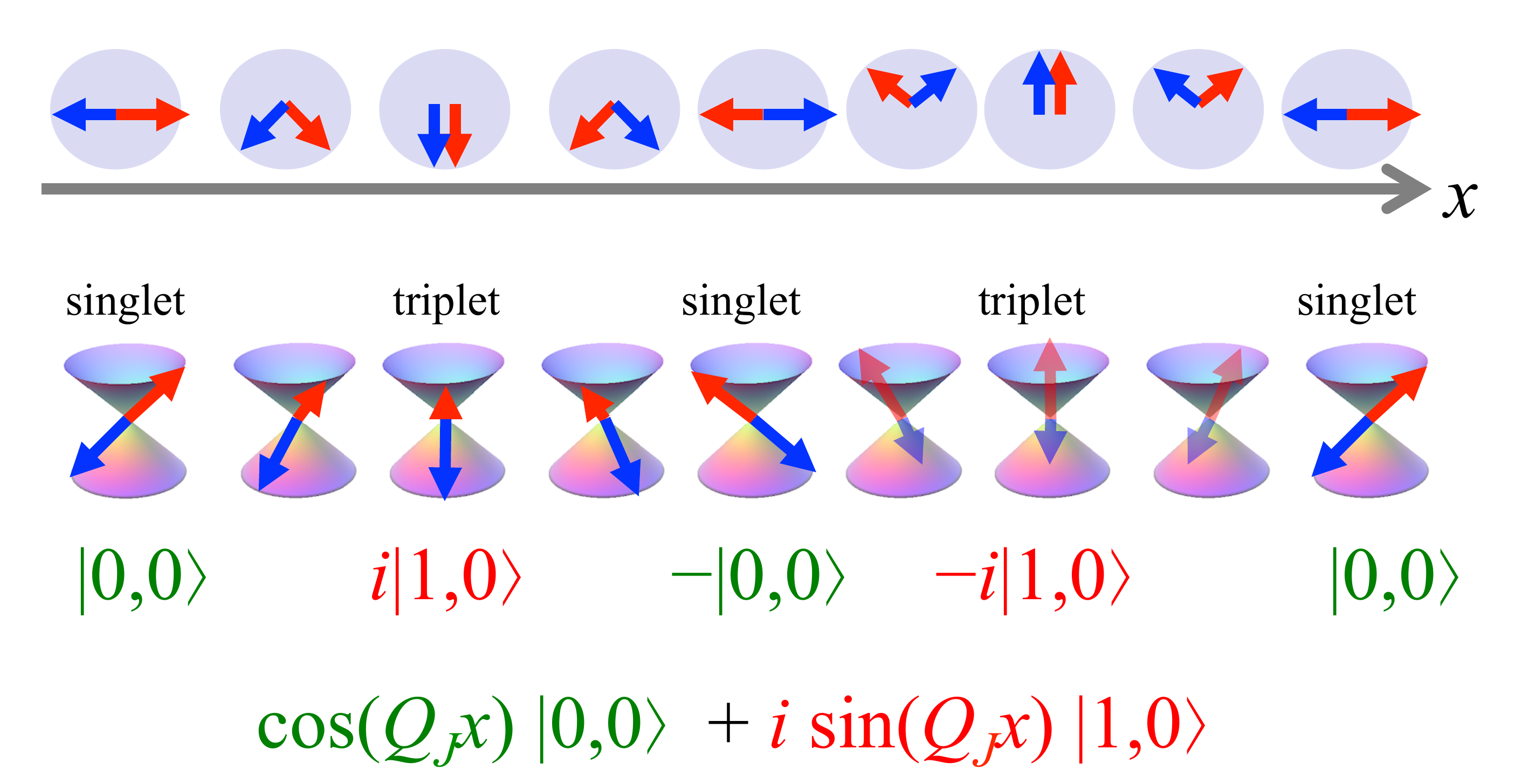}
\end{center}
\caption{
Singlet-triplet mixtures between $|S=0,S_z=0\rangle$ and $|S=1,S_z=0\rangle $ spin angular momentum states in an exchange field (or in an applied external magnetic field) in $z$-direction. Singlet and triplet spin states oscillate spatially out of phase relative to each other. Within the spin-vector representation the relative azimuthal angle between the two spin vectors within a pair varies as $\pi+2Q_Jx$, and its deviation from 0 (triplet) and $\pi$ (singlet) is a measure for the degree of singlet-triplet mixing.
The top row shows a top-view of the bottom row.
}
\label{SingTrip1}
\end{figure}

A similar mechanism works for a weakly ferromagnetic normal metal, as for 
example a ferromagnetic alloy (see Figure \ref{Andreev2}) \cite{Demler97}.
In this case, the electronic bands are
spin-split due to the exchange interaction by an amount of 
$\pm J_{\rm }\ll E_{f}$.
The Andreev reflection mechanism now requires that the $x$-component of
the momentum of the incoming electron with spin $\uparrow $ is 
$p_{1x}=p_{fx}+(\epsilon+J_{\rm })/v_{fx}$, which pairs
with an electron with momentum 
$p_{2x}=-p_{fx}+(\epsilon+J_{\rm })/v_{fx}$. This leads
to a Cooper pair momentum of $2(\epsilon +J_{\rm })/v_{fx}$.
On the contrary, if the incoming electron has spin $\downarrow$, it pairs with
a spin $\uparrow $ electron, and the sign of $J_{\rm }$ reverses in the
above expressions, leading to a center of mass momentum of 
$2(\epsilon -J_{\rm })/v_{fx}$. Denoting $Q_J=2J_{\rm }/\hbar v_{fx}$
and $Q_\epsilon = 2\epsilon/\hbar v_{fx}$,
we see that instead of a singlet Cooper pair
$(\uparrow \downarrow - \downarrow \uparrow ) e^{iQ_\epsilon x}$ as in the
case of a non-magnetic normal metal, a Cooper pair of the form
\be
(\uparrow \downarrow e^{iQ_Jx} - \downarrow \uparrow e^{-i Q_Jx}) e^{iQ_\epsilon x}
\ee
is created. 
This is exactly the FFLO type of pair, as it was considered in the 
original work. Only it is now induced as a proximity amplitude instead 
of a bulk phase.
It can be decomposed into a spin singlet
$(\uparrow \downarrow - \downarrow \uparrow ) \equiv
|S=0,S_z=0\rangle$ and a spin triplet 
$(\uparrow \downarrow + \downarrow \uparrow ) \equiv
|S=1,S_z=0\rangle$ (where $S$ is the total spin of the pair and $S_z$ its projection on the $z$-axis), leading to (see figure \ref{SingTrip1})
\be
\left[\cos(Q_Jx) |0,0\rangle + i \sin(Q_Jx) |1,0\rangle \right]e^{iQ_\epsilon x}.
\ee
The proximity amplitudes in the ferromagnet are such mixtures.
The triplet component also penetrates over a coherence length into the
singlet superconductor, thus leading to singlet-triplet mixing
in the superconductor close to the interface. 
The excess momentum during the Andreev reflection process is transferred partially to the interface and partially to the condensate (with the rate determined by the continuity equation for total charge conservation). 

In the diffusive limit the wave vector is given by $Q_\pm=(i+1)\sqrt{(\epsilon \pm J)/\hbar D}$ (where a small positive imaginary part of $\epsilon $ determines the root), i.e. the pair amplitudes oscillate and decay on the same length scale.

\subsection{Josephson effect}

If two superconducting materials are brought in contact via a sufficiently narrow non-superconducting region (or via a weak link), the macroscopic wave functions of the two superconductors may overlap. Under the condition that a phase difference exists between the superconductors, a supercurrent may flow between them directly through the non-superconducting region even under zero applied voltage. This effect was discovered by Josephson in 1962 \cite{Josephson62,Josephson64} and first observed by Anderson and Rowell in 1963 \cite{Anderson63}. 
The Josephson effect is a hallmark of macroscopic quantum coherence, proving the macroscopic character of the pair wave function as originally predicted by Ginzburg and Landau \cite{Ginzburg50}.
Originally, the effect described the zero resistance tunneling of Cooper pairs between two superconductors through an insulating barrier, an SIS junction. 
The Josephson current from superconductor 2 to superconductor 1 (i.e. electrons are flowing from 1 to 2), $I$,
is given by the set of Josephson relations
\numparts
\bea
I &=& F(\Delta \chi) , \quad F(\Delta \chi+2\pi)=F(\Delta \chi), \label{j1}\\
\Delta \chi &=& \chi_2-\chi_1-\frac{q^\ast}{\hbar } \int_1^2 \Av \cdot d\rv ,\label{j2}\\
\frac{\partial \Delta \chi }{\partial t}&=&-\frac{q^\ast }{\hbar } (\iPhi_2-\iPhi_1)
\label{j3}
\eea
\endnumparts
where
$q^\ast=2e=-2|e|$, 
$\chi_1$ and $\chi_2$ are the phases of the superconducting condensate wave function on either side of the contact,
$\Av$ is the electromagnetic vector potential, and $q^\ast\iPhi $ is the pair electrochemical potential ($\iPhi_2-\iPhi_1=V$ is the voltage across the junction). The function $F$ describes the current-phase relation.
The critical Josephson currents in positive and negative flow directions are defined via
\bea
I_{c+} &=&\max_{\Delta \chi} F(\Delta \chi), \qquad I_{c-} =-\min_{\Delta \chi} F(\Delta \chi) \label{j4} .
\eea
For non-magnetic barriers in the {\it tunneling limit}, $I_{c+} =I_{c-} =I_c$ and $I=I_c\sin (\Delta \chi)$, where $I_c $ is positive.
The relations (\ref{j1})-(\ref{j3}) are invariant against gauge transformations $\Av'=\Av+\grad \zeta $, $\iPhi'=\iPhi -\partial_t \zeta $, $\chi'=\chi+\frac{q^\ast}{\hbar }\zeta$. 
The function $F(\Delta \chi )$ may not be single valued, for example for the case of a weak link multiple solutions can exist, and a hysteresis when sweeping $\Delta \chi $ may appear.
If a flux $\Phi$ penetrates the junction, a characteristic Fraunhofer pattern appears when $I_c$ is plotted vs. the flux.
A multitude of interesting effects and devices based on the Josephson relations exist, and the reader is referred to review articles, e.g.
\cite{KulikYanson72,Likharev78,Barone82,Golubov04,Posazhennikova09}. 

Here we are mainly concerned with the dc Josephson effect for the case when the non-superconducting material is an extended region of normal metal or metallic ferromagnet. With the help of the proximity effect via Andreev reflections, the Josephson effect can also occur in such devices. The current passing through the normal region is carried by the Andreev states, which build bound states below the superconducting gaps within the normal conducting region. The number and distribution of the bound states depend on details such as interface transmission, mean free path, and length of the normal metal. In general, there is a characteristic energy, the Thouless energy \cite{Thouless72}, given by $\hbar v_f/L$ for the clean limit, and by $\hbar D/L^2$ for the diffusive limit. In normal metals coupled to conventional superconductors, there will be a low-energy gap in the spectrum of Andreev states, which for long junctions approximately scales with the Thouless energy and the transmission probabilities between the superconductors and the normal metal (possibly further reduced by inelastic scattering processes). This is the so-called minigap, found first by McMillan \cite{McMillan68}. This minigap can be probed by scanning tunneling microscopy \cite{Gueron96,Pannetier00}. When a supercurrent flows across the junction, or when an external magnetic field is applied, the minigap is reduced and eventually closes.

In the ballistic limit, each Andreev bound state with energy $E_{\rm b.s.}$ disperses as function of phase difference $\Delta \chi$, and the current is given by $(q^\ast/\hbar) \partial E_{\rm b.s.}/\partial \Delta \chi $. Apart from the current carried by the Andreev bound states, there is also a contribution from continuum states above the gap.

The current-phase relation $F(\Delta \chi)$ is in general strongly dependent on the details of the Josephson junction. Sinusoidal behavior is only observed under special circumstances, like for example in the tunneling limit. 
The critical Josephson current depends strongly on the length of the junction.
For more details I refer the reader to the above mentioned review articles.

\section{Symmetry classification of Cooper pairs in superconductors}
\subsection{Consequences of Pauli principle and Fermi statistics}

Due to the Fermi statistics the pair correlation function, or the Gor'kov ``anomalous Green function'' \cite{Gorkov59}, fulfills fundamental symmetries following from fermionic anti-commutation relations of the field operators. 
For fermions, the definition of the anomalous Green function \cite{AGD,Landau} 
in Matsubara representation \cite{Matsubara55} is given by
\ber
&&-F^{\rm M}_{\alpha \beta} (\vec{r}_1,\vec{r}_2;\tau )=\langle \mbox{\bf T}_{\tau} \left\{\Psi_{\alpha }(\vec{r}_1,-i\tau) \Psi_{\beta }(\vec{r}_2,0)\right\}\rangle
\nonumber
\\
&&\equiv \theta(\tau) \langle \Psi_{\alpha }(\vec{r}_1,-i\tau) \Psi_{\beta }(\vec{r}_2,0)\rangle
-\theta(-\tau) \langle \Psi_{\beta }(\vec{r}_2,0) \Psi_{\alpha }(\vec{r}_1,-i\tau) \rangle 
\eer
where the $\theta $-function is defined as $\theta(\tau)=1$ for $\tau\ge 0$ and 0 for $\tau<0$,
and where $|\tau| < |\hbar/k_{\rm B}T|$.
The function $F^{\rm M}$ fulfills the fundamental identity
\ber
F^{\rm M}_{\alpha \beta} (\vec{r}_1, \vec{r}_2;\tau )&=&
-F^{\rm M}_{\beta \alpha } (\vec{r}_2, \vec{r}_1;-\tau ).
\eer
which expresses the fermionic nature of the constituents of a pair.
Going over to relative and center of mass coordinates, $\vec{r}=\vec{r}_1-\vec{r}_2$, $\vec{R}=(\vec{r}_1+\vec{r}_2)/2$, 
the corresponding relation reads 
\ber
F^{\rm M}_{\alpha \beta} (\vec{r}, \tau;\vec{R})&=&
-F^{\rm M}_{\beta \alpha } (-\vec{r}, -\tau;\vec{R}),
\eer
or after Fourier transformation in the relative coordinates $\vec{r}\to\vec{p}$, $\tau\to z$,
\ber
F^{\rm M}_{\alpha \beta} (\vec{p}, z ;\vec{R})&=&
-F^{\rm M}_{\beta \alpha } (-\vec{p}, -z;\vec{R}).
\eer
Kubo-Martin-Schwinger (KMS) boundary conditions \cite{Kubo57,Martin59} allow for anti-periodicity in 
$\tau $ with period $\hbar/k_{\rm B}T$:  
$F^{\rm M}_{\alpha \beta} (\vec{r}, \tau-i\hbar/k_{\rm B}T;\vec{R})=-F^{\rm M}_{\alpha \beta} (\vec{r}, \tau;\vec{R})$. 
This leads to discrete Matsubara energies $z=\epsilon_n=(2n+1)\pi k_{\rm B}T $, 
\ber
F^{\rm M}_{\alpha \beta} (\vec{p}, \epsilon_n ;\vec{R})&=&
-F^{\rm M}_{\beta \alpha } (-\vec{p}, -\epsilon_n ;\vec{R}).
\eer
We consider now spin-singlet (spin-antisymmetric) and spin-triplet (spin-symmetric) components:
\ber
F^{\rm M}_{s/t} = \frac{1}{2}\left( F^{\rm M}_{\alpha \beta}\mp F^{\rm M}_{\beta \alpha}\right)
\eer
(suppressing the spin indices hereafter),
which fulfill
\numparts
\ber
F^{\rm M}_{s} (\vec{p}, \epsilon_n ;\vec{R})&=& \quad F^{\rm M}_{s} (-\vec{p}, -\epsilon_n ;\vec{R}) \\
F^{\rm M}_{t} (\vec{p}, \epsilon_n ;\vec{R})&=& -F^{\rm M}_{t} (-\vec{p}, -\epsilon_n ;\vec{R}).
\eer
\endnumparts
Each of these we can classify according to parity. We define even-parity and odd-parity
functions
\ber
F^{\rm M}_{\pm} (\epsilon_n ;\vec{R})&=& 
\frac{1}{2}\left(
F^{\rm M} (\vec{p}, \epsilon_n ;\vec{R}) \pm 
F^{\rm M} (-\vec{p}, \epsilon_n ;\vec{R})
\right)
\eer
(we suppress the momentum argument hereafter)
and find
\numparts
\ber
F^{\rm M}_{s+} ( \epsilon_n ;\vec{R})&=& \quad F^{\rm M}_{s+} ( -\epsilon_n ;\vec{R}) \\
F^{\rm M}_{s-} ( \epsilon_n ;\vec{R})&=& -F^{\rm M}_{s-} ( -\epsilon_n ;\vec{R}) \\
F^{\rm M}_{t+} ( \epsilon_n ;\vec{R})&=& -F^{\rm M}_{t+} ( -\epsilon_n ;\vec{R})\\
F^{\rm M}_{t-} ( \epsilon_n ;\vec{R})&=& \quad F^{\rm M}_{t-} ( -\epsilon_n ;\vec{R}).
\eer
\endnumparts
Thus, $F^{\rm M}_{s+} ( \epsilon_n ;\vec{R})$ and $F^{\rm M}_{t-} ( \epsilon_n ;\vec{R})$ are even in Matsubara frequency, whereas $F^{\rm M}_{s-} ( \epsilon_n ;\vec{R})$ and $F^{\rm M}_{t+} ( \epsilon_n ;\vec{R})$ are odd.

After it was realized that in diffusive heterostructures with ferromagnets the type $F^{\rm M}_{t+} (\epsilon_n ;\vec{R})$ appears, which is according to the above a 
so-called odd-frequency pair amplitude (for a review see \cite{Bergeret05a}), a systematic classification according to symmetry of pairing correlations was undertaken along the lines explained above. 
Pair amplitudes are classified into four types according to their behavior with respect to frequency, momentum (parity), and spin (see Fig. \ref{Symmetry}) \cite{Eschrig07}:
\begin{figure}
\begin{center}
\includegraphics[width=0.7\linewidth]{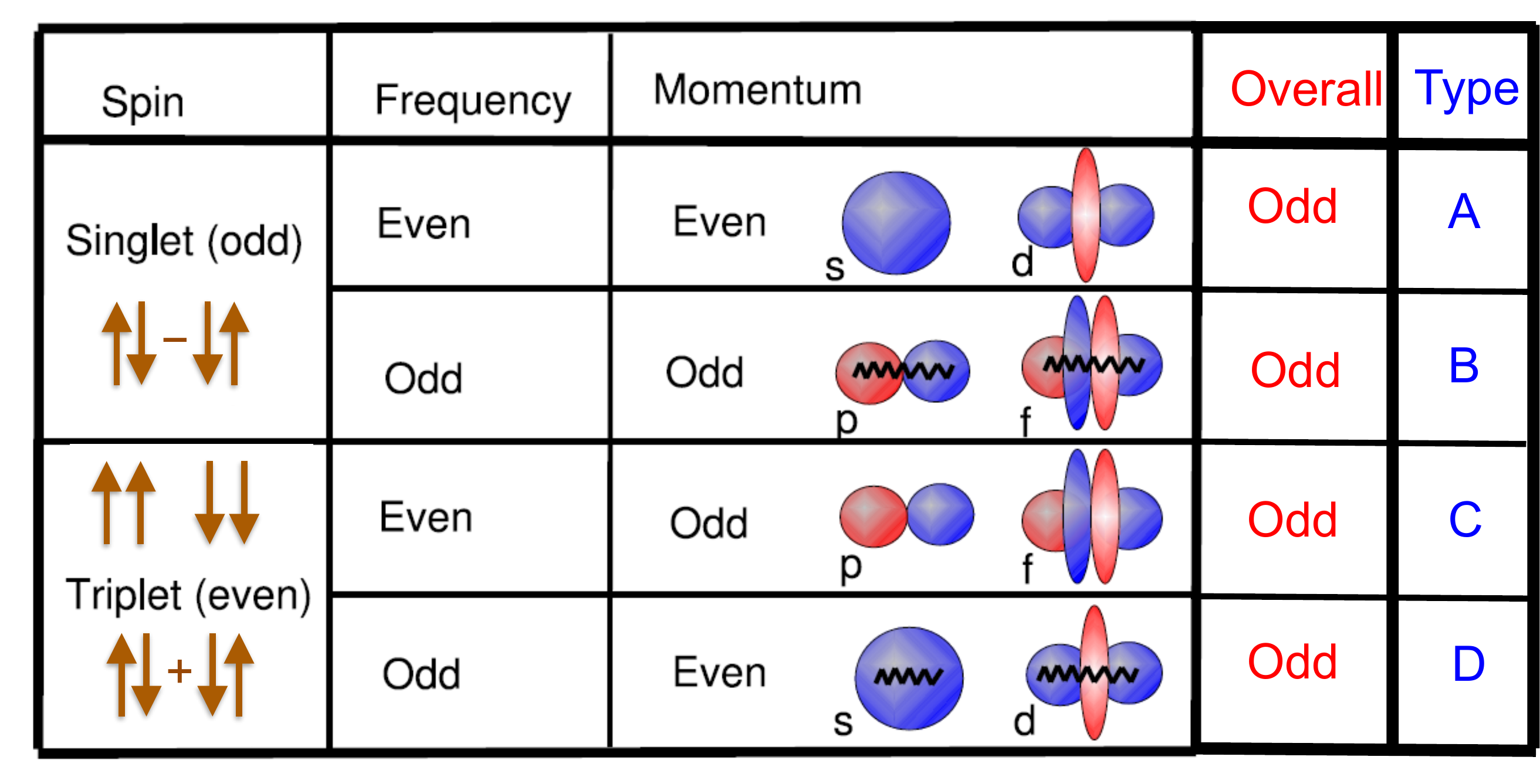}
\end{center}
\caption{
Symmetry classification of Cooper pairs in inhomogeneous systems. 
The wavy lines symbolize odd-frequency symmetry of the pair amplitude.
After \cite{Eschrig07,Eschrig08}.
With kind permission from Nature Publishing Group, and from
Springer Science and Business Media.
\label{Symmetry}
}
\end{figure}
\begin{eqnarray}
\mbox{Type A:}&\quad &\mbox{spin singlet, even frequency, even parity} \nonumber \\
\mbox{Type B:}&\quad &\mbox{spin singlet, odd frequency, odd parity} \nonumber \\
\mbox{Type C:}&\quad &\mbox{spin triplet, even frequency, odd parity} \nonumber \\
\mbox{Type D:}&\quad &\mbox{spin triplet, odd frequency, even parity} .\nonumber 
\end{eqnarray}
An identical classification scheme was independently proposed in Refs. \cite{Tanaka07,Tanaka07a}.
The four symmetry states above exhaust all possibilities compatible with Fermi statistics and the Pauli exclusion principle.\footnote{If more discrete degrees of freedom than the ones considered here are relevant (e.g. orbital) then the classification must be extended accordingly.}
The usual spin singlet, $s$-wave Bardeen-Cooper-Schrieffer (BCS) superconductor \cite{Bardeen57} is of type A, while the spin triplet, $p$-wave superfluid formed in $\,^3$He \cite{Leggett75,Vollhardt90} is of type C. Type D was first considered by Berezinski\u{i} \cite{Berezinskii74} in connection with early research on superfluid $\,^3$He. Finally, type B was introduced in connection with unconventional superconductors by Balatsky, Abrahams and others \cite{Balatsky92,Abrahams95,Dahal09}.

Suggestions for realization of Type B superconductivity include generalized one-dimensional $t-J$ and Hubbard models \cite{Balatsky93,Shigeta09}, two-channel Kondo models and Kondo lattices \cite{Emery92,Emery93,Coleman93,Jarrell97}, and quantum critical points \cite{Fuseya03}.
Realizations suggested for the Type D state include disordered two-dimensional electron fluids in semiconductors \cite{Kirkpatrick91,Belitz92,Belitz99},
triangular antiferromagnets \cite{Vojta99}, composite spin and orbital triplet superconductivity in two-channel Anderson lattices \cite{Anders02}, Hund's coupled pairing in double orbital Hubbard models \cite{Sakai04}, strong-coupling triplet superconductivity in Holstein Hubbard models \cite{Kusunose11a}, and one-dimensional systems with strong charge fluctuations \cite{Shigeta11}.

\subsection{Odd-frequency pairing amplitudes}

To date an odd-frequency superconductor has not been found in nature. However,
in contrast to all cases discussed above, where the appearance of an order parameter due to spontaneous symmetry breaking (global phase symmetry in superconductors) was in the focus, in this review we are interested in the case of explicit symmetry breaking taking place locally in some spatial region, e.g. near interfaces, line defects, or inclusions. For this case
an overwhelming number of experiments support the picture of the existence of proximity induced pairing states of type D. In its pure form (i.e. without additional components of different symmetry), it has been first predicted theoretically for a superconductor-ferromagnet proximity structure with a spiral inhomogeneous magnetization near the interface in Ref. \cite{Bergeret01}. Previous work had predicted the appearance of similar odd-frequency triplet anomalous functions in a model of coexistence of a superconducting phase and a helical ordering of localized spins, motivated by experiments on the superconductor ErRh$_4$B$_4$ \cite{Bulaevskii80}.

In fact, the Pauli principle requires odd-frequency amplitudes to be present in any inhomogeneous superconducting state, not necessarily spin-polarized \cite{Eschrig07}. For example, the case of a normal metal coupled to a superconductor involves pair amplitudes of  Type B.
Thus, odd-frequency amplitudes appear in heterostructures naturally due to breaking of translational symmetry at interfaces (similarly as 
Rashba spin-orbit coupling 
appears
at interfaces in semiconductor heterostructures). They have been present in all 
treatments of inhomogeneous superconductivity for a long time, however were not explicitly named as such. 
Similarly, pair amplitudes of Type D appear as soon as spin rotational symmetry is broken, e.g. by an external magnetic field via the Zeeman coupling. If both spin rotational symmetry and parity are broken, e.g. at an interface between a superconductor and a ferromagnet, all four types of pair amplitudes are generated at the interface \cite{Eschrig07,Eschrig08}.
Similar conclusions have been reached in Refs. \cite{Tanaka07,Tanaka07a}. 
The important issue in discussing odd-frequency amplitudes in Ref. \cite{Bergeret01} is not the presence of Type D correlations per se, but their only presence. Thus, in diffusive structures those $s$-wave odd-frequency triplet pair amplitudes have a definite symmetry, and thus are of special interest for fundamental research. In regions where all amplitudes are mixed, simply all symmetries are absent, and a symmetry classification is not very useful. However, once such regions are coupled to reservoirs in which symmetries are asymptotically (far away from the spatial regions where the symmetry breaking takes place) re-established, then it is useful to discuss processes at interfaces in the light of those asymptotic symmetries.

To date it remains a great challenge to find direct experimental verification of the odd-frequency symmetry. There have been a number of proposals for indirect experimental verification via tunneling density of states studies, by inducing odd-frequency triplet superconducting correlations in a normal metal \cite{Linder09,Linder10a,Cottet11}.

The odd-frequency pairing aspect in superconductor ferromagnet heterostructures was reviewed in Ref. \cite{Bergeret05a}.
A recent review dealing with odd-frequency pairing states and their relation to topological edge states can be found in Ref. \cite{Tanaka12}. Newer developments in finding odd-frequency order parameters include odd-frequency pairing states in multi-band superconductors \cite{Black13,Aperis15}, at the surface of topological insulators \cite{Black12},
in a controllable double-quantum dot system \cite{Sothmann14},
and in two-particle Bose-Einstein condensates \cite{Balatsky14}, as well as odd-frequency density wave states \cite{Pivovarov01,Kedem15}.

Odd-frequency symmetry requires that the equal time pair correlator vanishes, i.e. electrons in a pair avoid each other in time.
It has been argued that in such a case certain particle-hole symmetries based on a Green function approach are not appropriate and a Lagrangian formalism must be used \cite{Solenov09,Kusunose11}. In particular, these arguments were brought forward for homogeneous odd-frequency states with an odd-frequency pair potential appearing due to spontaneous symmetry breaking.
However, these arguments do not take into account that symmetries following from the 
full many body Hamiltonian of the system are fundamental. Should they be violated in a model Lagrangian approach, for which a corresponding model Hamiltonian cannot be found, then this simply means that the model Lagrangian is not appropriate for the problem in question \cite{Fominov15}. 
It has also been pointed out that spatially homogeneous odd-frequency states might be thermodynamically unstable in reality \cite{Heid95}.

\section{Magnetically active interfaces}
\subsection{Scattering phase delays}
Let us consider the basic quantum mechanical problem of a (quasi)-particle being reflected from an insulating region (which we assume at $x>0$, potential $V$). 
Let us assume the particle has energy $0<E<V$ with $E=(k^2+k_{||}^2)/2m= V+(-\kappa^2+k_{||}^2)/2m$, and thus $\kappa(E)=[2m(V-E)+k_{||}^2]^{1/2}$, $k(E)=[2mE-k_{||}^2]^{1/2}$. It is described by a wave function
$\Psi (x,\vec{r}_{||})=e^{i\vec{k}_{||}\vec{r}_{||}} (e^{ikx}+r\;e^{-ikx})$ at $x<0$ and 
$\Psi (x, \vec{r}_{||})= t e^{i\vec{k}_{||}\vec{r}_{||}} e^{-\kappa x}$ at $x>0$.
The reflection amplitude for such a process is $r=(k-i\kappa)/(k+i\kappa) = e^{-2i\arctan (\kappa/k)}$, where $k$ is the momentum component perpendicular to the interface and $\kappa $ determines the exponential decay of the wave function in the insulating region.
In the limit for large $\kappa $ the phase of the reflected wave is flipped by $\pi$ with respect to the incoming wave. For finite $\kappa $ there is a {\it phase delay} with respect to this, given by
$\phi(E)=\pi-2\arctan (\kappa/k)$.
This phase delay results from the fact that the particle penetrates the insulator over a length scale $\hbar /\kappa$, and it corresponds to a {\it time delay} between the maximum of an incoming wave packet and the corresponding reflected wave packet \cite{Eisenbud54} of $\tau_d=\hbar \partial_E \phi(E)=2m\hbar /(\kappa k)$ (accompanied by a Goos-H\"anchen shift along the interface of $s=2\hbar k_{||}/(\kappa k)$ \cite{Hora60}).

\subsection{Spin-mixing angle and spin-dependent scattering phase shifts}

\begin{figure}
\begin{center}
\includegraphics[width=0.7\linewidth]{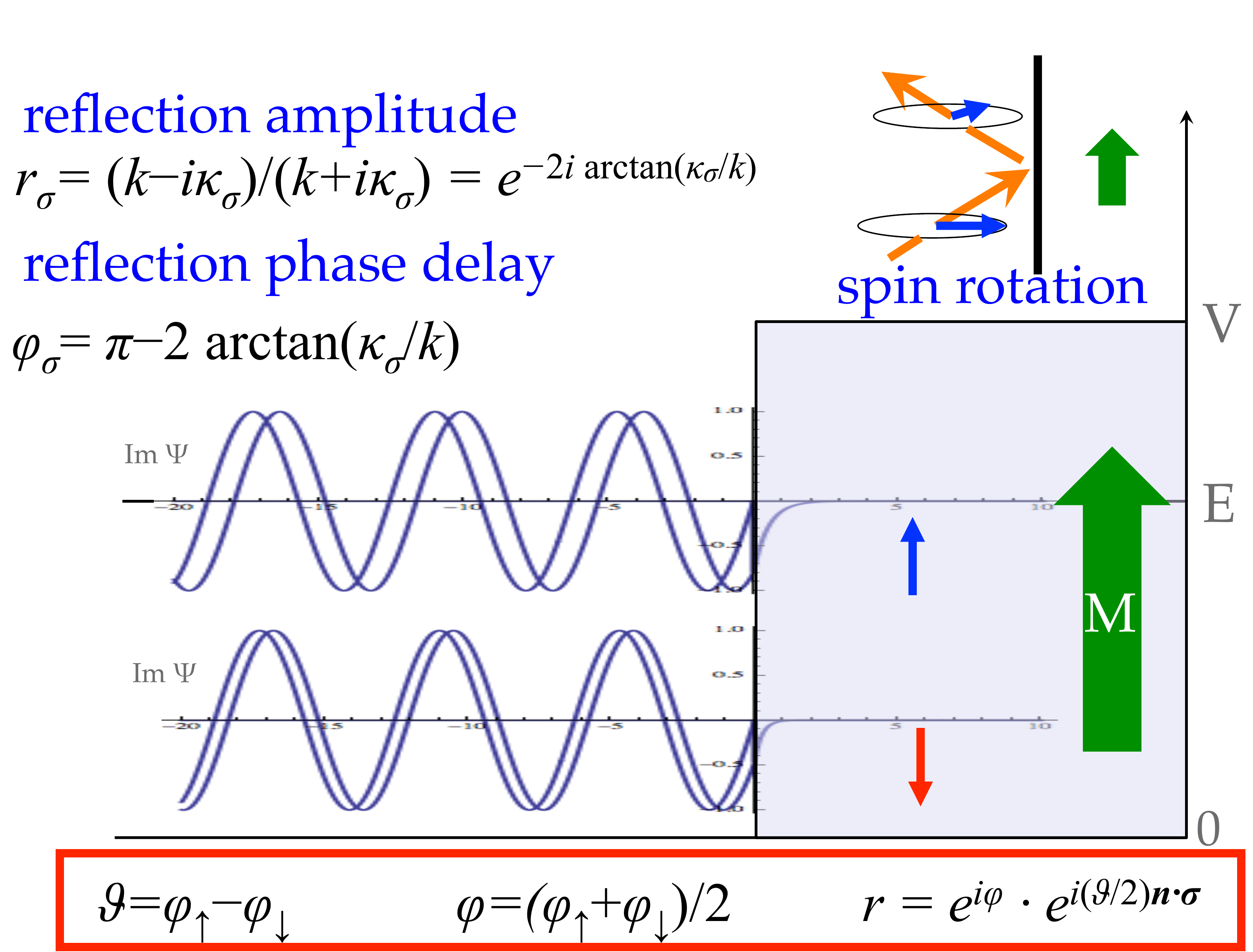}
\end{center}
\caption{When a Bloch wave is reflected from a ferromagnetic insulator, a phase delay appears between the reflected waves and the incoming waves, which differs for both spin directions. As a result, the spin-down reflected wave is delayed with respect to the spin-up reflected wave. If the incoming wave is polarized perpendicular to the ferromagnet's magnetization, then the reflected wave has its spin rotated with respect to the spin of the incoming wave (see top part of the drawing).}
\label{spinmixing}
\end{figure}

We now assume that an exchange field $\vJ$ is present, leading to a contribution to the Hamiltonian given by ${\cal H}_{\rm exch}=-\vJ \cdot \vsigma$. This can also be written as ${\cal H}_{\rm exch}=-\mu \; \vB_{\rm eff} \cdot \vsigma$, with $\vB_{\rm eff}=\vJ/\mu$, and $\mu $ is the (effective) magnetic moment of the charge carriers (negative for free electrons).
Let us now consider a ferromagnetic insulating region, so that the two spin directions $\sigma $ have different reflection amplitudes $r_\sigma =(k-i\kappa_\sigma )/(k+i\kappa_\sigma ) = e^{-2i\arctan (\kappa_\sigma /k)}$, with $\Psi_\sigma \sim e^{-\kappa_\sigma x}$ in the ferromagnetic insulating region (see Figure \ref{spinmixing}). 
As throughout this review, we take the spin quantization axis along the exchange field,
i.e. $\kappa_\sigma (E)=[2m(V-E-\sigma J)+k_{||}^2]^{1/2}$, 
such that the majority Fermi surface is assigned spin projection $\sigma=1$.\footnote{For free electrons (negative magnetic moment) 
this quantization axis would point opposite to the effective magnetic field. For a quantization axis in field direction, spin up and down would simply switch their roles.}
The scattering phase delays now are spin dependent, 
$\phi_\sigma (E)=\pi-2\arctan (\kappa_\sigma /k)$,
and we can define a quantity $\vartheta=\phi_\uparrow-\phi_\downarrow $, which is called {\it spin-mixing angle} \cite{Tokuyasu88} or {\it spin dependent interface scattering phase shift} \cite{Cottet09} in the literature (it is associated with a delay time
$\tau_{\vartheta}=\hbar \partial_E \vartheta(E)=2m\hbar /(\kappa_\uparrow k)-2m\hbar /(\kappa_\downarrow k)$ between the two spin components).
For our example we have $\kappa_\uparrow < \kappa_\downarrow $, and consequently $\vartheta > 0$, $\tau_\vartheta >0$.
Together with the spin-independent averaged phase $\phi=(\phi_\uparrow+\phi_\downarrow)/2 $, we write the reflection amplitude in a more general way as \cite{Tokuyasu88}
\begin{equation}
r(\phi,\vartheta,\hat{\mathbf{n}})=-e^{i\phi} \cdot e^{i\frac{\vartheta }{2}\hat{\mathbf{n}} \cdot \mbfsc{\sigma}}
\end{equation}
with respect to a spin quantization axis $\hat{\mathbf{n}}$. 
The spin-mixing angle also describes the rotation angle of the spin components perpendicular to the axis $\hat{\mathbf{n}}$ under reflection. 
This can be interpreted as a precession that the spins undergo around the axis $\hat{\mathbf{n}}$ as a result of quantum mechanical penetration into the magnetic and insulating region.

\begin{figure}
\includegraphics[width=1.05\linewidth]{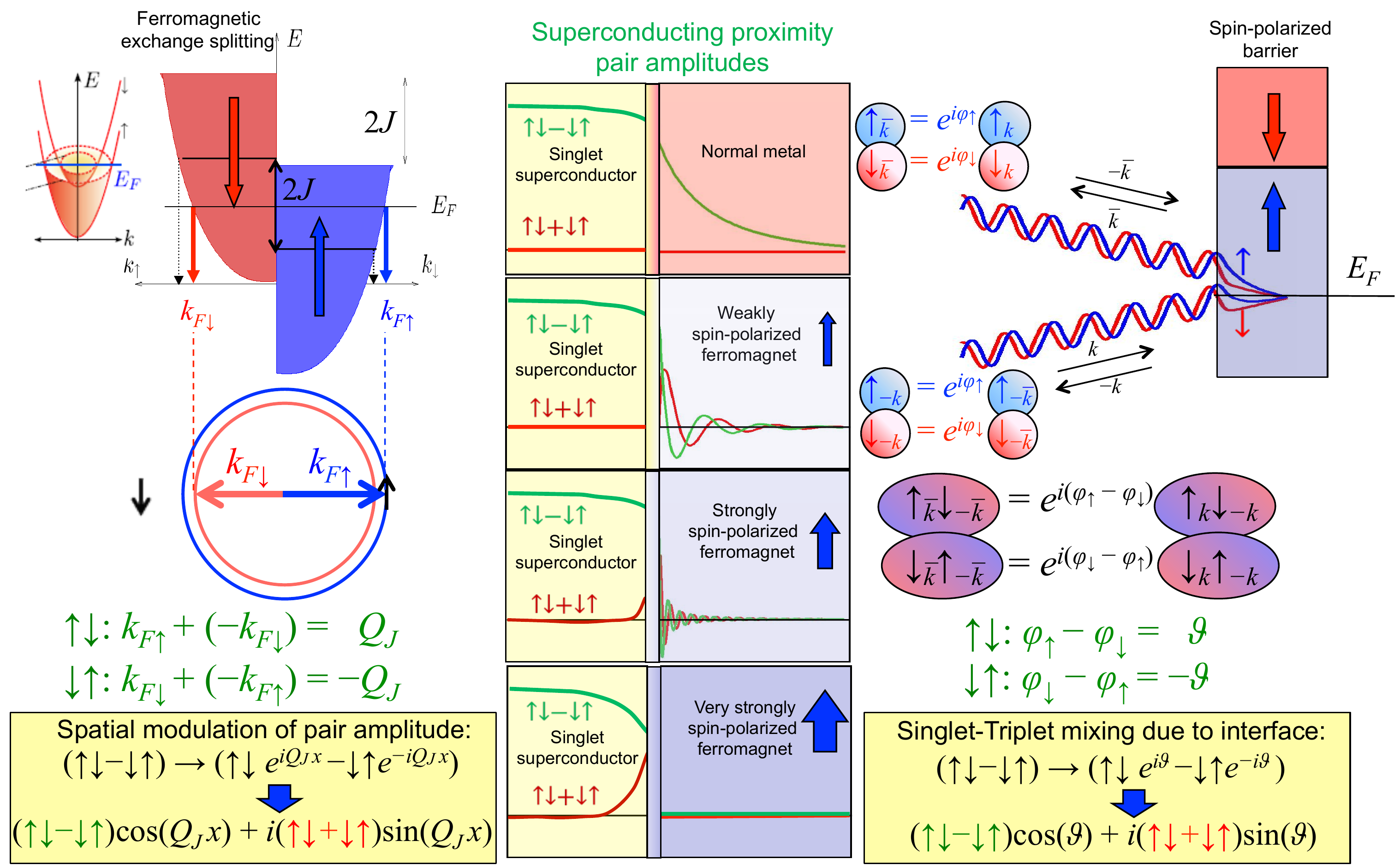}
\caption{Creation mechanisms for triplet pair correlations by direct and inverse proximity effect. See text for explanation. 
Reproduced and modified with permission from \cite{Eschrig11}. 
Copyright (2011) American Institute of Physics.
}
\label{Fig1}
\end{figure}

If one considers the Cooper instability in the presence of an interface with a ferromagnetic insulator, it becomes clear that pairing near the interface will be affected by these spin-mixing angles. In particular, a spin-up electron with momentum  pointing towards the interface will have a relative phase shift with respect to a spin-down electron with momentum pointing away from the interface if the Cooper pair volume overlaps with the interface region. As the Cooper pair's size is determined by the superconducting coherence length, Cooper pairs in a layer that extends a coherence length from the interface into the bulk will feel the spin-mixing phase shifts of the interface. This is shown in Figure \ref{Fig1} on the right. Near the interface, only a singlet-triplet mixed Cooper pair of the form
\ber
(\uparrow \downarrow e^{i\vartheta} - \downarrow \uparrow e^{-i \vartheta})\equiv
\cos(\vartheta) |0,0\rangle + i \sin(\vartheta) |1,0\rangle 
\label{mixing}
\eer
can be present. According to this, we can consider the parameter $\vartheta $ also as the parameter that governs the {\it degree of singlet-triplet mixing} at a spin-active interface.
As seen in Figure \ref{Fig1}, there is a certain analogy between the creation of FFLO correlations in a ferromagnetic metal and the creation of singlet-triplet mixtures near a magnetically active interface in a superconductor. The role of the phase $Q_Jx$ in the former is played by the spin-mixing angle $\vartheta $ in the latter. As demonstrated in the middle column of Figure \ref{Fig1}, with increasing spin polarization of the ferromagnet in a superconductor-ferromagnet heterostructure, a shift from the creation of FFLO correlations in the ferromagnet to the creation of singlet-triplet mixtures in the superconductor takes place. This is because for very weak spin polarizations, the spin-mixing angle of the interface usually is of similarly small order and should be neglected in a consistent expansion in small parameters of the theory. For strong spin-polarization it is important and must be taken into account. The FFLO correlations in the ferromagnet show an inverse behavior: they are important for weak spin polarizations, however are restricted to atomically small distances from the interface for strongly spin polarized ferromagnets.

The presence of spin-mixing angles has many consequences. For example, it leads to Andreev bound states at the interface, as predicted theoretically \cite{Fogelstrom00,Eschrig03,Krawiec04,Annett06,Lofwander10,Metalidis10}, and verified experimentally \cite{Huebler12}. It also is responsible for giant thermoelectric effects in non-local setups \cite{Machon13,Machon14} and is the main ingredient for creating triplet supercurrents in strongly spin-polarized ferromagnets \cite{Eschrig03,Eschrig08}. It also 
crucially affects point contact spectra \cite{Perez04,Kopu04,Lofwander10,Grein10,Piano11,Kupferschmidt11,Wilken12,Yates13}. Finally, it is the main cause of the inverse proximity effect and the magnetic proximity effect in strongly spin polarized hybrid structures \cite{Eschrig03,Eschrig08,Grein13,Flokstra15}.

\subsection{Scattering matrix and induced triplet correlations}

The interface scattering matrix connects incoming Bloch waves with outgoing Bloch waves at an interface between two itinerant electron materials (metals, half metals, itinerant ferromagnets), or at a surface of such a material with an insulator.
Relevant for transport are Bloch waves with energy close to the Fermi energy and momentum close to the Fermi momentum, in which case the Bloch waves are assumed to describe quasiparticle excitations. ``Incoming'' and ``outgoing'' refers to the projection of the quasiparticle's Fermi velocity on the surface normal. We distinguish between electron-like and hole-like quasiparticles. 
For electron-like quasiparticles the projection of the group velocity on the Fermi momentum is positive, for hole-like quasiparticles it is negative. This means that electron-like and hole-like quasiparticles associated with the same Fermi momentum have opposite group velocity projections on the surface normal and consequently have different scattering matrices. For atomically ordered interfaces
the crystal momentum component parallel to the interface, $\vp_{||}$, is conserved, and then the relation between hole (h) and electron (e) like scattering matrices is
$S^\dagger_{\rm h}(\vp_{||})= S_{ \rm e}^\ast (-\vp_{||})$.

In relation to superconductivity, it is sufficient to consider the normal state scattering matrix for quasiparticles at Fermi momentum and Fermi energy. This results from a systematic classification of all terms during a perturbation expansion in the small phase space volume associated with quasiparticles, which leads to quasiclassical theory of superconductivity \cite{Serene83,Sauls93}.
In terms of the normal-state scattering matrix, 
\begin{equation}
\label{scatt00}
\hat {\bf S}= \left( \begin{array}{cc} \hat R_1 & \quad \hat T_1\\ \hat T_2& -\hat R_2 \end{array} \right)
\end{equation}
we can study the superconducting amplitudes induced by the presence of a singlet pair potential
on either side of the interface.
To simplify algebra and to gain some intuition we limit ourselves to the case of small pair amplitudes, in which case one can linearize the boundary conditions in the pair amplitudes.
In our notation, $\hat R_1$ describes reflection into the superconductor, and $\hat R_2$ describes reflection into the ferromagnet. All reflection and transmission parameters in Eq. (\ref{scatt00}) are 2x2 spin matrices.
We consider the ballistic case, for which the reflected and transmitted pair amplitudes $\hat {f}^{\rm SC}_{\rm out}$, $\hat {f}^{\rm FM}_{\rm out} $ are given in terms of the incoming singlet pair amplitude $\hat {f}^{\rm SC}_{\rm in}=f_0 (i\sigma_y)$ by boundary conditions at the interface between the superconductor and the magnetic material, which in linear order read
\begin{eqnarray}
\label{RsR}
\hat {f}^{\rm SC}_{\rm out} &=& f_0 \; \hat R_1 \; (i\hat \sigma_y) \; \hat R_1^\ast \\
\hat {f}^{\rm FM}_{\rm out} &=& f_0 \; \hat T_2  \; (i\hat \sigma_y) \; \hat T_1^\ast .
\label{TsT}
\end{eqnarray}

The simplest case is that of total reflection from an interface with a magnetic insulator. 
In this case we have (choosing the quantization axis appropriately, and with a scalar phase $\psi$)
\begin{equation}
\label{scatt0}
\hat  R_1=
e^{i\psi} \; \left( \begin{array}{cc} e^{\frac{i}{2}\vartheta} &0\\ 0&e^{-\frac{i}{2}\vartheta} 
\end{array}
\right).
\end{equation}
The reflected amplitudes for this case follow from Eq.~(\ref{RsR}) and are given by
\begin{eqnarray}
\hat {f}^{\rm SC}_{\rm out}&=& 
\left[ f_{0}\cos(\vartheta ) + i f_{0} \sin (\vartheta ) \hat \sigma_z \right] i \hat \sigma_y .
\end{eqnarray}
This justifies the expression obtained by the simple arguments leading to Eq. (\ref{mixing}).

Next, we consider the case of arbitrarily large pair amplitudes in a ballistic superconductor with spatially constant pair potential. This is justified for sufficiently small spin-mixing angles $\vartheta $.
Incoming and reflected directions are parameterized by the polar angle $\theta_{\bvpfsc} $, measured from the surface normal. The cosine of this angle is called $\mu $.
Defining even parity (symmetric) and odd parity (antisymmetric) functions with respect to the propagation direction, $f^a=[f(\mu)-f(-\mu )]/2$ and $f^s=[f(\mu)+f(-\mu )]/2$, we obtain all four possible symmetry components as summarized in Table \ref{Table}. 
It can be seen that the corrections to the singlet amplitudes are $\sim \sin^2 (\vartheta/2)$. Thus,
to linear order in $\vartheta $ the pair potential stays unaffected. With increasing $\vartheta $ the singlet pair potential is reduced, leading to corrections to the expressions in Table \ref{Table}.
\begin{table}[t]
\begin{center}
\begin{tabular}{|r|c|c|}
\hline
\hline
symmetry & even frequency & odd frequency \\
\hline
even parity&
$\begin{array}{ll}
{
\small
\mbox{Type A, singlet} |0,0\rangle }\\
 f^{s}_s= 
\frac{\pi \Omega_n \Delta \left(1-\sin^2 \!\frac{\vartheta}{2}\right)}{\Omega_n^2-|\Delta |^2 \sin^2 \!\frac{\vartheta}{2} } 
\end{array}$
& 
$\begin{array}{ll}
{
\small
\mbox{Type D, triplet} |1,0\rangle }\\
 f_{t_0}^s= 
\frac{ i\pi \vepsilon_n \Delta \; \frac{1}{2}\sin \vartheta }{\Omega_n^2-|\Delta |^2 \sin^2 \!\frac{\vartheta}{2}}  
\end{array}$
\\
\hline
odd parity &
$\begin{array}{ll}
{
\small
\mbox{Type C, triplet} |1,0\rangle }\\
f_{t_0}^a= 
\frac{-i\pi 
\Omega_n s_\mu \Delta \; \frac{1}{2}\sin \vartheta }{\Omega_n^2-|\Delta |^2 \sin^2 \!\frac{\vartheta}{2}} 
\end{array}$
&
$\begin{array}{ll}
{
\small
\mbox{Type B, singlet} |0,0\rangle }\\
f_s^{a}= 
\frac{\pi \vepsilon_n s_\mu \Delta \sin^2 \!\frac{\vartheta}{2}
}{\Omega_n^2-|\Delta |^2 \sin^2 \!\frac{\vartheta}{2}} 
\end{array}$
\\
\hline
\hline
\end{tabular}
\end{center}
\caption{Symmetry components of the interface amplitudes 
inside a ballistic superconductor in contact with a ferromagnetic insulator, assuming a constant 
singlet order parameter $\Delta$.
Here, $\Omega_n=\sqrt{|\Delta|^2+\vepsilon_n^2}$, $\vartheta\equiv \vartheta(\mu)$, and 
$s_\mu={\rm sign} (\mu )$ ($\mu$ is the cosine of the impact angle). From Supplementary Material of Ref. \cite{Eschrig08}.}
\label{Table}
\end{table}

\begin{figure}[b]
\begin{center}
\includegraphics[width=0.7\linewidth]{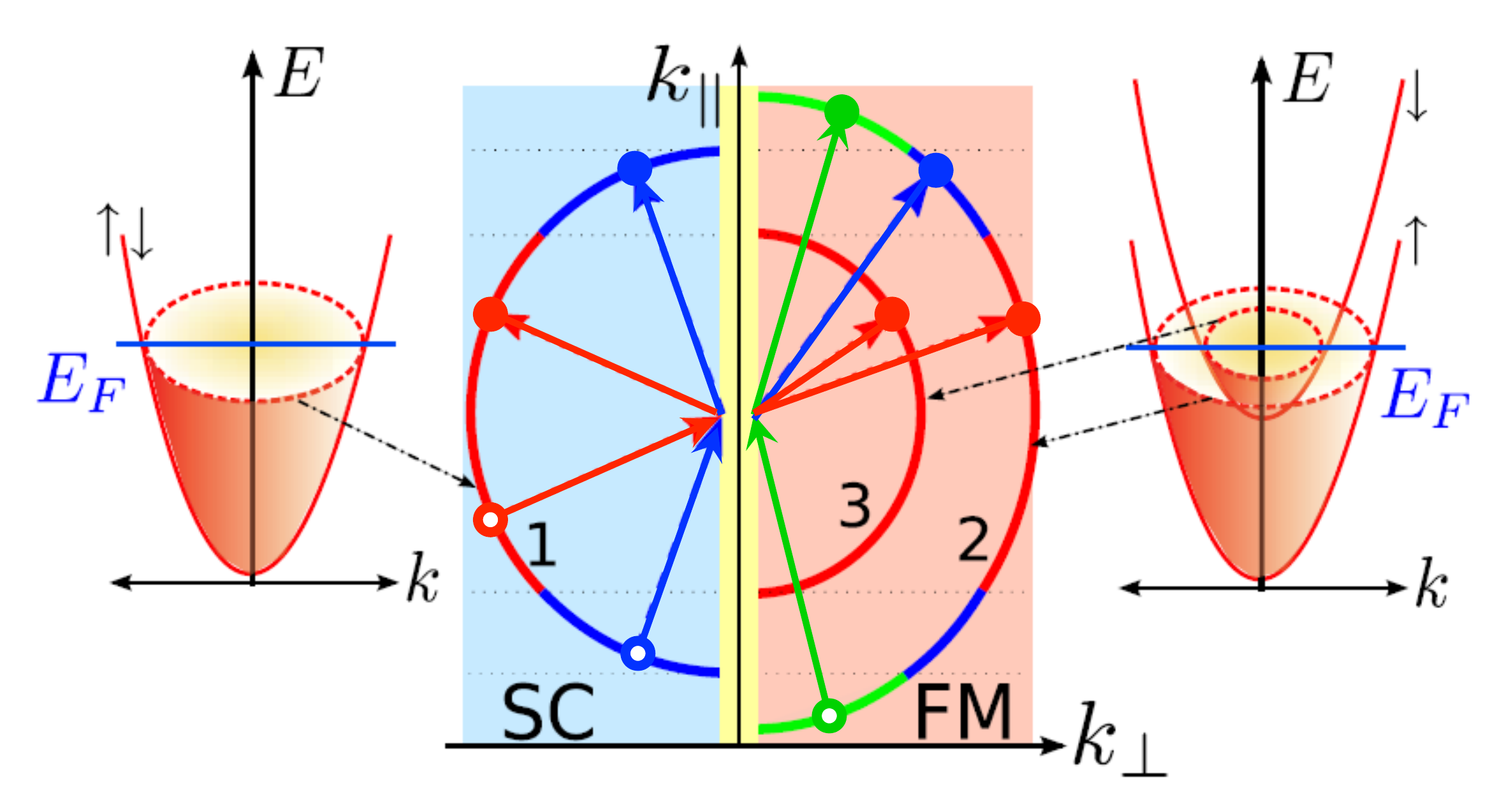}
\end{center}
\caption{
Fermi surface geometry for a superconductor (left)-ferromagnet (right) interface.  Various types of scattering events include total reflection (green), involvement of only one (blue) or of both spin bands (red).
After \cite{Grein09}, Copyright (2009) by the American Physical Society.
}
\label{Boundary2}
\end{figure}

For the case of an interface with a strongly spin-polarized ferromagnet, one has to consider Fermi surface geometry, due to the presence of different Fermi surfaces for the two spin projections in the ferromagnet. Various cases can occur, depending on the scattering channel (parameterized in the ballistic case by $\vp_{||}$, see figure \ref{Boundary2}): 
for example for reflection from the interface on the superconducting side,
there can occur
total reflection from the ferromagnet, total reflection of only one spin component from the ferromagnet (so-called half-metallic channels), or partial reflection for both spin components. Depending on the geometry of the Fermi surface in the superconductor, one, two, or all three cases may occur.

For illustrative purposes we will consider 
a simple model of an interface potential of width $d$ between a superconductor and a ferromagnet, with corresponding (free) electronic dispersions $V_{\rm S}+\vk^2/2m$ in the superconductor, $V_{\rm I}-\sigma J +\vk^2/2m$ in the interface region, and $V_{\rm FM}-\sigma J +\vk^2/2m$ in the ferromagnet.
If the magnetization in the interface region is collinear with that in the ferromagnet, there will be no transmitted pair amplitudes, as for a strongly spin-polarized ferromagnet only $\uparrow\uparrow $ and $\downarrow\downarrow $ pairs are supported, which both are not generated in such a system. There is, however, an induced triplet component on the superconducting side, with spin projection zero on the spin quantization axis. 
\begin{figure}
\begin{overpic}[width=0.495\linewidth]{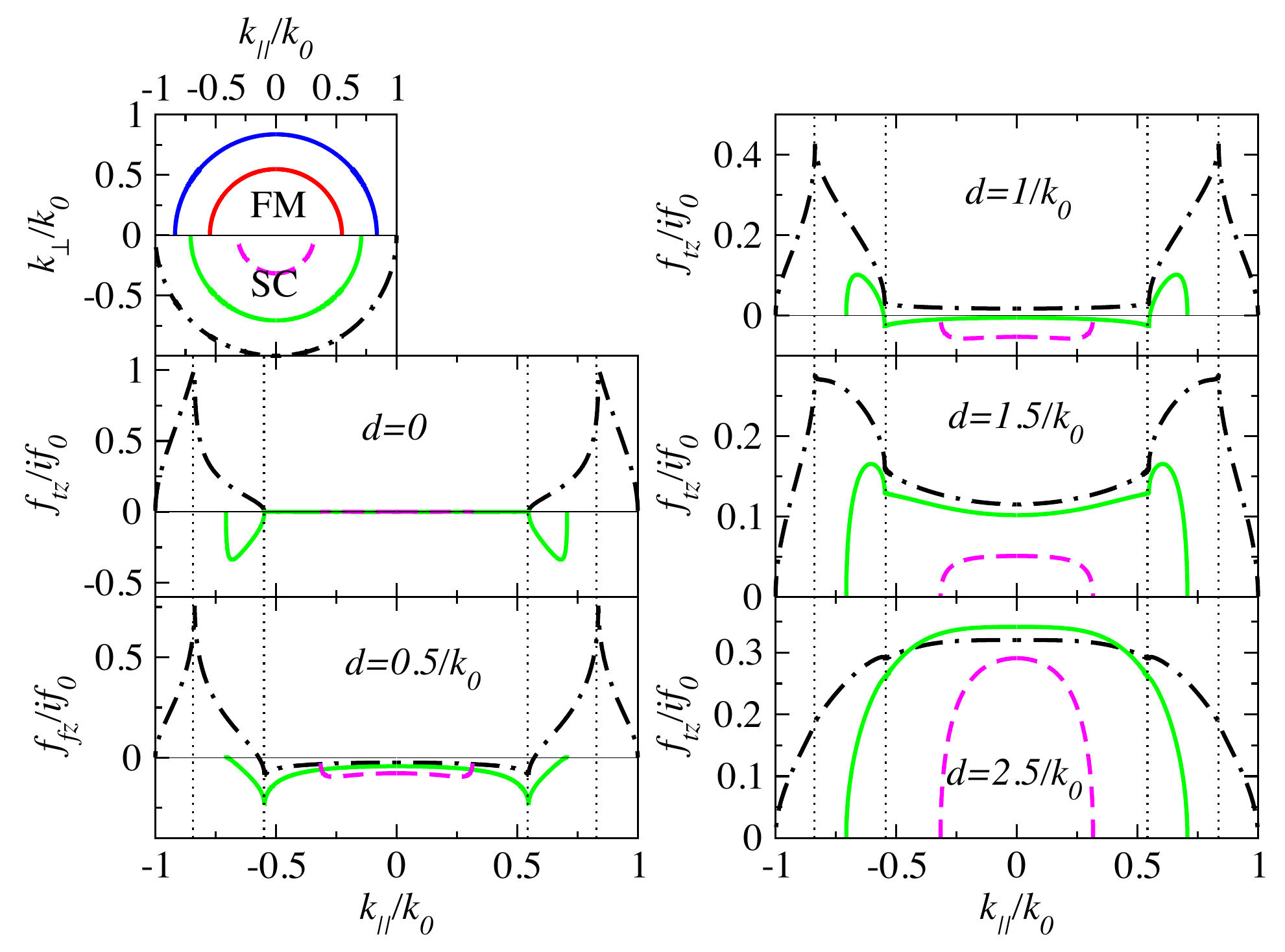}
\put(40,60){\makebox(0,2){(a)}}
\end{overpic}
\begin{overpic}[width=0.495\linewidth]{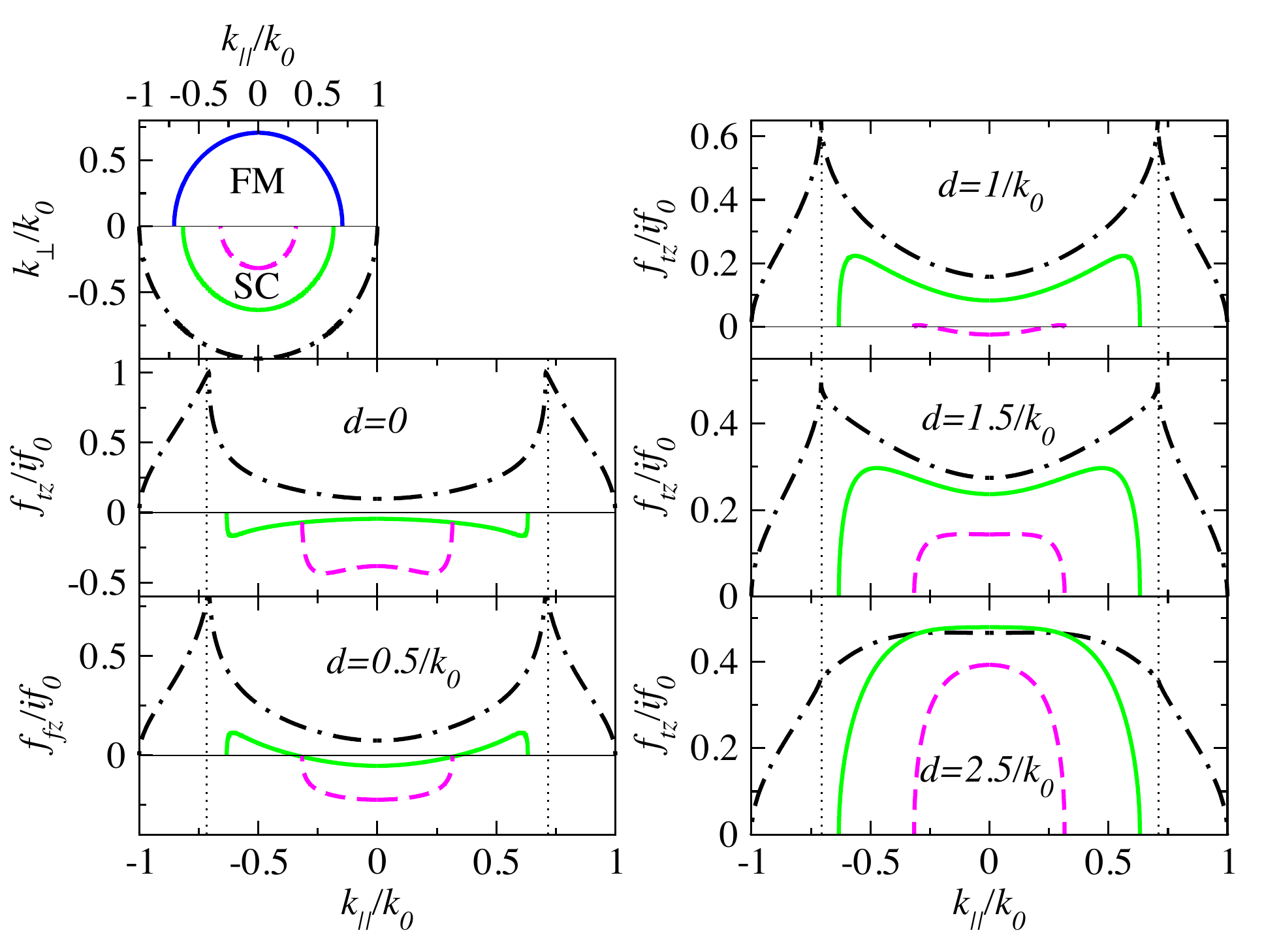}
\put(40,60){\makebox(0,2){(b)}}
\end{overpic}
\caption{
Induced triplet components at a superconductor-ferromagnet interface with an interface barrier of width $d$.  Model free-electron dispersions are: $V_{\rm S}+\vk^2/2m$ (superconductor), $V_{\rm I}-\sigma J +\vk^2/2m$ (barrier), $V_{\rm FM}-\sigma J +\vk^2/2m$ (ferromagnet). The singlet amplitude in the superconductor is $f_0$, the induced triplet component is $f_{tz}$ (and $f_{tz}/f_0$ is purely imaginary).
The Fermi wavevectors for the spin bands in the ferromagnet are fixed in (a) and (b).
The various curves correspond to various Fermi wavevectors in the superconductor (as illustrated in the top left diagram in each panel). Wavevectors are normalized to $k_0=\sqrt{2mE_F}$.
For (a) $V_{\rm I}=1.3$, $V_{\rm FM}=0.5$, $J=0.2$, $V_{\rm S}=0$ (black dashed-dotted), 0.5 (green full line), 0.9 (magenta dashed);
for (b) $V_{\rm I}=1.4$, $V_{\rm FM}=0.8$, $J=0.3$, $V_{\rm S}=0$ (black dashed-dotted), 0.6 (green full line), 0.9 (magenta dashed);
energies are in units of the Fermi energy $E_F$. 
(b) corresponds to a half-metallic ferromagnet, with only one spin projection itinerant (i.e. a Fermi surface exists only for $\sigma=1$).
}
\label{Fig_alpha0}
\end{figure}
It is shown in Figure~\ref{Fig_alpha0}, for the case of a strongly spin polarized ferromagnet in (a), and for the case of a half-metallic ferromagnet (with only spin projection $\sigma=1$ itinerant) in (b). When the barrier is absent, there is no induced triplet component for the regions connecting wave vectors that support Bloch waves for both spin projections. Only wave vectors for which maximally one spin projection in the ferromagnet supports Bloch waves contribute to the induced triplet amplitudes in the superconductor. This shows that for high-transmission contacts half-metallic ferromagnets are most effective for including triplet correlations in the superconductor. If a spin-polarized barrier is present, triplet amplitudes are created also in the barrier region, and their sign depends on the relative size of the Fermi surfaces in the superconductor and in the ferromagnet. In the tunneling limit, the creation of triplet correlations happens mainly in the barrier region, such that they become insensitive to the Fermi surface geometry.

\begin{figure}
\includegraphics[width=0.495\linewidth]{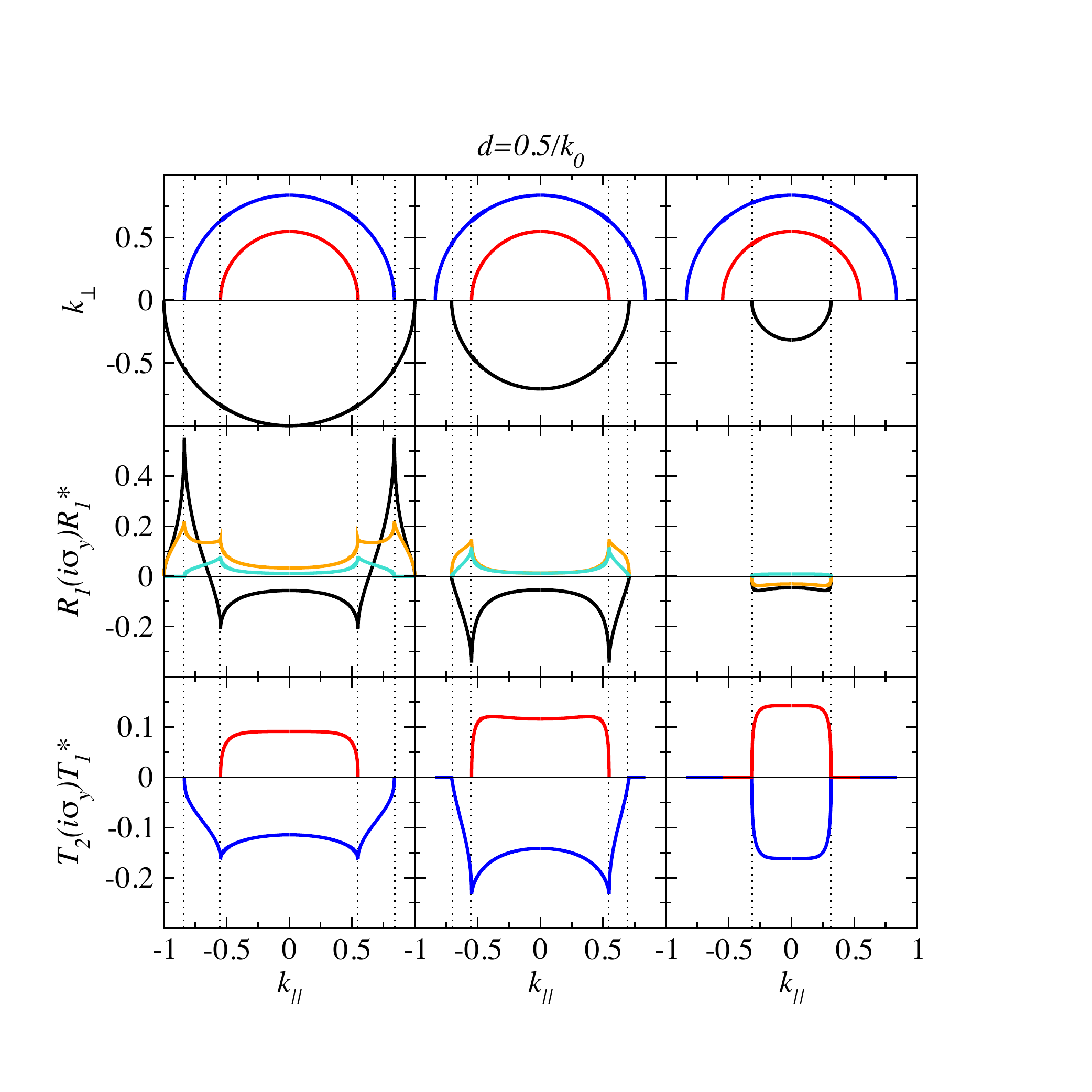}
\includegraphics[width=0.495\linewidth]{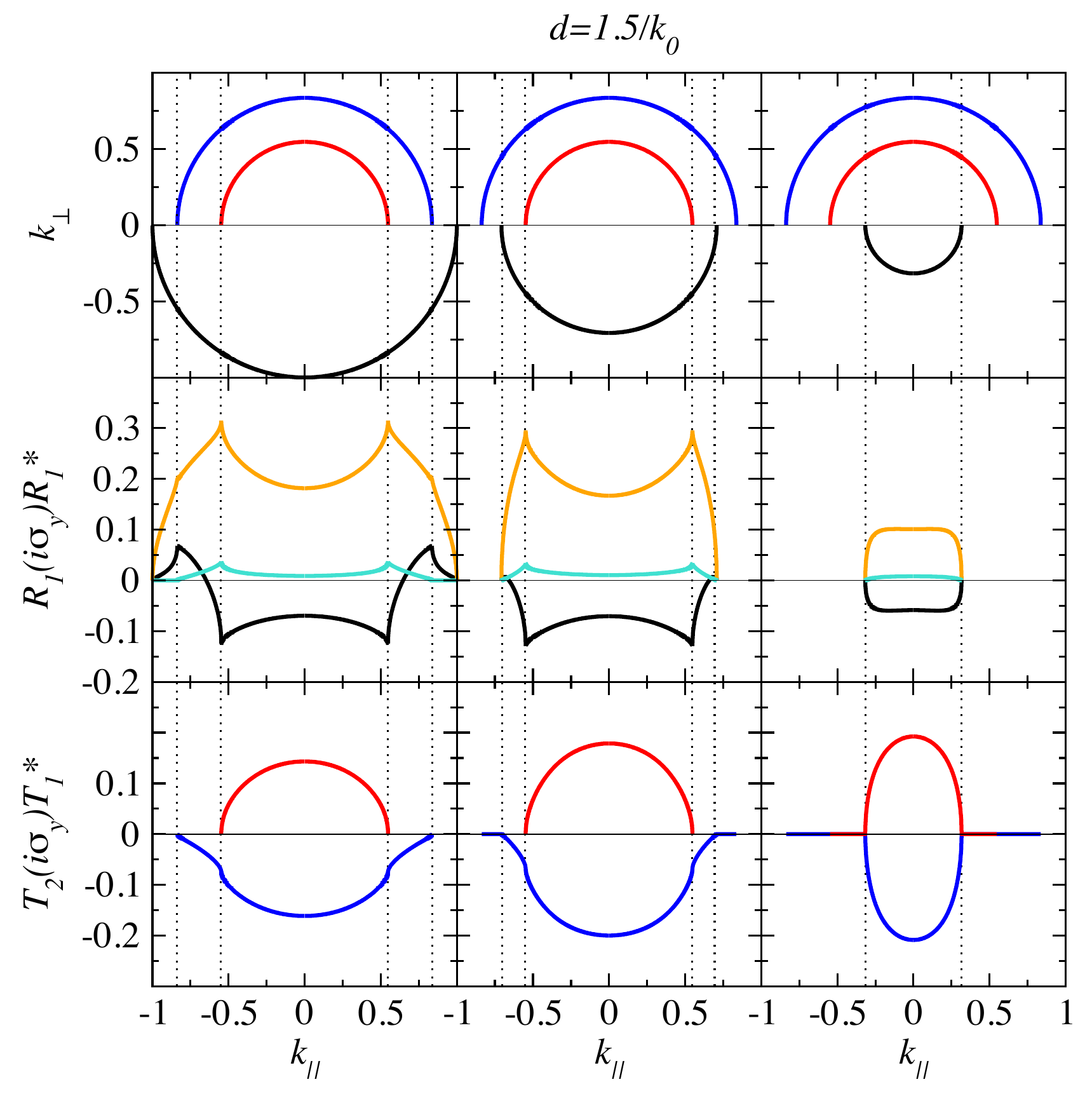}
\caption{
Induced triplet components at a superconductor-ferromagnet interface with an interface barrier of width $d$, spin polarized in $x$-direction, perpendicular to the ferromagnet's spin polarization in $z$-direction. 
Other parameters are as in Fig.~\ref{Fig_alpha0} (a), with $V_{\rm S}=0,0.5 E_F,0.9 E_F$ for each thickness $d$.
The components shown are: in the superconductor $f^{\rm SC}_{tz}/(if_0)=f^{\rm SC}_{\frac{1}{2}(\uparrow\downarrow+\downarrow\uparrow)}/(if_0)$ (black); $f^{\rm SC}_{tx}/(if_0)=f^{\rm SC}_{\frac{1}{2}(-\uparrow\uparrow+\downarrow\downarrow)}/(if_0)$ (orange); $f^{\rm SC}_{ty}/f_0=f^{\rm SC}_{\frac{1}{2i}(\uparrow\uparrow+\downarrow\downarrow)}/f_0$ (turquoise); in the ferromagnet $f^{\rm FM}_{\uparrow\uparrow}/(if_0)$ (blue); $f^{\rm FM}_{\downarrow\downarrow}/(if_0)$ (red).
}
\label{Fig_alpha1}
\end{figure}

In order to allow for proximity induced triplet amplitudes in the ferromagnet, one needs an inhomogeneous non-collinear magnetization in the interface region. We study a simple model where the interface barrier is magnetized in direction perpendicular to the magnetization in the ferromagnet. 
We again assume a spin quantization axis in direction of the exchange field in the ferromagnet, which we take as the $z$-axis, and a barrier exchange field pointing along the $x$-axis in spin space. In Figure~\ref{Fig_alpha1} we show for 
two 
values of barrier thickness all triplet amplitudes generated in the superconductor and in the ferromagnet. For each value of barrier thickness we show three representative Fermi surface geometries. In the ferromagnet the spin bands support only triplet amplitudes of the form $f^{\rm FM}_{\uparrow\uparrow}$ and $f^{\rm FM}_{\downarrow\downarrow}$, whereas in the superconductor we show all three triplet amplitudes $f^{\rm SC}_{tz}=f^{\rm SC}_{\frac{1}{2}(\uparrow\downarrow+\downarrow\uparrow)}$, $f^{\rm SC}_{tx}=f^{\rm SC}_{\frac{1}{2}(-\uparrow\uparrow+\downarrow\downarrow)}$, and $f^{\rm SC}_{ty}=f^{\rm SC}_{\frac{1}{2i}(\uparrow\uparrow+\downarrow\downarrow)}$. 

It is interesting to note that for intermediate barrier thicknesses, also a component $f^{\rm SC}_{ty}$ is generated, although the magnetization always lies in the $x-z$ plane. This is due to the fact that spins polarized in $z$-direction, precessing around the $x$-direction, develop a $y$-component (and similarly for spins polarized in $x$-direction and precessing around the $z$-direction). 
In a real gauge of the singlet pair potential, $f^{\rm SC}_{tz}$ and $f^{\rm SC}_{tx}$ are purely imaginary, and $f^{\rm SC}_{ty}$ is purely real. 
As a consequence, only the former two contribute to the magnetization in the superconductor. 

Furthermore, depending on the barrier thickness, the vector of the triplet amplitudes in the superconductor 
changes direction rapidly as function of impact angle with respect to the surface normal. Only for tunneling barriers the direction of the triplet vector is dominated by the barrier exchange field.

The equal-spin triplet amplitudes generated in the ferromagnet have opposite sign for the two spin projections. This reflects the fact that a triplet component with zero spin projection along the exchange field in the barrier region ($x$-direction) decomposes into equal-spin pairs with respect to the $z$-direction with equal magnitude and opposite sign: $(\uparrow\downarrow+\downarrow\uparrow)_x=(-\uparrow\uparrow+\downarrow\downarrow)_z$. 
Whereas in the tunneling limit the largest contributions to triplet amplitudes arise from near-normal impact, for thin barriers the dominating contributions arise from wavevectors that have for at least one spin projection evanescent solutions in the ferromagnet.

For a general orientation of the interface barrier exchange field we parameterize its direction by polar and azimuthal angles, $\alpha $ and $\phi $,\footnote{In the following the scalar scattering phase $\phi$ will not appear anymore in this review, such that no confusion with the azimuthal angle $\phi$ should arise.} respectively, measured from the $z$-axis in spin space. The well-known transformation formulas for basis vectors quantized along the direction $\alpha,\phi$ in terms of basis vectors quantized along the $z$-axis read
\numparts
\ber
\uparrow_{\alpha,\phi} &=& \quad \cos \frac{\alpha}{2} e^{-i\frac{\phi}{2}} \uparrow_z + \sin \frac{\alpha}{2} e^{i\frac{\phi}{2}} \downarrow_z \\
\downarrow_{\alpha,\phi} &=& -\sin \frac{\alpha}{2} e^{-i\frac{\phi}{2}} \uparrow_z + \cos \frac{\alpha}{2} e^{i\frac{\phi}{2}} \downarrow_z 
\eer
\endnumparts
and can be used to find the transformation for pair amplitudes, for example
\numparts
\ber
(\uparrow\downarrow-\downarrow\uparrow)_{\alpha,\phi} &=& 
(\uparrow\downarrow-\downarrow\uparrow)_z\\
(\uparrow\downarrow+\downarrow\uparrow)_{\alpha,\phi} &=& 
-\sin \alpha \left[e^{-i\phi} (\uparrow\uparrow)_z - e^{i\phi} (\downarrow\downarrow)_z\right]
+ \cos \alpha (\uparrow\downarrow+\downarrow\uparrow)_z .
\label{triplet_z}
\eer
\endnumparts
It can be seen from this that if a triplet component of the form (\ref{triplet_z}) is created in the barrier region, it gives rise to $f^{\rm FM}_{\uparrow\uparrow}$ and $f^{\rm FM}_{\downarrow\downarrow}$ correlations in the ferromagnet with relative phase of $\pi+2\phi$. 
This leads to a non-trivial current-phase relation in Josephson devices, in which two singlet superconductors are connected by the two spin bands of the ferromagnet, each with their own separate current-phase relation \cite{Grein09}. The azimuthal angle $\phi$ plays a crucial role in such devices \cite{Eschrig07,Eschrig08,Braude07}. 

For the case of a long-wavelength magnetic inhomogeneity one has to deal with the spatial variation of the pair amplitudes, which can be done within the framework of quantum transport equations, which we discuss in the next section.
In general, the scattering matrix should be calculated from first principles or from microscopic models, taking into account disorder, band anisotropy, and micromagnetics. 

\section{Theoretical tools}

Theoretical treatments of superconductor-ferromagnet heterostructures can be separated into two types: microscopic treatments and quasiclassical treatments.
There are two main methods, which both can be applied to the two types of theories:
wave-function methods and Green function methods.
Green function methods are
based on an asymptotic expansion of the fundamental many-body Hamiltonian, whereas wave function methods work preferably with an effective mean-field Hamiltonian. 
Green function methods allow in an easier way for the treatment of disorder, and for generalization to strong coupling with bosonic excitations and to non-equilibrium situations. 
\begin{table}[t]
\begin{center}
\begin{tabular}{|l|l|l|}
\hline
\hline
& {\bf\small microscopic model} & {\bf \small quasiclassical}\\
\hline
\hline
{\bf \small wave function} & \small Bogoliubov-de Gennes & \small Andreev equations\\
{\bf \small methods:} & \small equations & \\
\hline
{\bf \small Green function} & \small Gor'kov equations & \small Eilenberger-Larkin-\\
{\bf \small methods:} &&\small Ovchinnikov equations\\
\hline
\hline
\end{tabular}
\end{center}
\caption{Theories for inhomogeneous superconductivity can be classified into four types, as shown in this table. }
\label{Table2}
\end{table}
The resulting four theoretical frameworks with their corresponding equations of motion
are shown in table \ref{Table2}. 
The Bogoliubov-de Gennes equations \cite{deGennes66,Bogolyubov58} lead in quasiclassical approximation to the Andreev equations for the envelopes of the waves \cite{Andreev64}. Alternatively, one can start from the microscopic Nambu-Gor'kov matrix Green functions obeying the Gor'kov equations \cite{Gorkov59,Nambu60}. These lead in quasiclassical approximation to the Eilenberger-Larkin-Ovchinnikov equations \cite{Eilenberger68,Larkin68}, where the concepts of BCS pairing theory of superconductors \cite{Bardeen57} were merged with the concepts of Boltzmann transport equations within Landau's Fermi-liquid theory \cite{Landau56}.
An extension to non-equilibrium was developed by Eliashberg \cite{Eliashberg71} and by Larkin and Ovchinnikov \cite{Larkin75}. In quasiclassical approximation the Gor'kov equations result into envelope Green functions that vary on the superconducting coherence length scale and are free of irrelevant fine-scale structures on the Fermi wavelength scale.
Dynamical phenomena are described in Green function methods within the Keldysh technique \cite{Keldysh64}.

In this review we concentrate on two, which have been mostly used so far in the literature: microscopic wave function methods based on the Bogoliubov-de Gennes equations, and quasiclassical Green function methods, based on Eilenberger-Larkin-Ovchinnikov equations and their counterpart for diffusive systems, the Usadel equations.
The two other frameworks, not covered in this review, are formulated in terms of Andreev equations and in terms of Gor'kov equations.

\subsection{Bogoliubov-de Gennes equations}
The Hartree-Fock-Bogoliubov (HFB) mean-field Hamiltonian
\begin{eqnarray}
{\cal H}_{\rm HFB}&=& \int \int d^3\vec{r} d^3\vec{r}' \sum_{\sigma\sigma'\in \{\uparrow,\downarrow\} } \Big(
H_{\vec{r}\sigma,\vec{r}'\sigma'} \;
\psi^\dagger_{\vec{r}\sigma} \psi_{\vec{r}'\sigma'}
\Big.\nonumber \\ 
&&\qquad \qquad \Big.+\frac{1}{2} \Delta_{\vec{r}\sigma,\vec{r}'\sigma'}\psi^\dagger_{\vec{r}\sigma} \psi^\dagger_{\vec{r}'\sigma'}+ \frac{1}{2} \Delta^\ast_{\vec{r}\sigma,\vec{r}'\sigma'}
\psi_{\vec{r}'\sigma'}
\psi_{\vec{r}\sigma} 
\Big)
\end{eqnarray}
is the basis for the Bogoliubov-de Gennes theory.
Hermiticity requires $H_{\vec{r}\sigma,\vec{r}'\sigma'}=H^\ast_{\vec{r}'\sigma',\vec{r}\sigma}$, and Fermi statistics leads to $\Delta_{\vec{r}\sigma,\vec{r}'\sigma'}=-\Delta_{\vec{r}'\sigma',\vec{r}\sigma}$. 
The Hamiltonian can be rewritten in a compact form using the 4$\times$4 matrix
\begin{eqnarray}
\hat H(\vec{r},\vec{r}')\equiv
\left( \begin{array}{cc}
\quad [H_{\vec{r}\sigma,\vec{r}'\sigma'}]_{2\times 2} & \quad [\Delta_{\vec{r}\sigma,\vec{r}'\sigma'}]_{2\times 2} \\
-[\Delta^\ast_{\vec{r}\sigma,\vec{r}'\sigma'}]_{2\times 2} & -[H^\ast_{\vec{r}\sigma,\vec{r}'\sigma'}]_{2\times 2}
\end{array}\right) 
\end{eqnarray}
where, for example,
\begin{eqnarray}
[H_{\vec{r}\sigma,\vec{r}'\sigma'}]_{2\times 2}=
\left( \begin{array}{cc}
H_{\vec{r}\uparrow,\vec{r}'\uparrow}&
H_{\vec{r}\uparrow,\vec{r}'\downarrow}\\
H_{\vec{r}\downarrow,\vec{r}'\uparrow}&
H_{\vec{r}\downarrow,\vec{r}'\downarrow}
\end{array}\right). 
\end{eqnarray}
With the definitions
$\hat \Psi^\dagger(\vec{r})=(\psi^\dagger_{\vec{r}\uparrow},\psi^\dagger_{\vec{r}\downarrow},\psi_{\vec{r}\uparrow},\psi_{\vec{r}\downarrow})$, $\hat \Psi(\vec{r})=(\psi_{\vec{r}\uparrow},\psi_{\vec{r}\downarrow},\psi^\dagger_{\vec{r}\uparrow},\psi^\dagger_{\vec{r}\downarrow})^{\rm T}$ (here T denotes a transpose) the Hartree-Fock-Bogoliubov Hamiltonian reads
\begin{eqnarray}
{\cal H}_{\rm HFB}&=& \frac{1}{2} \int \int d^3\vec{r} d^3\vec{r}' 
\left(\hat \Psi^\dagger (\vec{r})\; \hat H(\vec{r},\vec{r}') \; \hat \Psi (\vec{r}')\right) + \mbox{const.}
\end{eqnarray}
This Hamiltonian
is diagonalized by a Bogoliubov-Valatin transformation \cite{Bogoliubov58,Valatin58}. 
To achieve this, we introduce
the notation for the Bogoliubov amplitudes ($\alpha \in \{+,-\}$)
\begin{eqnarray}
U_{n\alpha}(\vec{r}) \equiv
\left( \begin{array}{c}
u_{\vec{r}\uparrow,n\alpha}\\u_{\vec{r}\downarrow,n\alpha}\\v_{\vec{r}\uparrow,n\alpha}\\v_{\vec{r}\downarrow,n\alpha} 
\end{array}\right)
,\quad
\bar U_{n\alpha}(\vec{r}) \equiv
\left( \begin{array}{c}
v^\ast_{\vec{r}\uparrow,n\alpha}\\v^\ast_{\vec{r}\downarrow,n\alpha}\\u^\ast_{\vec{r}\uparrow,n\alpha}\\u^\ast_{\vec{r}\downarrow,n\alpha} 
\end{array}\right).
\end{eqnarray}
Here, the index $\alpha $ may, e.g., refer to another spin basis, or to a helicity basis if strong spin-orbit interactions are present. The quantum numbers $n,\alpha$ fully characterize the eigenstates of the set of Bogoliubov-de Gennes equations given by
\numparts
\begin{eqnarray}
\int d^3\vec{r}' 
\hat H (\vec{r},\vec{r}') U_{n\alpha}(\vec{r}') &=& \quad E_{n\alpha} U_{n\alpha}(\vec{r}) \\
\int d^3\vec{r}' 
\hat H (\vec{r},\vec{r}') \bar U_{n\alpha}(\vec{r}') &=& -E_{n\alpha} \bar U_{n\alpha}(\vec{r}) .
\end{eqnarray}
\endnumparts
For each eigenvalue $E_{n\alpha}>0$ there exists an eigenvalue $-E_{n\alpha}$, with corresponding eigenvectors obtained by the substitution $u_{\vec{r}\sigma,n\alpha} \to v_{\vec{r}\sigma,n\alpha}^\ast$, $v_{\vec{r}\sigma,n\alpha} \to u_{\vec{r}\sigma,n\alpha}^\ast$. 
One can divide the quantum numbers $n $ in disjunct sets 
${\cal Z}$, ${\cal P}$ and ${\cal N}$
(and relabel them if necessary)
such that for $n \in {\cal Z}$ we have $E_{n\alpha }=0$,
for $n\in {\cal P} $ we have $E_{n\alpha }>0$ and for $n\in {\cal N} $ we have $E_{n\alpha}<0$.
The eigenvectors 
can be chosen to build an orthonormal set, e.g.
for $n,n'$ within ${\cal P}$ 
\numparts
\begin{eqnarray}
\int d^3\vec{r} \sum_\sigma \left(u_{\vec{r}\sigma,n\alpha}u_{\vec{r}\sigma,n'\alpha'}^\ast+v_{\vec{r}\sigma,n\alpha}v_{\vec{r}\sigma,n'\alpha'}^\ast \right)&=& \delta_{nn'} \delta_{\alpha\alpha'} , \\
\int d^3\vec{r} \sum_\sigma \left(u_{\vec{r}\sigma,n\alpha}v_{\vec{r}\sigma,n'\alpha'}+v_{\vec{r}\sigma,n\alpha}u_{\vec{r}\sigma,n'\alpha'} \right)&=& 0.
\end{eqnarray}
\endnumparts
We build for $n\in {\cal P} $ a 4$\times$4 matrix $\hat U_n(\vec{r})$ from the four orthogonal eigenvectors $U_{n+}(\vec{r}),U_{n-}(\vec{r}),\bar U_{n+}(\vec{r}),\bar U_{n-}(\vec{r})$ (or a 4$\times$2 matrix if only one value for $\alpha $ belongs to $n$).
Similarly, for $n\in {\cal Z} $ we build a 4$\times$2 matrix $U_{n+}(\vec{r}),U_{n-}(\vec{r})$ (or a 4$\times$1 matrix if only one value for $\alpha $ belongs to $n$).
Then the completeness relation
\begin{eqnarray}
\sum_{n\in \bar{\cal P}} \hat U_n(\vec{r}) \hat U^\dagger_n(\vec{r}') = \delta(\vec{r}-\vec{r}') \hat 1
\end{eqnarray}
(with the 4$\times$4 unit matrix $\hat 1$, and with $\bar{\cal P}={\cal Z}\cup {\cal P}$)
holds, which reads explicitly  
\numparts
\begin{eqnarray}
\sum_{n\in \bar{\cal P},\alpha}
\left(u_{\vec{r}\sigma,n\alpha} u^\ast_{\vec{r}'\sigma',n\alpha} +
s_{n }v^\ast_{\vec{r}\sigma,n\alpha } v_{\vec{r}'\sigma',n \alpha }\right) &=& \delta_{\sigma\sigma'} \delta(\vec{r}-\vec{r}')\\
\sum_{n\in \bar{\cal P},\alpha}
\left(u_{\vec{r}\sigma,n\alpha} v^\ast_{\vec{r}'\sigma',n\alpha} +
s_{n }v^\ast_{\vec{r}\sigma,n \alpha } u_{\vec{r}'\sigma',n\alpha } \right) &=& 0 
\end{eqnarray}
\endnumparts
($s_{n }=0$ for $n \in {\cal Z}$, $s_{n }=1$ else).
With this, the Hartree-Fock-Bogoliubov Hamiltonian is diagonalized by
\begin{eqnarray}
\label{BT}
\hat \Psi (\vec{r})= \sum_{n\in \bar{\cal P} } \hat U_n(\vec{r}) \hat \gamma_n, \quad \hat \gamma_n= \int d^3\vec{r} \; \hat U^\dagger_n(\vec{r}) \hat \Psi (\vec{r}), 
\end{eqnarray}
with, e.g.,
$\hat \gamma_{n}^\dagger =(\gamma^\dagger_{n+},\gamma^\dagger_{n-},\gamma_{n+},\gamma_{n-})$, $\hat \gamma_{n}=(\gamma_{n+},\gamma_{n-},\gamma^\dagger_{n+},\gamma^\dagger_{n-})^{\rm T}$ denoting Bogoliubov quasiparticle operators
and $ \hat E_n\equiv \mbox{diag}(E_{n+} , E_{n-}, -E_{n+} , -E_{n-} )$ corresponding energies,
leading to
\begin{eqnarray}
\hat H(\vec{r},\vec{r}')&=& \sum_{n\in {\cal P} } \hat U_n(\vec{r}) \hat E_n \hat U^\dagger_n(\vec{r}') \\
{\cal H}_{\rm HFB} &=& \frac{1}{2} \sum_{n\in {\cal P} } \hat \gamma^\dagger_n \hat E_n
\hat \gamma_n.
\end{eqnarray}
Explicitly, Eq. (\ref{BT}) reads
\numparts
\begin{eqnarray}
\label{BT1}
\psi_{\vec{r}\sigma} = 
\sum_{n\in \bar{\cal P},\alpha } \left(
u_{\vec{r}\sigma,n\alpha} \gamma_{n\alpha}+ s_{n }v^\ast_{\vec{r}\sigma,n\alpha} \gamma^\dagger_{n\alpha} 
\right) \\
\gamma_{n\alpha}= \int d^3\vec{r} \sum_\sigma
\left(u^\ast_{\vec{r}\sigma,n\alpha} \psi_{\vec{r}\sigma} + v^\ast_{\vec{r}\sigma,n\alpha} \psi^\dagger_{\vec{r}\sigma} \right).
\end{eqnarray}
\endnumparts
The Majorana property $\gamma_{n\alpha}=\gamma_{n\alpha}^\dagger$ holds if $v_{\vec{r}\sigma,n\alpha} = u_{\vec{r}\sigma,n\alpha}^\ast$; for $n\in {\cal Z}$ a solution of the Bogoliubov-de Gennes equations can always be chosen of this form provided this is compatible with the boundary conditions.
The single-particle part $H^{\rm 0}_{\vec{r}\sigma,\vec{r}'\sigma'}$
of $H_{\vec{r}\sigma,\vec{r}'\sigma'}$ has typically the form 
\begin{equation}
H^{\rm 0}_{\vec{r}\sigma,\vec{r}'\sigma'}=\delta(\vec{r}-\vec{r}') \left\{ \varepsilon(\hat \vec{P}) -\mu  +
\vec{g}(\hat \vec{P}) \cdot \bsp 
- \vec{h}(\vec{r}) \cdot \bsp \right\}_{\sigma \sigma'} 
\label{H0}
\end{equation}
with $\hat \vec{P}=-i\hbar \vec{\nabla}_{\vec{r}} -e \vec{A}(\vec{r})$. Here, 
$\varepsilon(\hat \vec{P})$ is even in $\hat \vec{P}$ and $\vec{g}(\hat \vec{P}) $ is odd in $\hat \vec{P}$. The last two terms in Eq. (\ref{H0}) describe spin-orbit coupling and Zeeman coupling, and $\mu $ is the electrochemical potential. 
The order parameter is given self-consistently by
\begin{eqnarray}
\Delta_{\vec{r}\sigma,\vec{r}'\sigma'}= \int \int d^3\vec{r}_1d^3\vec{r}_1' \sum_{\sigma_1\sigma_1'} V_{\sigma \sigma'\sigma_1\sigma_1'}(\vec{r},\vec{r}',\vec{r}_1,\vec{r}_1')
\langle \psi_{\vec{r}_1'\sigma_1'} \psi_{\vec{r}_1\sigma_1} \rangle .
\end{eqnarray}
with
\begin{eqnarray}
&&\langle \psi_{\vec{r}'\sigma'} \psi_{\vec{r}\sigma} \rangle = 
-\frac{1}{2}\sum_{n\in {\cal P},\alpha}
\Big\{ u_{\vec{r}\sigma,n\alpha} v^\ast_{\vec{r}'\sigma',n\alpha} -
v^\ast_{\vec{r}\sigma,n \alpha } u_{\vec{r}'\sigma',n\alpha } \Big\}
\tanh \frac{E_{n\alpha}}{2k_{\rm B}T} .
\end{eqnarray}
The spin-dependence of these equations simplifies when 
the pairing interaction can be split into singlet and triplet parts, 
\begin{eqnarray}
V_{\sigma \sigma'\sigma_1\sigma_1'}= \frac{1}{2}(i\sigma_2)_{\sigma \sigma'} V^s (i\sigma_2)^\ast_{\sigma_1 \sigma_1'} +\frac{1}{2}(\bsp i\sigma_y)_{\sigma \sigma'}\cdot \tVt \cdot (\bsp i\sigma_y)^\ast_{\sigma_1 \sigma_1'} ,
\end{eqnarray}
and similarly for the order parameter,
\begin{eqnarray}
\Delta_{\vec{r}\sigma,\vec{r}'\sigma'} = 
\Delta^s_{\vec{r},\vec{r}'} (i\sigma_2)_{\sigma \sigma'} + 
\Dt^t_{\vec{r},\vec{r}'} \cdot (\bsp i\sigma_y)_{\sigma \sigma'}.
\end{eqnarray}
The gap equation for spin singlet and triplet ferromagnetic superconductors has been scrutinized by Powell, Annett, and Gy\"orffy \cite{Powell03}.

In order to study odd-frequency amplitudes one uses a Heisenberg representation
\begin{eqnarray}
\psi_{\vec{r}\sigma}(t) = 
\sum_{n \in \bar{\cal P},\alpha } \left(
u_{\vec{r}\sigma,n\alpha} e^{\frac{i}{\hbar }E_{n\alpha}t}\gamma_{n\alpha}+ s_{n }v^\ast_{\vec{r}\sigma,n\alpha} e^{-\frac{i}{\hbar }E_{n\alpha}t}\gamma^\dagger_{n\alpha} 
\right)
\end{eqnarray}
and evaluates the pair correlation function \cite{Halterman08}
\begin{eqnarray}
&&F_{\vec{r}\sigma,\vec{r}'\sigma'}(t)=\langle \psi_{\vec{r}\sigma}(t) \psi_{\vec{r}'\sigma'} (0)\rangle 
\end{eqnarray}
using $\langle \gamma^\dagger_{n\alpha} \gamma_{n\alpha}\rangle = f(E_{n\alpha})$, 
$\langle \gamma_{n\alpha} \gamma^\dagger_{n\alpha} \rangle= 1-f(E_{n\alpha})$,
with $f(E_{n\alpha})=[1-\tanh(E_{n\alpha}/2k_{\rm B}T)]/2$ the fermionic distribution function.
Concentrating on local correlation functions, one obtains \cite{Halterman08,Fritsch14}
\numparts
\begin{eqnarray}
&&F_\mp(\vec{r},t)\equiv
F_{\vec{r}\uparrow,\vec{r}\downarrow}(t)\mp F_{\vec{r}\downarrow,\vec{r}\uparrow}(t) =
\nonumber\\
&&\quad \sum_{n\in {\cal P},\alpha }
\Big( u_{\vec{r}\uparrow,n\alpha} v^\ast_{\vec{r}\downarrow,n\alpha} \mp
u_{\vec{r}\downarrow,n\alpha} v^\ast_{\vec{r}\uparrow,n\alpha} 
\Big) \zeta^\mp_{n\alpha}(t,T)\\
&&F_{\sigma\sigma}(\vec{r},t)\equiv F_{\vec{r}\sigma,\vec{r}\sigma}(t)=
\sum_{n\in {\cal P},\alpha}
u_{\vec{r}\sigma,n\alpha} v^\ast_{\vec{r}\sigma,n\alpha} \zeta^+_{n\alpha}(t,T)
\end{eqnarray}
\endnumparts
for the singlet ($F_{-}$) and the three triplet ($F_{+},F_{\uparrow\uparrow},F_{\downarrow\downarrow}$) components,
where
\numparts
\begin{eqnarray}
\zeta^{-}_{n\alpha}(t,T)\equiv
\cos\left(\frac{E_{n\alpha}t}{\hbar}\right)\tanh \left(\frac{E_{n\alpha}}{2k_{\rm B}T}\right) -i \sin \left(\frac{E_{n\alpha}t}{\hbar}\right),\\
\zeta^{+}_{n\alpha}(t,T)\equiv
\cos\left(\frac{E_{n\alpha}t}{\hbar}\right) -1-i \sin \left(\frac{E_{n\alpha}t}{\hbar}\right) \tanh \left(\frac{E_{n\alpha}}{2k_{\rm B}T}\right)  .
\end{eqnarray}
\endnumparts
The triplet correlations show a time dependence according to $\zeta^+_{n\alpha}$ and consequently vanish identically for $t=0$, whereas the singlet correlations, governed by $\zeta^{-}_{n\alpha}$, survive. This expresses the odd-frequency nature of the local triplet pair correlations in contrast to the even-frequency nature of the local singlet pair correlations. In particular, for $t\to -i\tau$ with $\hbar/k_{\rm B}T > \tau \ge 0$, and applying KMS boundary conditions $\zeta^\mp_{n\alpha}(-i\tau+i\hbar/k_{\rm B}T ,T)=
-\zeta^\mp_{n\alpha}(-i\tau ,T)$, the relations
$\zeta^\mp_{n\alpha}(i\tau,T)=\pm \zeta^\mp_{n\alpha}(-i\tau,T)$ follow.

Similarly, measurable quantities like magnetization, current density, and charge density can be expressed in terms of Bogoliubov quasiparticle operators using Eq.~(\ref{BT1}) and evaluated using the solutions of the Bogoliubov-de Gennes equations.

\subsection{Quasiclassical theory of superconductivity}
\label{secQCl}

The treatment of a superconductor in the presence of a Zeeman spin-splitting induced by an external magnetic field has been theoretically investigated in detail by Alexander, Orlando, Rainer, and Tedrow in 1985 \cite{Alexander85} within quasiclassical theory of superconductivity. This theory, a generalization of previous studies for superfluid $^3$He \cite{Leggett,Leggett75,Serene83}, includes Fermi liquid effects self-consistently, as well as impurity scattering (including ballistic and diffusive limits), internal exchange fields, and spin-orbit effects\footnote{The derivation in this work is more rigorous than 
in the later work by Demler {\it et al.} in 1997 \cite{Demler97}.}.
It is formulated in terms of generalized Landau parameters $A^{s,a}(\bvpf,\bvpf')$ as well as singlet and triplet pairing interactions $V^{s,t}(\bvpf,\bvpf')$, based on Leggett's treatment for clean systems \cite{Leggett,Leggett75}, and combines previous theories of Fulde \cite{Fulde73} and Buchholtz and Zwicknagl \cite{Buchholtz81}.
First experimental tests of these theories were performed by Tedrow and Meservey \cite{Tedrow79}.
Generalized Landau parameters
lead for example to anisotropic renormalizations of the quasiparticle magnetic moment by coupling to electrons outside the phase space regions where quasiparticles live (``high-energy electrons'').

For reviews of quasiclassical theory of superconductivity see Refs.
\cite{Schmid75,Schmid81,Serene83,Larkin86,Rammer86,Sauls93,Muzikar94,Mineev99,Belzig99,Eschrig01,Kopnin09}.
Quasiclassical theory is an expansion in quantities like temperature or gap divided by Fermi energy, or Fermi wavelength divided by superconducting coherence length \cite{Serene83,Rainer86,Rainer95,Eschrig99b,Eschrig12}. It predicts its own breakdown in low dimensions \cite{Eschrig94}.

The central quantity in quasiclassical theory of superconductivity is the quasiclassical $4\times 4$ matrix propagator or Green function $\hat g (\bvpf,\Rv;\epsilon)$, 
which depends on the spatial coordinate $\Rv $,
on energy $\epsilon $, 
and  on the momentum directions $\bvpf$ on the Fermi surface
(out of equilibrium there is in addition a time dependence and the formalism can be extended to $8\times 8$ Keldysh matrices \cite{Keldysh64}).
The $2\times2$ Nambu-Gor'kov matrix structure of $\hat g$ reflects the particle-hole degree of freedom \cite{Gorkov59,Nambu60}. 
Its matrix elements are $2\times2$ spin matrices (if the spin-splitting of the energy bands is not comparable to the energy band width, otherwise they are scalar). We will use a notation of unit matrices and Pauli matrices in spin space and particle hole space, where $(\sigma_0,\sigma_1,\sigma_2,\sigma_3)$ refers to spin and $(\hat \tau_0,\hat \tau_1,\hat \tau_2,\hat \tau_3)$ to particle-hole degrees of freedom.
We expand the elements in Nambu-Gor'kov space into spin-scalars and spin-vectors,
\be
\hat g=
\left(
\begin{array}{cc}
g\sigma_0+\gv\cd\bsp&
(f\sigma_0+\fv\cd\bsp)i\sigma_2 \\
i\sigma_2(\tilde f\sigma_0-\tilde \fv\cd\bsp)&
\sigma_2(\tilde g\sigma_0-\tilde \gv\cd\bsp)\sigma_2
\end{array}\right).
\label{matrixg}
\ee
with e.g. $\gv =(g_{x},g_{y},g_{z})$ the vector part of $\hat g_{11}$, and $g$ its scalar part. Of particular interest for this review are $\fv $, $\tilde \fv$, describing spin-triplet correlations, and $\gv$, $\tilde \gv$ describing the spin magnetization and the spin current that develop as a result of the generation of triplet correlations. 

\subsection{Eilenberger equations}
\label{secEil}
We discuss first the case when the spin splitting of the energy bands is small (comparable to the superconducting energy scales), so that it can be treated as a perturbation around the un-split Fermi surface. In this case one can 
integrate out the energy dependence of the Green's functions identically for the spins $\uparrow$ and $\downarrow$ \cite{Kopu04} and
assume spin-independent Fermi velocities $\bvvf(\bvpf )$.
The propagator $\hat{g}$ obeys the Eilenberger transport equation \cite{Eilenberger68,Larkin68}
\begin{equation}
\left[
\varepsilon \hat{\tau}_{3} - \hat{h}, \, \hat{g} \right]+ 
\qcgrad \hat g 
=\hat 0 \label{E}
\end{equation}
with the normalization condition \cite{Eilenberger68}
\begin{equation}
\hat{g}^{2}=- \pi^{2} \hat{\tau}_{0} \label{No},
\end{equation}
where $\bvvf (\bvpf )$ is the Fermi velocity for the direction
$\bvpf$ at the Fermi surface, and with
\begin{equation} 
\hat{h}= \hat{v}_{\rm ext}+\hat{\sigma}_{\rm mf}+\hat{\sigma}_{\mathrm{imp}}
\end{equation}
where $\hat{v}_{\rm ext}$ are external potentials, $\hat{\sigma}_{\rm mf}$ are Fermi liquid mean fields (both diagonal and off-diagonal), and $\hat{\sigma}_{\mathrm{imp}}$ is the impurity self energy.
The matrix $\hat h(\bvpf,\Rv;\epsilon)$ entering the Eilenberger
equation (\ref{E}) has a similar structure as $\hat g$,
\be
\hat h
=\left(
\begin{array}{cc}
\nu\sigma_0+\nut\!\cdot\!\bsp&(\Ds\sigma_0+\Dt\!\cdot\!\bsp)i\sigma_2\\
i\sigma_2 (\Ts\sigma_0-\Tt\!\cdot\!\bsp)&\sigma_2 (\tilde \nu\sigma_0-\tut\!\cdot\!\bsp)\sigma_2
\end{array}
\right)
\label{selfenergy}
\ee
The Nambu-Gor'kov matrix structure contains some redundancy, which results into symmetries between the particle and hole elements \cite{Serene83}. They are expressed by the tilde operation $\tilde q(\bvpf,\Rv;\epsilon)= q(-\bvpf,\Rv;-\epsilon^{*})^{*}$ for $q\in \{g,f,\gv,\fv,\nu, \Ds, \nut, \Dt \}$.
Matsubara propagators are obtained by $(\epsilon\rightarrow i\varepsilon_n)$
(where $\varepsilon_n=\pi k_{\rm B}T(2n+1)$ is the Matsubara energy),
retarded propagators
by $(\epsilon\rightarrow \epsilon+i\delta)$, and advanced propagators
by $(\epsilon\rightarrow \epsilon-i\delta)$.
Further fundamental symmetry relations are: $g(\varepsilon^\ast)=g(\varepsilon)^\ast$, $\gv(\varepsilon^\ast)=\gv (\varepsilon)^\ast$, $f(\varepsilon^\ast)=\tilde f (\varepsilon)^\ast$, $\fv(\varepsilon^\ast)=-\tilde{\!\fv}(\varepsilon)^\ast$ (we omitted here for brevity the other arguments).

The self-consistency equations for the
impurity self energy $\hat \sigma_{\rm imp}$ for impurity types $i$ with
impurity potential $\hat v_{i}$ and impurity concentration $n_{i}$ is
\be
\hat \sigma_{\rm imp}(\bvpf,\Rv;\epsilon)=\sum_i n_{i} \hat t_{i}(\bvpf,\bvpf,\Rv;\epsilon) \\
\label{impurities}
\ee
with the quasiclassical T-matrix equation
\ber
&&\hat t_{i}(\bvpf,\bvpf^\prime,\Rv;\epsilon) =\hat v_{i}(\bvpf,\bvpf^\prime) + \nonumber \\
&&\qquad +\langle N_f(\bvpf^{\prime\prime})\hat v_{i}(\bvpf,\bvpf^{\prime\prime})
\hat g(\bvpf^{\prime\prime},\Rv;\epsilon)
\hat t_{i}(\bvpf^{\prime\prime},\bvpf^\prime,\Rv;\epsilon)\rangle_{\bvpfsc^{\prime\prime}}
\label{tmatrix}
\eer
$\langle \dots \rangle_{\bvpfsc}=\int d^2\bvpf/(2\pi\hbar)^2 \ldots$ denotes a Fermi-surface integral,
and $N_f(\bvpf)=(2\pi\hbar |\bvvf|)^{-1}$ is for a fixed Fermi surface point the (one-dimensional) density of states in perpendicular direction to the Fermi surface, per spin projection, in the normal state.

The Fermi-liquid mean-field self energies $\hat \sigma_{\rm mf}$ 
are self-consistently determined from
\numparts
\ber
\nu_{\rm mf}(\bvpf,\Rv)=k_{\rm B}T\sum_{\varepsilon_n}
\big\langle  N_f(\bvpf^\prime)W^s(\bvpf,\bvpf^\prime) g(\bvpf^\prime,\Rv;\varepsilon_n) \big\rangle_{\bvpfsc^\prime}\\
\label{landausymm}
\nut_{\rm mf}(\bvpf,\Rv)=k_{\rm B}T\sum_{\varepsilon_n}
\big\langle N_f(\bvpf^\prime)\tWa(\bvpf,\bvpf^\prime) \gv(\bvpf^\prime,\Rv;\varepsilon_n) \big\rangle_{\bvpfsc^\prime} 
\label{landauanti}
\\
\Ds_{\rm mf}(\bvpf,\Rv)=k_{\rm B}T\sum_{\varepsilon_n}\big\langle N_f(\bvpf^\prime)V^s(\bvpf,\bvpf^\prime)f(\bvpf^\prime,\Rv;\varepsilon_n)
\big\rangle_{\bvpfsc^\prime}\\
\label{singletop}
\Dt_{\rm mf}(\bvpf,\Rv)=k_{\rm B}T\sum_{\varepsilon_n}\big\langle N_f(\bvpf^\prime)\tVt(\bvpf,\bvpf^\prime)\fv(\bvpf^\prime,\Rv;\varepsilon_n)
\big\rangle_{\bvpfsc^\prime} .
\label{tripletop}
\eer
\endnumparts
Fermi-liquid interactions are parameterized by dimensionless Fermi-liquid parameters $A^s(\bvpf,\bvpf^\prime)=N_fW^s(\bvpf,\bvpf^\prime)$
and $\tAa(\bvpf,\bvpf^\prime)=N_f\tWa(\bvpf,\bvpf^\prime)$, 
with $N_f=\langle N_f(\bvpf) \rangle_{\bvpfsc}$ the density of states per spin projection at the Fermi level,
and the superconducting pair potentials 
by singlet and triplet pairing interactions $V^s$
and $\tVt$ (where $\tWa$, $\tAa$, and $\tVt$ are in general tensors).

\begin{figure}[t]
\begin{center}
\begin{minipage}{0.5\linewidth}
\begin{center}
\begin{overpic}[width=1.0\linewidth]{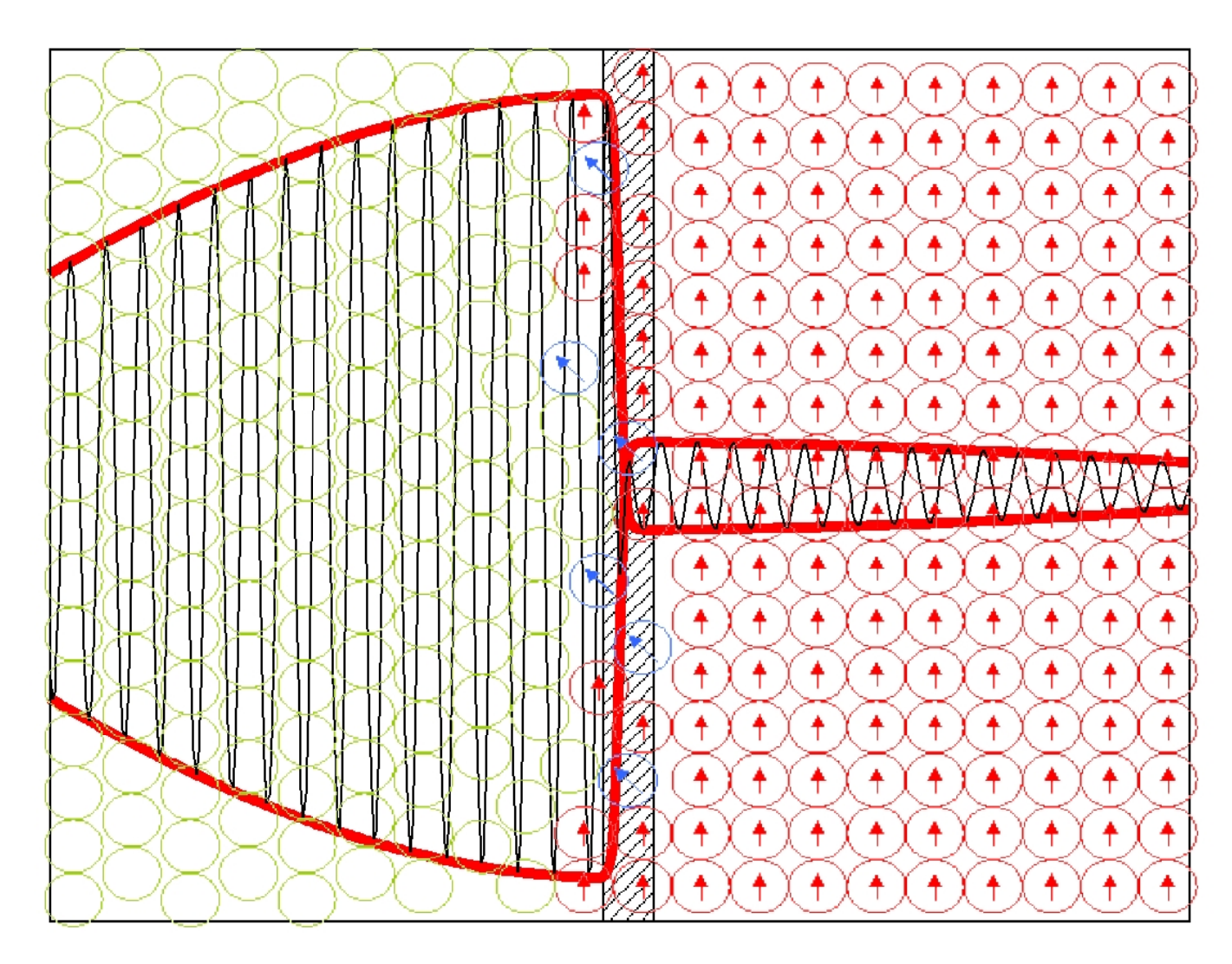}
\end{overpic}
\end{center}
\end{minipage}
\end{center}
\caption{
In quasiclassical approximation boundary conditions involve relations between the envelope functions of Bloch waves on the two sides of an interface. These envelope functions in general show a jump, even when the Bloch wave functions themselves are continuous. 
}
\label{Boundary}
\end{figure}

Finally, $\hat v_{\rm ext}$  contains the coupling to external fields, for example
the electro-magnetic coupling to a magnetic vector potential $\Av(\Rv)$,
\be
\hat v_{\rm orbital}(\bvpf,\Rv)=-e\bvvf(\bvpf)\cd\Av(\Rv)\hat \tau_3
\label{orbital}
\ee
($e=-|e|$), and the Pauli-coupling to the quasiparticle spin
\be
\hat v_{\rm spin}(\bvpf,\Rv)=-\Bv(\Rv)\cdot \tmu_{\rm eff}(\bvpf)\cd \hat \Sv
\label{pauli}
\ee
with the effective quasiparticle magnetic moment\footnote{$\hat 1$ denotes a 3$\times$3 unit tensor.}
\be
\tmu_{\rm eff}(\bvpf)= \big[\hat 1-\tAa(\bvpf)\big] \mu_{\rm e}
\ee
where $\tAa(\bvpf)=N_f^{-1}\langle N_f(\bvpf^\prime)\tAa(\bvpf,\bvpf^\prime) \rangle_{\bvpfsc^\prime}$,
and the spin matrix in particle-hole space
\be
\hat \Sv=\left(\begin{array}{cc}\bsp&0\\0&-\sigma_2\bsp\sigma_2\end{array}\right).
\ee
$\mu_{\rm e}$ is the electron magnetic moment ($\mu_{\rm e} = -|\mu_{\rm e}|$).\footnote{$\mu_{\rm e}=-g\mu_{\rm B}/2$, with the Bohr magneton $\mu_{\rm B}=|e|\hbar/2m_ec$ and $g=2.0$}
The vector potential $\Av$ and hence the magnetic field ($\Bv=\nabla\times\Av$) are calculated
from the current density as
\be
\nabla\times\nabla\times\Av(\Rv)=\mu_0\jv(\Rv)
\label{ampere}
\ee
with the permeability of free space $\mu_0$.
The scalar electrochemical potential follows from local charge quasi-neutrality, and is 
zero in equilibrium.
The charge current density is obtained from $\hat g(\bvpf,\Rv;\varepsilon_n)$ via
\be
\jv(\Rv)=2e k_{\rm B}T\sum_{\varepsilon_n} \big\langle N_f(\bvpf)\bvvf(\bvpf) g(\bvpf,\Rv;\varepsilon_n) \big\rangle_{\bvpfsc} ,
\label{current}
\ee
the spin current density for spin projection along the axis $\ev_\alpha $ via
\be
\Jv_\alpha (\Rv) = 2e k_{\rm B}T\sum_{\varepsilon_n} \big\langle N_f(\bvpf)\bvvf(\bvpf) [\ev_\alpha \cdot \gv(\bvpf,\Rv;\varepsilon_n)] \big\rangle_{\bvpfsc}
\label{spincurrent}
\ee
and the spin magnetization via
\be
\Mv(\Rv)=\tchi_{\rm n} \Bv(\Rv) +2 k_{\rm B}T\sum_{\varepsilon_n}
\big\langle N_f(\bvpf)\tmu_{\rm eff}\!\!(\bvpf)\gv(\bvpf,\Rv;\varepsilon_n) \big\rangle_{\bvpfsc},
\label{magnitization}
\ee
where $\tchi_{\rm n} = 2 N_f (\hat 1-\tAan)\mu_{\rm e}^2$ is the normal state spin susceptibility, defined in terms of the parameter 
$\tAan=N_f^{-1}\langle N_f(\bvpf) \!\!\tAa(\bvpf) \rangle_{\bvpfsc}$. 
The local density of states for a given spin direction $\ev$
is calculated at real energies as
\be
N_{\evsc}(\Rv;\epsilon )
=-\frac{1}{\pi}{\rm Im}\langle N_f(\bvpf)[g(\bvpf,\Rv;\epsilon_+)+\ev\cd
\gv(\bvpf,\Rv;\epsilon_+)]\rangle_{\bvpfsc} 
\label{spindos}
\ee
with $\epsilon_+=\epsilon+i \delta$ ($\delta>0$ infinitesimal).
The Eilenberger equation for $\hat g$, its normalization condition, the
equations for the self energies $\hat \sigma$, and Maxwell's equation for $\Av$,
must be solved self consistently by iteration together
with the appropriate boundary conditions imposed on the propagator and the vector potential \cite{Zaitsev84,Shelankov85,Zhang87,Millis88,Yip97}. Boundary conditions in quasiclassical theory are notoriously difficult to derive, as the quasiclassical propagators show in general jumps at interfaces (see figure \ref{Boundary}), in contrast to microscopic propagators which are continuous.

A powerful way to implement boundary conditions and to solve Eilenberger equations is the Riccati method \cite{Schopohl95,Eschrig99,Eschrig00,Eschrig09}.
Modern versions of boundary conditions for Eilenberger equations using these techniques are presented e.g. in Refs. \cite{Eschrig00,Fogelstrom00,Zhao04,Eschrig09}. 

Instead of applying an external field $\Bv (\Rv )$, a superconductor can be spin-polarized via the inverse (or magnetic) proximity effect when in contact with a ferromagnetic material. This effect already appears when the ferromagnet is insulating.
A theory for the spin polarization and the associated induced exchange field in the superconductor has been developed by Tokuyasu, Sauls, and Rainer in 1988 \cite{Tokuyasu88}. Although differing in details, the general
mechanism presented there is essentially the same as for all subsequent studies of the inverse proximity effect in superconductor/ferromagnet heterostructures in the last decade: 
the appearance of a Cooper pair spin polarization, or of triplet pair correlations, in the superconductor
creates a finite spin magnetization inside the superconducting region, extending roughly a coherence length away from the interface. This decay length is dictated by the decay of triplet pair correlations as the bulk of the
singlet superconductor is approached.

\subsection{Normalization condition and transport equation in spin-space}
\label{secNorm}

The interrelation between triplet amplitudes and induced magnetic moment can be understood within quasiclassical theory already by the normalization condition for the propagator.
The normalization condition (\ref{No}) encodes important information about the spin structure of the diagonal and off-diagonal propagators \cite{Champel05}.
In equilibrium, physical solutions require the condition
\begin{equation}
\mathrm{Tr }\left(\hat{g} \right)=0,  \hspace*{0.4cm} \mathrm{i.e.} \hspace*{0.4cm} \tilde{g}_{0}=-g_{0} \label{hs}
\end{equation}
which also ensure local charge neutrality to leading order in the Fermi liquid expansion parameters.
It is useful to introduce the notation $\gv_\pm= (\gv \pm \gtv)/2$, $\nut_\pm = (\nut \pm \tut )/2$, $\nu_\pm = (\nu \pm \tilde\nu )/2$.
The external and internal fields are distributed according to: $\hat v_{\rm orbital } $ in $\nu_-$; $\hat v_{\rm spin}$ in $\nut_+$, which also contains the internal exchange field $\Jv$; spin-orbit band splitting in $\nut_-$.
In terms of measurable quantities, $\gv_+$ determines the magnetization, $\gv_-$ the spin current density, and $g_0$ the charge current density.
One obtains from (\ref{No}) and (\ref{hs}) for the spin components in (\ref{matrixg}) \cite{Champel05}
\numparts
\begin{eqnarray}
g_{0}^{2}+\pi^{2}&=&f \tilde{f}-\fv \cdot \ftv- \gv_+^{2}-\gv_-^{2}\label{norm2}\\
2g_{0} \gv_-&=& i \, \ftv \times \fv \label{rel1}\\
2g_{0} \gv_+&=& \tilde{f} \fv-f \ftv \label{rel2}
\end{eqnarray}
\endnumparts
with $\gv_\pm^{2}=g_{x,\pm}^{2}+g_{y,\pm}^{2}+g_{z,\pm}^{2}=\gv_\pm \cdot \gv_\pm$.
It follows immediately that $\gv_+\cdot \gv_-=0$, $f\gv_-=i \fv \times \gv_+$ and $\fv \cdot \gv_-= \ftv \cdot \gv_-=0$.
According to (\ref{rel1}), $\gv_-={\bf 0}$ when $\fv \parallel \ftv $.
Moreover, from (\ref{rel1})-(\ref{rel2}) it follows that  $\gv_+=\gv_-={\bf 0}$ when $\fv =\ftv ={\bf 0}$. 
The Eilenberger equations read 
\numparts
\begin{eqnarray}
\frac{1}{2}\, \qcgrad f+(\epsilon-\nu_-) \, f &=& \nut_+ \cdot \fv -g \Ds - \gv_+ \cdot \Dt \label{Eils0}\\
\frac{1}{2}\, \qcgradbr \fv+(\epsilon-\nu_-) \, \fv &-&i \nut_- \times \fv = 
\nonumber \\
&=&\nut_+ f -\gv_+ \Ds -g\Dt - i\gv_- \times \Dt ,
\label{Eilt0}
\end{eqnarray}
\endnumparts
which must be solved together with (\ref{norm2}), (\ref{rel1}) and (\ref{rel2}).
These equations have to be complemented by self-consistency equations for the self-energies. Near $T_{\rm c}$ all off-diagonal quantities ($f$, $\fv$, $\Ds$, $\Dt$) are small, and then from (\ref{norm2})-(\ref{rel2}) follows that $\gv_\pm $ can be neglected being second order in $\fv, \tilde\fv$, and $g$ can be replaced by its normal state value [$-i\pi \mbox{sign} (\varepsilon_n)$ in Matsubara representation].
If one introduces the matrices 
\be
\tV=\left(\begin{array}{cccc}
\nu_- &\nu_{+,x}& \nu_{+,y} &\nu_{+,z}\\
\nu_{+,x}& \nu_-&i\nu_{-,z}&-i\nu_{-,y}\\
\nu_{+,y}& -i\nu_{-,z}&\nu_-&i\nu_{-,x}\\
\nu_{+,z}& i\nu_{-,y}&-i\nu_{-,x}&\nu_-
\end{array}\right)
\ee
and
\be
\tG=\left(\begin{array}{cccc}
g &g_{+,x}& g_{+,y} &g_{+,z}\\
g_{+,x}& g&ig_{-,z}&-ig_{-,y}\\
g_{+,y}& -ig_{-,z}&g&ig_{-,x}\\
g_{+,z}& ig_{-,y}&-ig_{-,x}&g
\end{array}\right)
\ee
as well as the vectors
\be
\vec{F}=\left(\begin{array}{c}
f\\f_x\\f_y\\f_z
\end{array}\right), \quad 
\vec{\Delta }= \left(\begin{array}{c}
\Delta \\ \Delta _x\\ \Delta_y\\ \Delta_z
\end{array}\right), \quad 
\ee
then the system (\ref{Eils0})-(\ref{Eilt0}) can be written compactly as
\be
(\qcgrad +2\epsilon -2\tV ) \vec{F}=2\tG \vec{\Delta} .
\ee
The four eigenvalues of $2(\epsilon -\tV)/\hbar v_f$ determine the wavevectors in the system, and can be calculated as
\be
\left[2(\epsilon -\nu_-)\pm \sqrt{\nu_x^2+\nu_y^2+\nu_z^2}\pm \sqrt{\tilde\nu_x^2+\tilde\nu_y^2+\tilde\nu_z^2} \right]/(\hbar v_f).
\ee

For a singlet superconductor in proximity contact with a ferromagnet with exchange field $\Jv$, assuming for the following discussion $\nu_-=0$, $\nut_-={\bf 0}$, $\nut_+=-\Jv$, $\Dt=0$, the Eilenberger equations simplify to
\numparts
\begin{eqnarray}
\left(\mbox{$\frac{1}{2}$}\, \qcgrad +\bar\epsilon \right)\, f &=& -\Jv \cdot \fv -g \Ds \label{Eils}\\
\left(\mbox{$\frac{1}{2}$} \, \qcgrad +\bar\epsilon \right)\, \fv 
&=&-\Jv f -\gv_+ \Ds.
\label{Eilt}
\end{eqnarray}
\endnumparts
where we have introduced $\bar \epsilon =\epsilon+i\alpha_{\rm s}$ for Im$(\epsilon)>0$ and 
$\bar \epsilon =\epsilon-i\alpha_{\rm s}$ for Im$(\epsilon)<0$
to account for possible spin-flip scattering with rate $\alpha_{\rm s}=\hbar/\tau_{\rm s}$, and $\tau_{\rm s}$ is the spin-flip scattering time.
The wavevectors are then given by
\be
k_{1,2}=2(\bar\epsilon \pm |\Jv| )/\hbar v_f , \quad k_{3,4}=2\bar\epsilon/\hbar v_f .
\label{kvec}
\ee
In a superconductor-ferromagnet hybrid structure,
the superconducting order parameter vanishes in the ferromagnetic regions, $\Delta_{\rm mf}=0$.
The proximity effect manifests itself in nonzero pair amplitudes $f, \fv \neq 0$ in the ferromagnet.
In the superconductor, the exchange field vanishes, $\Jv={\bf 0}$,
expressing the non-coexistence of superconducting and ferromagnetic orders.
\Eref{kvec} defines the propagation and decay of Cooper pairs into the ferromagnet. 
If one replaces $\epsilon $ by the first Matsubara energy $i\pi k_{\rm B}T$, then
one obtains an exponential decay on the length scale $\xi_T=\hbar v_f/2(\pi k_{\rm B}T+\alpha_{\rm s})$. In addition, two of the components oscillate on the length scale $\xi_J=\hbar v_f/2|\Jv|$ (with wavelength $2\pi \xi_J$) along the direction $\bvvf $. After averaging over all directions, this leads to an algebraic decay $\sim 1/x$ in $x$-direction away from the interface, before on a larger length scale $\xi_T$ exponential decay sets in \cite{Eschrig07}. 
The components with wavevectors $k_{1,2}$, oscillating on the scale $\xi_J$, 
correspond to eigenvectors with
$\fv= \pm f \Jv/|\Jv| $ (i.e. $\fv\times \Jv={\bf 0}$),
and represent an equal weight superposition of the singlet amplitude ($S=0,S_z=0$) and the triplet amplitude with zero spin-projection to $\Jv$ ($S=1,S_z=0 $). The (degenerate) components with wavevectors $k_{3,4}$, insensitive to $\Jv$,
correspond to an eigenvector where $f=0$, $\fv \cdot \Jv=0$, and represent 
equal-spin pairing amplitudes with respect to the quantization axis $\Jv$ ($S=1,S_z=\pm1 $). 
For increasing exchange field, the pre-factor for the proximity amplitudes oscillating with wavelength $\xi_J$ is of order $\Delta /|\Jv |$, and thus vanishes in the limit of large exchange splitting.

\subsection{Usadel equations}
\label{secUsadel}
In the
dirty (diffusive) limit, the transport equation (\ref{E}) can be greatly simplified.  A strong scattering by impurities averages the quasiclassical propagator over momentum directions.
The Green's function may be expanded in the small parameter $\tau  T_{c}$  ($\tau$ is the momentum relaxation time) following the standard procedure \cite{Usadel70,Alexander85}
\begin{eqnarray}
\hat{g}(\bvpf,\Rv,\varepsilon_{n}) \approx \hat{g}^{(0)}(\Rv,\varepsilon_{n}) + \hat{g}^{(1)} (\bvpf,\Rv,\varepsilon_{n}) \label{exp}
\end{eqnarray}
where the magnitude of $\hat{g}^{(1)}$ is small compared to that of $\hat{g}^{(0)}$. 
The impurity self-energy  is related to an (in general anisotropic) lifetime function $\tau(\bvpf',\bvpf)$ \cite{Alexander85}.
Substituting (\ref{exp}) into (\ref{E}) and (\ref{No}), 
multiplying with $N_f(\bvpf') v_{f,j}(\bvpf') \tau(\bvpf',\bvpf)$,
averaging over  momentum directions, and considering that $J \tau /\hbar \ll 1$, 
 one obtains (with the help of the normalization condition) $\langle N_f(\bvpf) v_{f,j}(\bvpf) \hat{g}^{(1)}(\bvpf)\rangle_{\bvpfsc} = N_f\sum_k
(D_{jk}/i\pi) \hat{g}^{(0)} \nabla_k \hat{g}^{(0)}$, where 
\be
D_{jk}= \frac{1}{N_f^2}\Big\langle\Big\langle 
N_f(\bvpf' ) v_{f,j}(\bvpf') \, \tau (\bvpf',\bvpf)\, N_f(\bvpf) v_{f,k}(\bvpf) 
\Big\rangle_{\bvpfsc}\Big\rangle_{\bvpfsc'}
\ee
is the diffusion constant tensor.
The function $\hat{g}^{(0)}$ obeys the Usadel transport equation
(hereafter we omit the superscript (0) and apply the Einstein summation convention)
\begin{equation}
\left[\epsilon \hat{\tau}_{3} - \hat{h}_0, \, \hat{g} \right]+ \frac{\hbar D_{ij}}{\pi} \nabla_{i}\left(  \hat{g} \nabla_{j}   \hat{g}\right)=0 \label{Usa1}
\end{equation}
where $\hat h_0=\langle N_f(\bvpf) [\hat v_{\rm ext}(\bvpf)+\hat \sigma_{\rm mf}(\bvpf)]\rangle_{\bvpfsc}/N_f$,
together with the normalization condition
\begin{equation}
\hat{g}^{2}=- \pi^{2} \hat{\tau}_{0} .
\end{equation}
The diffusion constant is a material property. It is spin independent in agreement with the approximation of treating all exchange energy effects as perturbation.
In the diffusive limit, the different components of $g$ and $f$ in the spin space 
are related through (\ref{norm2}), (\ref{rel1}) and (\ref{rel2}), since the Green's function  $\hat{g}$ averaged over momentum fulfills the same normalization condition as before averaging.
A vector potential enters in a gauge invariant manner by replacing the spatial derivative operators by (see e.g. \cite{Tanaka09})
\begin{equation}
\nabla_{i} \hat X \to \hat \partial_i \hat X \equiv \nabla_{i} \hat X -i \left[ \frac{e}{\hbar }\hat \tau_3 A_i,\hat X\right].
\end{equation}
The charge current density is obtained from $\hat g(\Rv;\varepsilon_n)$ via
\be
j_k=\frac{2e N_f D_{kj}}{\pi}k_{\rm B}T\sum_{\varepsilon_n} 
\mbox{Im} \left[f^\ast \; \overline\partial_j f
-\fv^\ast \; \overline\partial_j \fv
\right],
\label{diffcurrent}
\ee
with $\overline\partial_j = \nabla_j -\frac{2ie}{\hbar } A_j$,
and the spin-vector current density in spatial $k$-direction
\be
\vJ_{k}=\frac{2e N_f D_{kj}}{\pi}k_{\rm B}T\sum_{\varepsilon_n} 
\mbox{Re} \left[ \gv \times \nabla_j \gv + \fv^\ast \times \overline\partial_j \fv \right].
\label{diffspincurrent}
\ee
Note that from Eqs.~(\ref{rel1})-(\ref{rel2}) follows that 
$\gv=\left[\mbox{Im}(f^\ast \fv)+i\mbox{Re}(\fv) \times \mbox{Im}(\fv)\right]/\mbox{Im}(g_0)$ 
at Matsubara frequencies.
As for the case of Eilenberger equations, an important part of the problem is the formulation of appropriate boundary conditions \cite{Kupriyanov88,Nazarov99}. In quasiclassical theory this is usually a highly non-trivial task, as boundary conditions are in general non-linear. 
A powerful way to implement boundary conditions and to solve Usadel equations is also here the Riccati method \cite{Eschrig04,Konstandin05,Cuevas06}. Modern versions of boundary conditions for 
Usadel equations are found in Ref.~\cite{Cottet09} for weak spin polarization, small spin-dependent scattering phase shifts, and small transmission. A general formulation appropriate for arbitrary transmission, spin polarization, and spin-dependent phase shifts has been derived in Ref.~\cite{Eschrig15}. The tunneling limit of this boundary condition has been used in Ref.~\cite{Machon13}, and was independently introduced in Ref.~\cite{Bergeret12}.\footnote{
A recently suggested alternative boundary condition \cite{Machon15} 
assumes the simplified case of coinciding spin-quantization axes for reflection and transmission matrix elements of each scattering channel within a channel-conserving approximation, 
and has a considerably narrower range of applicability. }

Near the critical temperature $T_{c}$,
the pair amplitudes $f$ and ${\fv}$ are small and the Green's function $g$ deviates only slightly from its value 
in the normal state, so that the Usadel equations can be linearized and take the simpler form \cite{Bergeret05a,Champel05a,Buzdin05}
\begin{eqnarray}
\left(\mbox{$\frac{\hbar}{2i}$}\nabla_j D_{jk}\nabla_k +\bar \epsilon \right)&f=& 
-{\bf J} \cdot \fv
+i\pi \Delta
\label{fs}
\\
\left(\mbox{$\frac{\hbar}{2i}$} \nabla_j D_{jk}\nabla_k +\bar \epsilon \right)&\fv =&
-{\bf J}f, \label{coup}
\end{eqnarray}
where $\bar \epsilon $ is defined as after Eq.~(\ref{Eilt}).
The characteristic wavevectors for isotropic systems ($D_{ij}\to D\delta_{ij}$) are now given by (we chose the roots such that Im$(k)>0$)
\be
k_{1,2}=(1+i)\sqrt{(\bar \epsilon \pm |\Jv| )/\hbar D} , \quad k_{3,4}=(1+i)\sqrt{\bar \epsilon/\hbar D }.
\label{kvecd}
\ee
If we again substitute $\epsilon \to i\pi k_{\rm B}T$, this leads to exponential decays for all four solutions. However, there are two long-range solutions with decay length 
\be
\xi_T=\left(\frac{\hbar D}{2(\pi k_{\rm B} T+\alpha_{\rm s})}\right)^{\frac{1}{2}}, 
\ee
and two short-range solutions showing damped spatial oscillations with decay length \cite{Ryazanov01}
\be
\xi_{1}= \left(\frac{\hbar D}{\sqrt{(\pi k_{\rm B}T+\alpha_{\rm s})^2+|\Jv|^2}+ (\pi k_{\rm B}T+\alpha_{\rm s})}\right)^{\frac{1}{2}}
\label{xi1}
\ee
and inverse oscillation wavevector
\be
\xi_{2}= \left(\frac{\hbar D}{\sqrt{(\pi k_{\rm B}T+\alpha_{\rm s})^2+|\Jv|^2}- (\pi k_{\rm B}T+\alpha_{\rm s})}\right)^{\frac{1}{2}}.
\label{xi2}
\ee
Note that $\xi_1<\xi_2$ except when $T=0$ and $\hbar/\tau_{\rm s}=0$, thus the oscillations are strongly damped, so that one cannot expect to observe many oscillation periods.
This is in contrast to the clean case, where the oscillation wavelength is entirely decoupled from the decay length, which can become very long for low temperatures.
Thus, many oscillation periods are expected to be observable in experiment. For the diffusive case, on the other hand, both length scales are temperature dependent. This allows for oscillatory behavior as a function of temperature, an effect absent in the clean limit. 
For $|\Jv |\gg k_{\rm B}T, \alpha_{\rm s}$ both length scales merge and approach \cite{Oboznov06}
\begin{eqnarray}
\xi_{1,2}=\sqrt{\frac{\hbar D}{|\Jv|}} \left( 1 \mp \frac{\pi k_{\rm B}T+\alpha_{\rm s}}{2|\Jv|}\right).
\nonumber
\end{eqnarray}
With increasing exchange splitting, the decay length shrinks to zero and
the short-range proximity amplitudes are expected to
be suppressed to unmeasurable size, unless one finds a mechanism to create
new types of proximity amplitudes.

\subsection{General features of the S/F proximity effect}

It can be noticed from the transport equation (\ref{Eilt}) or (\ref{coup}) that the triplet vector $\fv $ obeys in the ferromagnet an inhomogeneous differential equation, implying that $\fv $ is necessarily non-zero if the singlet component $f$ penetrates in the ferromagnet.
For spatially constant exchange field $\Jv$,
the triplet vector aligns with $\Jv $.
The singlet component and the triplet component with the zero spin-projection on $\Jv $ coexist always in the ferromagnet near the S/F interface.
As they both involve electrons from both spin bands,
both are characterized by short-range penetration lengths in the ferromagnet.

On the contrary,  if the triplet vector $\fv $ is non-collinear with $\Jv $, triplet components with nonzero spin-projection on $\Jv $ are produced. Since these correspond to equal-spin pairing, they are not limited locally by the paramagnetic interaction with the local exchange field and have
long-range scales in the ferromagnet.
A misalignment between the triplet vector $\fv $ and the moment $\Jv $ occurs in presence of sudden changes in orientation of $\Jv $. The reason is that $\fv $ obeys a differential equation and its variations in orientation have thus to be relatively smooth.

The counterpart for the production of triplet components ($\fv  \neq {\bf 0}$) 
in the pair amplitudes
is the presence of $\gv \neq {\bf 0}$ 
in the diagonal components of the Green function \cite{Champel05}.
As a direct consequence, the density of states for the up and down spin projections differs [see \Eref{spindos}].
This feature of the S/F proximity effect has been
found numerically as early as in 1999 \cite{Fazio99}.
In the presence of long-range triplet components, the particle-hole diagonal Green function components
contain also off-diagonal terms in the spin-space, a signature of a spin-flip scattering process.

As a result of the spin splitting in the density of states  generated by the S/F proximity effect,
a spin magnetization $\delta \Mv$ is also induced near the S/F interface.
This magnetization leakage has been investigated in Ref. \cite{Bergeret04a} within a model considering a fixed exchange field.
In the dirty limit the spin magnetization induced by the proximity effect is given by \cite{Alexander85,Tokuyasu88}
\begin{equation}
\delta \Mv(\Rv )=2 N_f k_{\rm B}T \tmu_{\rm eff} \sum_{n} 
\mbox{Re} \left[\gv \left(\Rv , \varepsilon_{n}\right)\right], \label{aim}
\end{equation}
with $\tmu_{\rm eff}= (\hat 1-\!\!\!\tAan) \mu_{\rm e}$. 
Since the triplet vector $\fv $ is also induced in the superconductor near the S/F interface  via an inverse proximity effect,
the vector $\gv $ characterizing the magnetic correlations penetrates also in the superconductor according to the relation (\ref{rel2}), which for the diffusive limit reads Im$(g_0)$Re$(\gv)=$Im$(f^\ast \fv)$ (all functions of $\epsilon_n$).

It is possible to convince oneself that the sum over Matsubara frequencies in the expression (\ref{aim}) is nonzero.
Indeed, as noticed in Refs. \cite{Bergeret01,Volkov03,Bergeret03}, in the diffusive limit the triplet components are odd functions of the Matsubara frequencies $\varepsilon_{n}$, while the singlet amplitude $f$ is an even function of $\varepsilon_{n}$. 
Note that in the diffusive limit
\begin{eqnarray}
g_0(-\varepsilon_{n})=-g_0(\varepsilon_{n}), \quad
\gv(-\varepsilon_{n})=\tilde{\gv }(\varepsilon_{n}). \label{dirty}
\end{eqnarray}

For unitary pairing states, $\fv \times \tilde\fv={\bf 0}$,
the relations (\ref{rel1})-(\ref{rel2}) between the vectors $\gv $ and $\fv $ get simplified.
In this case,
it can be seen 
that
 if the gap $\Delta$ can be chosen real (and thus the singlet amplitude $f(\varepsilon_{n})$ is real), then the triplet vector $\fv(\varepsilon_{n}) $ is purely imaginary, i.e. $\tilde\fv=\fv^{\ast}=-\fv$
(taking into account that $\tilde{g}_0=-g_0$, i.e. $g_0$ is purely imaginary). As a result, one obtains the simpler relations
\begin{equation}
\gv =\tilde{\gv }, \hspace*{1cm} g_{0} \gv= f \fv. \label{simp}
\end{equation}
These simplifications arise also when $\fv \parallel \Jv$ \cite{Bergeret04a}. 
Combining the relations (\ref{simp}) and (\ref{dirty}), it follows that the spin-vector part $\gv $ is an even function of $\varepsilon_{n}$ in the diffusive limit, which demonstrates that the induced spin magnetization $\delta \Mv$ is in general non-zero.

Near $T_{c}$, the singlet amplitude $f$ and the triplet vector amplitude are small so that  $\gv $ and thus $\delta \Mv $ appear to be second order terms [see \Eref{rel2}].
Accordingly, the induced spin magnetization $\delta \Mv $ penetrating the superconductor, which is negligibly small near $T_{c}$, increases significantly by reaching temperatures well below $T_{c}$.

\subsection{Strongly spin-polarized ferromagnets}

In the case that the spin splitting of the spin bands in a ferromagnet is much larger than the superconducting gap in the adjacent superconductor, the above mentioned theories must be modified in various respects. First, Usadel approximation assumes that the inverse life time of quasiparticles due to impurity scattering averaged over the Fermi surface, $\hbar/\tau $, is much larger than all energy scales appearing in the Usadel equation, including the magnitude of the exchange field $|\vJ|$. If this is not the case, the spatial variation of superconducting correlations into the ferromagnet happens on length scales comparable to or shorter than the mean free path, and Usadel theory cannot be used within the ferromagnet on the same footing as the exchange field. One must resort to Eilenberger equations in this case. If the exchange splitting is, however, not much smaller than the Fermi energy scale, then even Eilenberger-Larkin-Ovchinnikov theory cannot accommodate the exchange field in the way described in the previous sections. A modified version of Eilenberger and Usadel equations can be used for the case that $|\vJ|\sim E_F$. In this case, there exist two well separated fully spin-polarized Fermi surfaces in the system, and the length scale associated with $\hbar/|\bvpfu-\bvpfd |$ is much shorter than the coherence length scale in the ferromagnet. Superconducting pairs of the type $\uparrow\uparrow $ and $\downarrow\downarrow $ are still long-ranged in such a system, however pairs of the type $\uparrow\downarrow $ and $\downarrow\uparrow$ are negligible within quasiclassical approximation. Both Fermi velocity and density of states at the Fermi level are spin-dependent. The same holds for diffusion constant, and coherence length. The quasiclassical propagator is then spin-scalar for each quasiclassical trajectory in the ferromagnet, 
\be
\hat g_{\uparrow \uparrow} = \left( \begin{array}{cc} g_{\uparrow\uparrow} &f_{\uparrow\uparrow}\\
\tilde f_{\uparrow\uparrow} & \tilde g_{\uparrow\uparrow} \end{array}\right),\quad
\hat g_{\downarrow \downarrow} = \left( \begin{array}{cc} g_{\downarrow\downarrow} &f_{\downarrow\downarrow}\\
\tilde f_{\downarrow\downarrow} & \tilde g_{\downarrow\downarrow} \end{array}\right),
\ee
and only has a particle-hole discrete degree of freedom.
Similarly, all mean field self energies have the same structure (spin-scalar, only 2x2 particle-hole matrices), and the Fermi liquid interactions are replaced by spin-scalar interactions, which in the simplest case do not mix the spin bands: $A^s, \tAa \to A_{\uparrow\uparrow},A_{\downarrow\downarrow}$, $V^s,\tVt \to V_{\uparrow\uparrow},V_{\downarrow\downarrow}$ (in a more general case, the Fermi liquid interactions could be of the form 
$A_{\uparrow\uparrow,\uparrow\uparrow}$, $A_{\uparrow\uparrow,\downarrow\downarrow}$ etc).
Eilenberger equation and Usadel equation have the same form as before for each separate spin band. 
The spin-resolved current densities are given in the ballistic case by
\be
\jv_{\uparrow}=e k_{\rm B}T\sum_{\varepsilon_n} 
\big\langle N_{f\uparrow}\bvvfu g_{\uparrow\uparrow} \big\rangle_{\bvpfscu} ,\quad
\jv_{\downarrow}=e k_{\rm B}T\sum_{\varepsilon_n} 
\big\langle N_{f\downarrow}\bvvfd g_{\downarrow\downarrow} \big\rangle_{\bvpfscd},
\label{currentstrong}
\ee
and the spin magnetization by
\ber
M_z&=&M_{\rm n}(B_z) + 
k_{\rm B}T\sum_{\varepsilon_n}
\left(
\big\langle N_{f\uparrow} \mu_{\rm eff\uparrow}g_{\uparrow\uparrow} \big\rangle_{\bvpfscu}
-\big\langle N_{f\downarrow}\mu_{\rm eff\downarrow}g_{\downarrow\downarrow} \big\rangle_{\bvpfscd}
\right).
\label{magnitizationstrong}
\eer
The expressions in the diffusive limit are modified accordingly: e.g. for the spin-resolved current density one obtains
\be
j_{k\uparrow}=-\frac{e }{\pi}k_{\rm B}T\sum_{\varepsilon_n} 
\mbox{Im} \left[N_{f\uparrow} D_{kj\uparrow}f_{\uparrow\uparrow}^\ast \; \overline\partial_j f_{\uparrow\uparrow} \right],\quad
\label{diffcurrentstrong}
\ee
and analogously for spin down.

\section{Pair amplitude oscillations and $\pi$-Josephson junctions}

\subsection{$0-\pi$ transitions in Josephson junctions}
Proximity-induced pairs in a superconductor-ferromagnet bilayer are subject to the FFLO effect, leading to 
spatially oscillating pair amplitudes in the ferromagnetic regions. These act 
back on the superconducting singlet pair potential in the superconducting region. 
How exactly this interaction between pair potential and FFLO amplitudes takes 
place depends on details of the interfaces and
must be determined numerically, however for a sufficiently thin superconducting layer the oscillating nature of the FFLO 
amplitudes ultimately leads to oscillations in the modulus of the pair potential of the superconductor 
(and thus in the transition temperature) as function of the geometric dimensions of the hybrid structure.
In addition, the oscillatory nature of the FFLO pairs in the ferromagnet affects the properties of 
superconductor-ferromagnet-superconductor Josephson devices, leading to 
the possibility of oscillations between junctions with a phase difference of zero and junctions with a phase difference of $\pi$ in the ground state when a certain control parameter is changed.

\begin{figure}
\centerline{(a) \hfill (b) \hfill}
\includegraphics[width=0.47\linewidth]{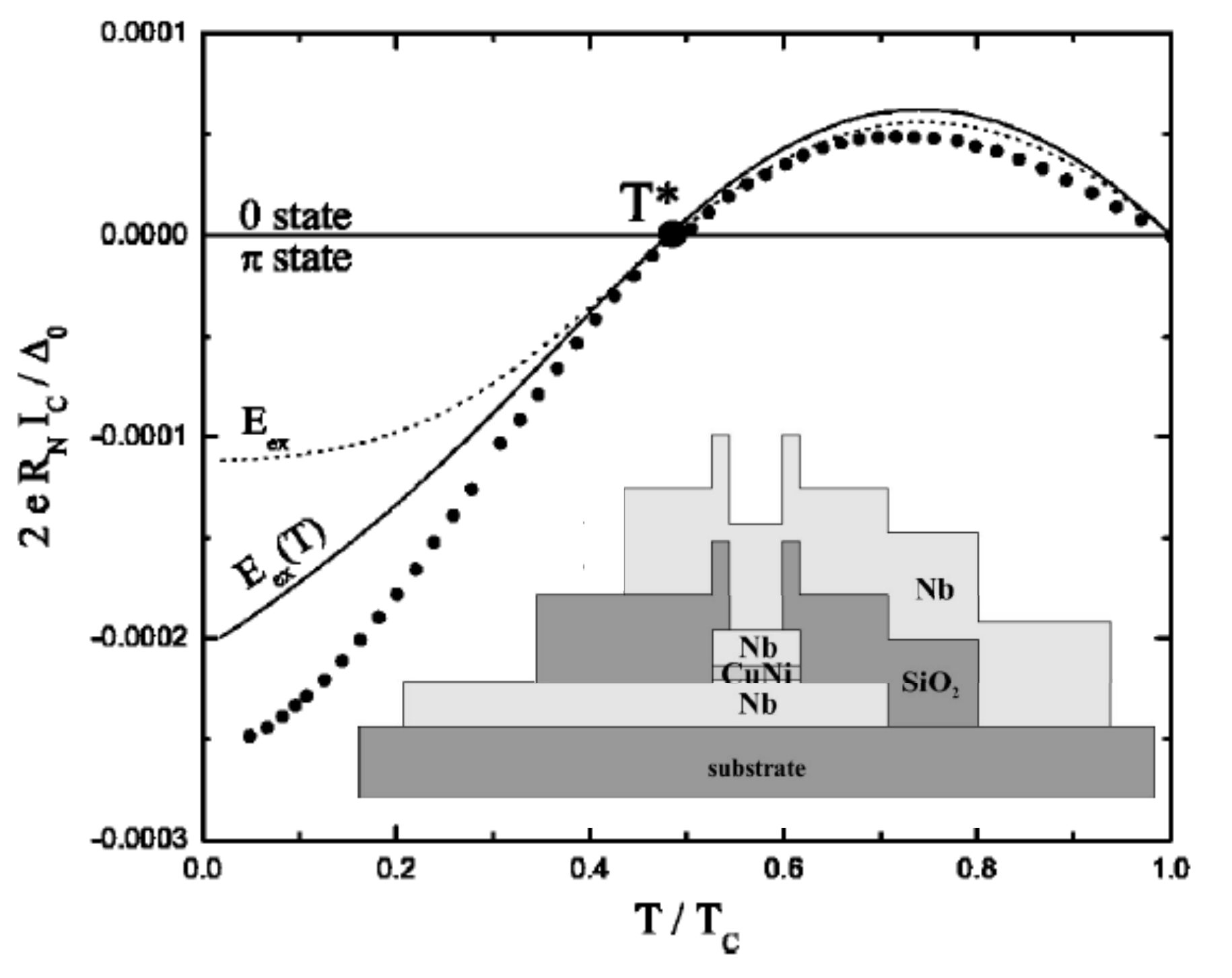}
\includegraphics[width=0.53\linewidth]{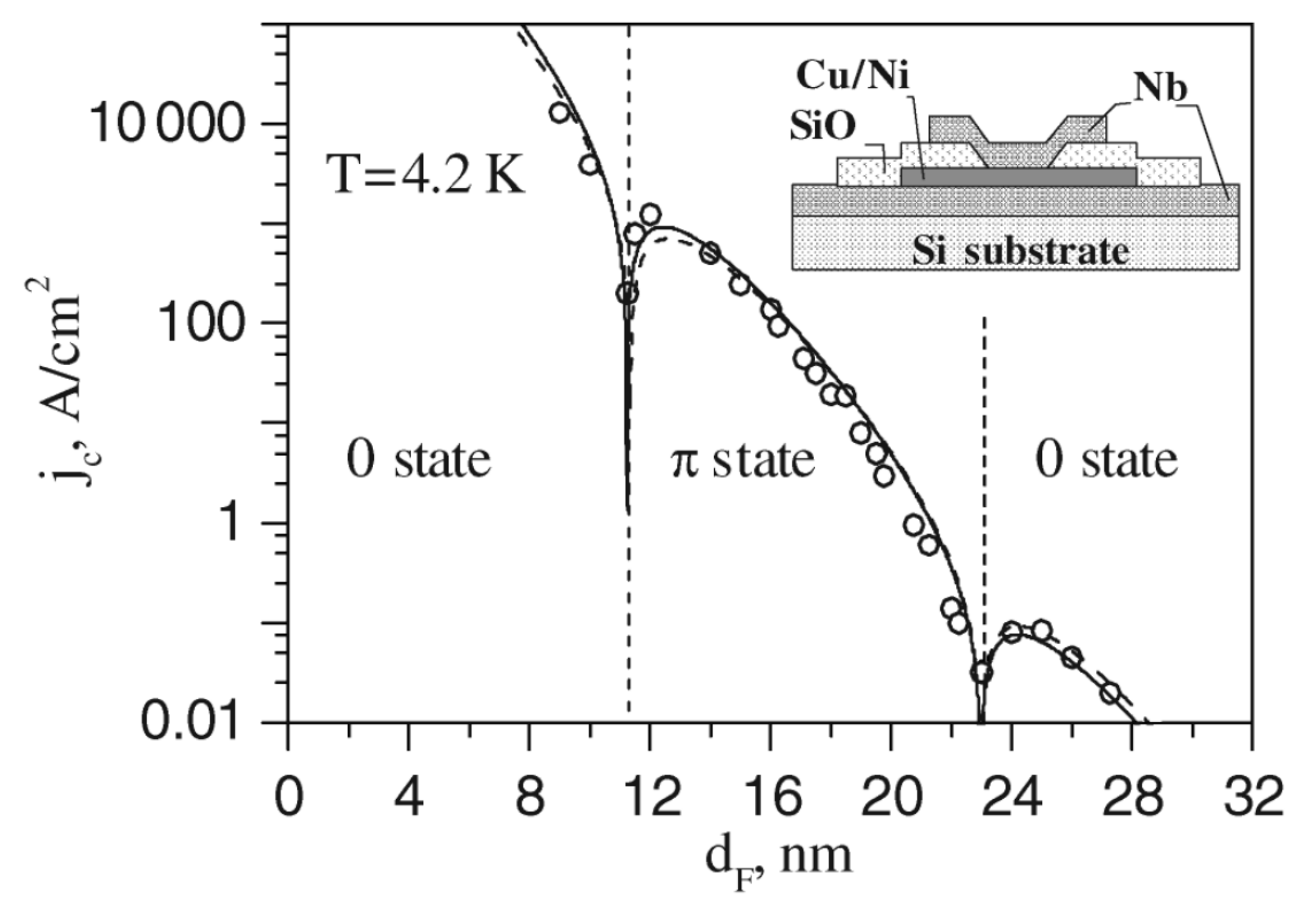}
\caption{
(a) Temperature dependence of the critical Josephson current in a Nb/Cu$_{52}$Ni$_{48}$/Nb trilayer device, shown in the inset, for a ferromagnet layer thickness $d=18$ nm. 
(The lines correspond to theoretical models with constant (dotted line) or temperature dependent (solid line) exchange energy.)
From Sellier {\it et al.} \cite{Sellier03}.
Copyright (2003) by the American Physical Society.
\label{Sellier03}
(b) The ferromagnet-layer thickness dependence of the critical current density for Nb/Cu$_{0.47}$Ni$_{0.53}$/Nb junctions at temperature
4.2 K. Open circles represent experimental results; solid and
dashed lines show model calculations.
The inset shows a schematic cross section of the SFS junctions.
From Oboznov {\it et al.} \cite{Oboznov06}. 
Copyright (2006) by the American Physical Society.
\label{Oboznov06}
}
\end{figure}

The idea of Josephson junctions with a $\pi $ phase difference in
its ground state when a magnetic impurity is inserted in the junction
was introduced by
Bulaevski\u{i}, Kuzi\u{i}, and Sobyanin in 1977 \cite{Bulaevskii77}. Similar ideas had also been proposed by Kulik in 1965 \cite{Kulik65}.
In 1982 it was also shown in a classical paper by 
Buzdin, Bulaevski\u{i}, and Panyukov 
that the pair amplitude oscillates in ferromagnets in contact with a superconductor
\cite{Buzdin82}. Further early studies of this effect followed in Refs.~\cite{Andreev91,Buzdin92,Chtchelkatchev01}.
One technological problem that became evident quickly was that weakly spin-polarized 
systems, like ferromagnetic Cu-Ni or Pd-Ni alloys, are better suited to observe
$0-\pi$ oscillations, as
otherwise the proximity amplitudes become so short ranged that extremely thin layers are necessary.
Typical ferromagnetic alloys that have been used are Cu$_{1-x}$Ni$_x$ (CuNi) and Pd$_{1-x}$Ni$_x$ (PdNi).
The real break-through came in the beginning of the 2000's with the experimental verification
of the switching between 0 and $\pi$-Josephson junctions based on the
effect predicted theoretically before. Pioneering work was done by
Ryazanov and co-workers
who studied transitions between zero- and $\pi $-states as function of temperature \cite{Veretennikov00,Ryazanov01,Ryazanov04,Oboznov06} and thickness of the ferromagnet layer \cite{Ryazanov04,Oboznov06} in Nb/CuNi/Nb trilayers (see Figure \ref{Oboznov06}).
Closely following were experiments by 
Kontos and co-workers
who studied the zero-$\pi$ transition as function of barrier thickness in Nb/Al/Al$_2$O$_3$/PdNi/Nb junctions \cite{Kontos02},
by Blum {\it et al.} \cite{Blum02}, who use Nb/Cu/Ni/Cu/Nb junctions and vary both temperature and thickness of the Ni layer (thus employing a much stronger spin-polarized ferromagnet than the ferromagnetic alloys that were used by the other groups),
and by 
Sellier {\it et al.} 
studying a temperature induced $0-\pi$ transition in Nb/CuNi/Nb (see Figure \ref{Sellier03}), as well as the appearance of half-integer Shapiro steps \cite{Sellier03,Sellier04}.

The critical Josephson current density as function of layer thickness in the limit $k_{\rm B}T,\alpha_{\rm s}\ll |\Jv|$ is described in the clean limit by formula \cite{Buzdin05}
\numparts
\be
I_c(d_{\rm F}) \sim \frac{
\sin\left(\frac{2|\Jvsc|}{\hbar v_f}(d_{\rm F}-d_0)\right)
}
{
\frac{2|\Jvsc|}{\hbar v_f}(d_{\rm F}-d_0)
}
\label{cleanIc}
\ee
where $d_0$ is a fit parameter which is usually assigned to a ``dead layer'',i.e. a layer of suppressed magnetism at the interface.
In the diffusive limit the corresponding expression for long ($d_{\rm F}\gg \xi_1$) junctions is
\be
I_c(d_{\rm F}) \sim e^{-d_{\rm F}/\xi_1}\cos\left(\frac{d_{\rm F}-d_0}{\xi_2}\right)
\label{dirtyIc}
\ee
\endnumparts
with the length scales $\xi_{1,2}$ as in (\ref{xi1}) and (\ref{xi2}).
A study by Pugach {\it et al.} \cite{Pugach11} bridging the two limiting cases showed that
the two length scales $\xi_1$ and $\xi_2$ 
may exhibit a non-monotonic dependence on the properties of the ferromagnetic layer such as exchange field or electron mean-free path.

\begin{figure}
\includegraphics[width=1.0\linewidth]{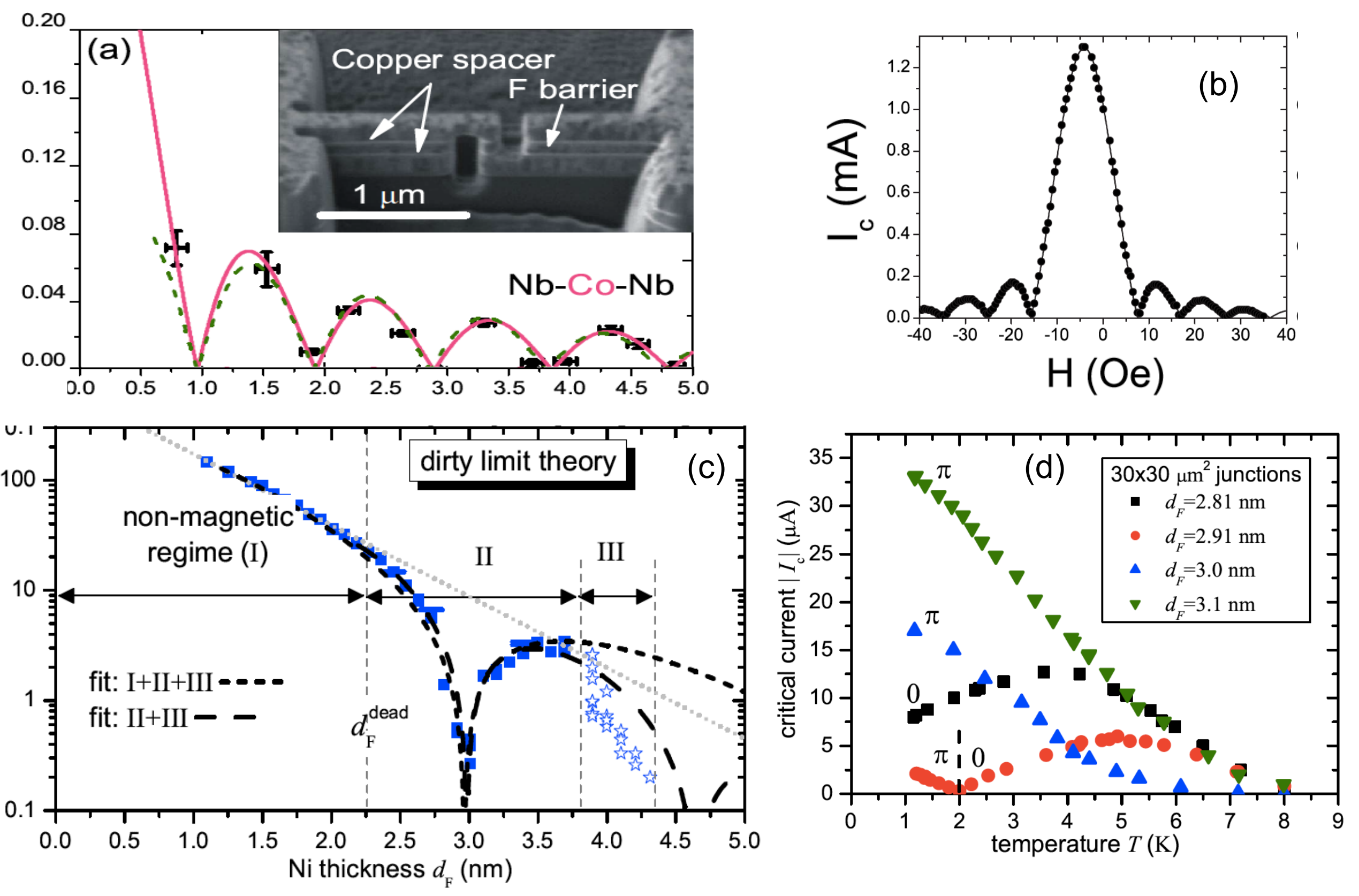}
\caption{
(a)
Characteristic voltage of a Nb/Co/Nb Josephson junction as a function of Co thickness at 4.2K. 
The dashed line is a fit to theory.
Inset: A focused ion beam micrograph of a typical Nb/Co/Nb Josephson junction. 
From Robinson {\it et al.} \cite{Robinson06}.
Copyright (2006) by the American Physical Society.
(b)
Critical current vs applied magnetic field obtained for
Nb/Cu/Co/Ru/Co/Cu/Nb circular Josephson junctions with total Co layers thickness of
6.1 nm.
From Khasawneh {\it et al.} \cite{Khasawneh09}.
Copyright (2009) by the American Physical Society.
(c)
$j_{\rm c}(d_{\rm F} )$ dependence for an SFS junction with F=Ni. 
Two fits appropriate for the dirty limit theory for
data from regimes I+II+III and just II+III are shown as dashed lines.
(d)
$I_{\rm c}(T)$ dependence for SFS Josephson junctions.
A temperature induced 0 to $\pi$  transition is observed (sample $d_{\rm F} = 2.91$ nm).
(c) and (d) from Bannykh {\it et al.} \cite{Bannykh09}.
Copyright (2009) by the American Physical Society.
\label{Khasawneh09}
}
\end{figure}
Further studies of strongly spin-polarized interlayers were performed, using various materials, among those Ni in Ref.~\cite{Shelukhin06}, Co, Ni, and Py (Ni$_{80}$Fe$_{20}$) in Ref.~\cite{Robinson06}, and Py (Fe$_{0.75}$Co$_{0.25}$) barriers in Ref.~\cite{Sprungmann09}.
A detailed study, presented in Ref.~\cite{Bannykh09}, with Ni interlayers using the formulas (\ref{cleanIc}), (\ref{dirtyIc}) above has been done recently, indicating that these structures are in the dirty limit. The thicknesses of the Ni layers ranged between 1 and 5 nm. In Figure \ref{Khasawneh09} examples for $0-\pi$ oscillations in a Nb/Co/Nb structure are shown.
In order to eliminate unwanted effects due to the variation of the electromagnetic vector potential across the junction due to the intrinsic magnetic flux in the junction, a special geometry was used where two Co layers were exchange coupled by a thin Ru layer in between to align antiferromagnetically. This cancels all the unwanted effects to a high degree, and a previously wildly fluctuating critical current vs. magnetic field curve turns into an excellent Fraunhofer pattern when the Ru interlayer present, as seen in  Figure \ref{Khasawneh09} (b) \cite{Khasawneh09}. With this improvement high-quality $0-\pi$ transitions are observed both as function of temperature and of ferromagnet thickness. These results complement the results shown in Figure \ref{Sellier03}, which are for ferromagnetic alloys. The main difference is a one order of magnitude smaller ferromagnet layer thickness necessary to observe the effect, which is an experimental challenge.

The effect of additional normal and/or insulating layers between the superconductor and the ferromagnet was studied theoretically in detail in Ref.~\cite{Heim13}. It was found that even a thin additional normal conducting layer may shift the 0-$\pi$ transitions to larger or smaller values of the thickness of the ferromagnet, depending on its conducting properties, and for certain parameter ranges a 0-$\pi$ transition can even be achieved by changing only the normal layer thickness. 

A promising setup are so-called double-proximity structures, in which two superconductors are connected by a weak link of a normal-metal/ferromagnet bilayer or a ferromagnet/normal-metal/ferromagnet trilayer forming a bridge between the superconducting banks.
The proximity effect acts here twice: first to provide pair amplitudes from the superconducting banks into the bilayer or trilayer, and second at the interface between the normal metal and ferromagnet. Such structures have been proposed by Karminskaya and Kupriyanov \cite{Karminskaya07}, and experimental realization in terms of hybrid planar Al-(Cu/Fe)-Al submicron bridges is reported
in Ref.~\cite{Golikova12}. The oscillation periods and damping lengths of the critical Josephson current can be much longer in such devices than in the conventional setup.
Various geometries have been theoretically studied in subsequent work, including (SN)-(NF)-(SN), (SNF)-(NF)-(SNF), (SNF)-N-(SNF), and S-(NF)-S junctions, in order to optimize practical performance \cite{Karminskaya10}.

\subsection{Phase sensitive measurements }
\begin{figure}
\includegraphics[width=0.55\linewidth]{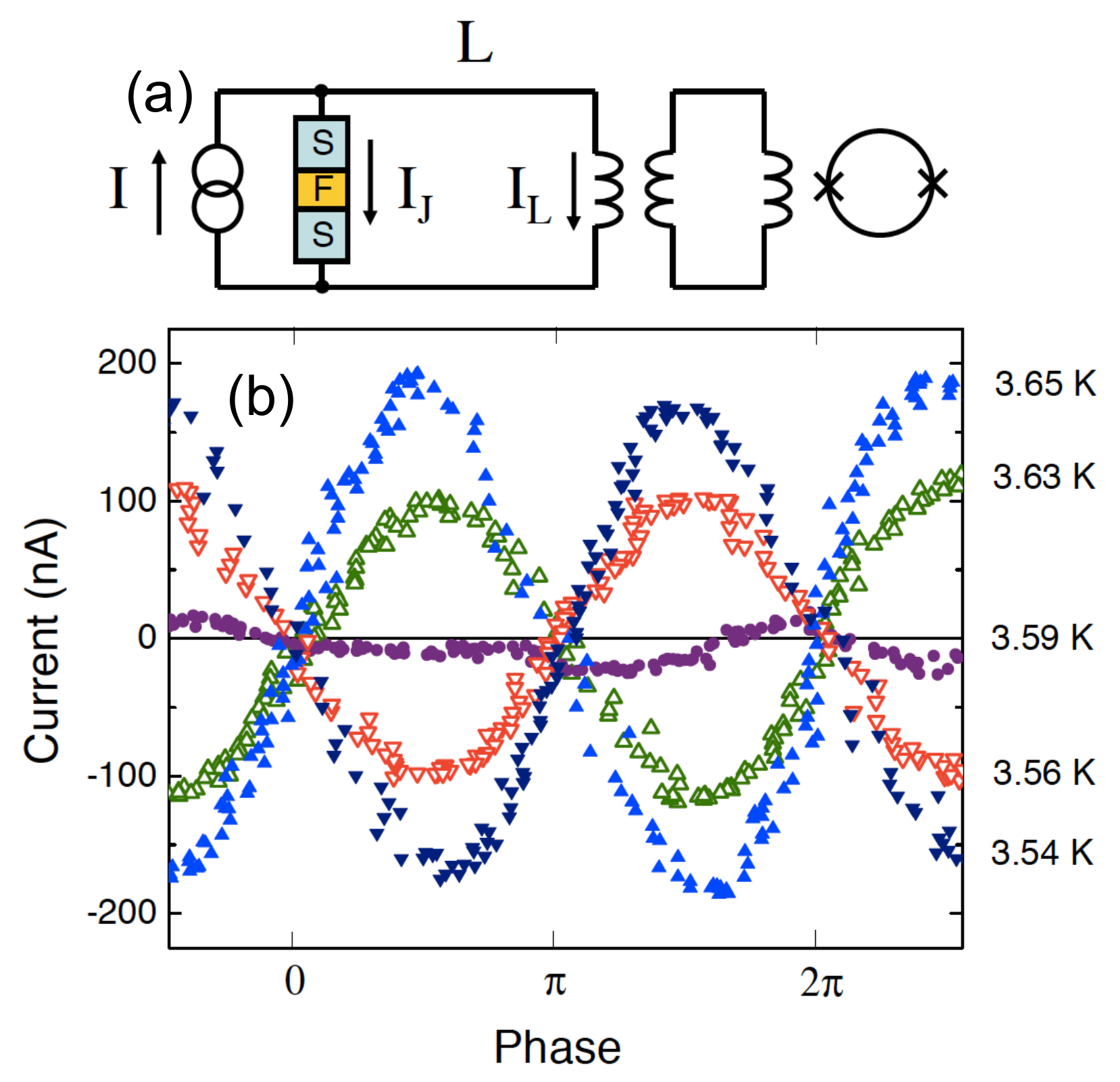}
\hfill
\includegraphics[width=0.45\linewidth]{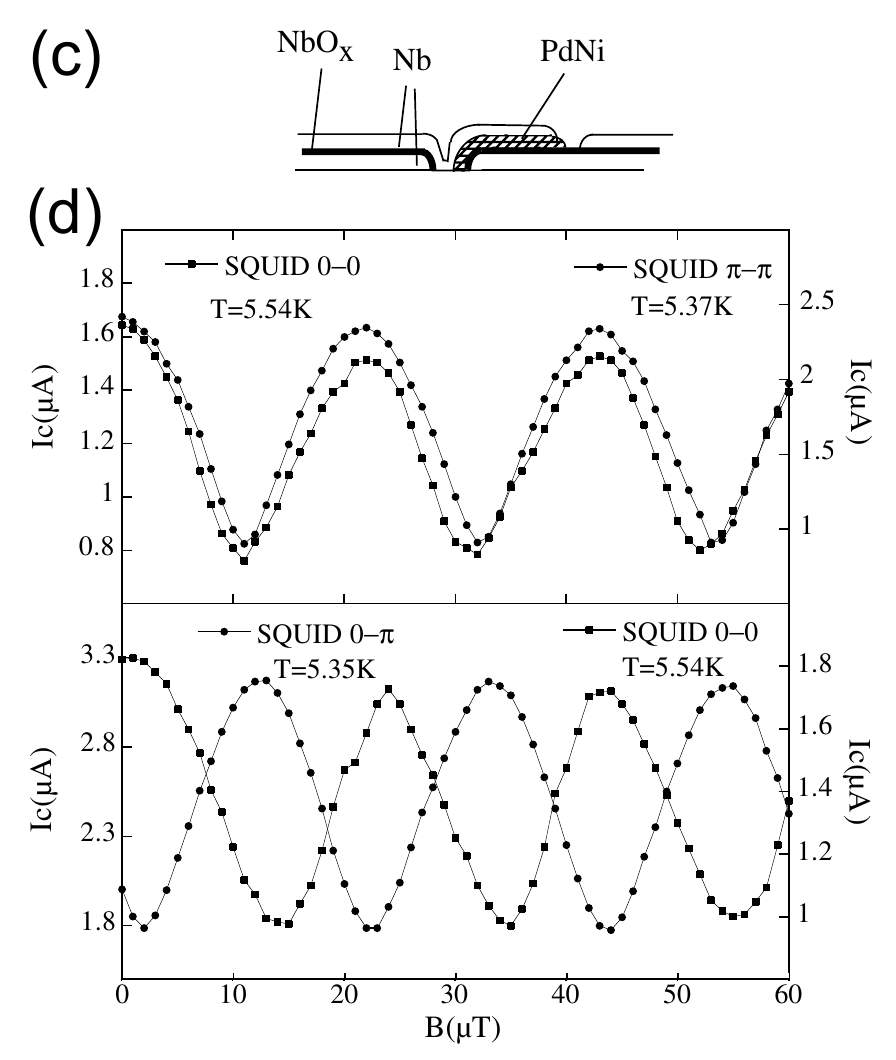}
\caption{
(a)Circuit for measuring the current-phase relations of a SFS junction. 
(b) Current-phase relation derived from rf SQUID
modulation curves showing the transition to a $\pi$ Josephson
junction as the temperature is lowered.
(a) and (b) from Frolov {\it et al.} \cite{Frolov04}.
Copyright (2004) by the American Physical Society.
\label{Frolov04}
(c) 
A drawing of the cross section of a single SQUID junction.
(d)
Critical current modulations showing the expected no-shift between a $0-0$ SQUID and a
$\pi$-$\pi$ SQUID and a shift of $\Phi_0/2$ between a $0-\pi$ SQUID and a $0-0$ SQUID.
As the $T_{\rm c}$ of the $0-0$, $0-\pi$, and $\pi$-$\pi$ SQUIDs are different
the $I_{\rm c}(B)$ curves are observed at different temperatures in order to measure in the same range of critical currents for each couple.
(c) and (d) from Guichard {\it et al.} \cite{Guichard03}.
Copyright (2003) by the American Physical Society.
\label{Guichard03}
}
\end{figure}

The current-phase relation (CPR) as a function of temperature in a Nb/CuNi/Nb junction was measured by Frolov {\it et al.}  (see Figure \ref{Frolov04}) \cite{Frolov04}. Using an rf SQUID, the change of the current-phase relation from zero to $\pi$-junction behavior is clearly visible. This allows to observe the {\it sign} of the supercurrent. A $\pi$-junction is a Josephson junction with a negative critical current $I_{\rm c}$, as its current-phase relation (CPR) is $I_J(\phi)=|I_{\rm c}| \sin (\phi+\pi)=-|I_{\rm c}| \sin (\phi)$.
As conventional measurement of the current-voltage characteristic of a Josephson junction is not sensitive to the sign of the supercurrent, it was necessary to include the Josephson junction in a multiply connected geometry.
The sign was in Ref.~\cite{Frolov04} observed in an rf SQUID configuration [see Figure \ref{Frolov04}(a)] by shorting the electrodes of the junction with a superconducting loop. If that loop contains a $\pi $ junction in zero external field it will exhibit a spontaneous circulating current, generating a flux of $\Phi_0/2$ which can be detected by a SQUID magnetometer or a Hall probe. In Figure \ref{Frolov04} (b) the transition between a zero and a $\pi$ state is unambiguously observed (although residual magnetic fields leading to shifts in the CPR curves did not allow to pin down if it was a 0-$\pi$ or a $\pi$-0 transition).

Phase sensitive measurements of the ground state of ferromagnetic Josephson junctions using a single dc SQUID have been performed 
by Ryazanov and co-workers \cite{Ryazanov01a} for Nb/NbO$_y$/CuNi/Nb junctions, and
by Guichard {\it et al.} \cite{Guichard03} for Nb/NbO$_y$/PdNi/Nb junctions, showing that the sign change of the Josephson coupling is observed as a shift of half of a flux quantum $\Phi_0=\frac{h}{2|e|}$ in the SQUID diffraction pattern (see Figure \ref{Guichard03}). A 0-0 and a $\pi$-$\pi$ junction show identical critical current modulations with flux, whereas a 0-$\pi$ junction exhibits a pattern shifted by $\Phi_0/2$ with respect to that for 0-0 and $\pi$-$\pi$ junctions. 
In this experiment a single dc SQUID with two SFS junctions embedded. The ferromagnetic layer thickness $d_{\rm F1}$ and $d_{\rm F2}$ can then be chosen to correspond to zero or $\pi$ coupling independently, leading to the possibility of a $0-0$, a $\pi$-$\pi$, a 0-$\pi$, or a $\pi$-0 SQUID. For a dc SQUID with negligible loop inductance and equal junction critical currents the modulation of the critical current with applied flux is given by \cite{Barone82}
\be
I_{\rm c} (\Phi_{\rm ext})=2I_0 \left| \cos \left( \pi \frac{\Phi_{\rm ext}}{\Phi_0} + \frac{\delta_{12}}{2}\right)\right|
\ee
where $\delta_{12} $ is the sum of the internal phases (0 or $\pi $) in the two junctions of the SQUID. Thus, for a $0-0$ and a $\pi$-$\pi$ SQUID the $I_{\rm c}(\Phi_{\rm ext})$ patterns are identical, whereas they are shifted by half a flux quantum if $\delta_{12}=\pi$, i.e. a $0-\pi$ or a $\pi$-0 SQUID. This shift is shown in Figure \ref{Guichard03} (d).

\begin{figure}
\includegraphics[width=1.0\linewidth]{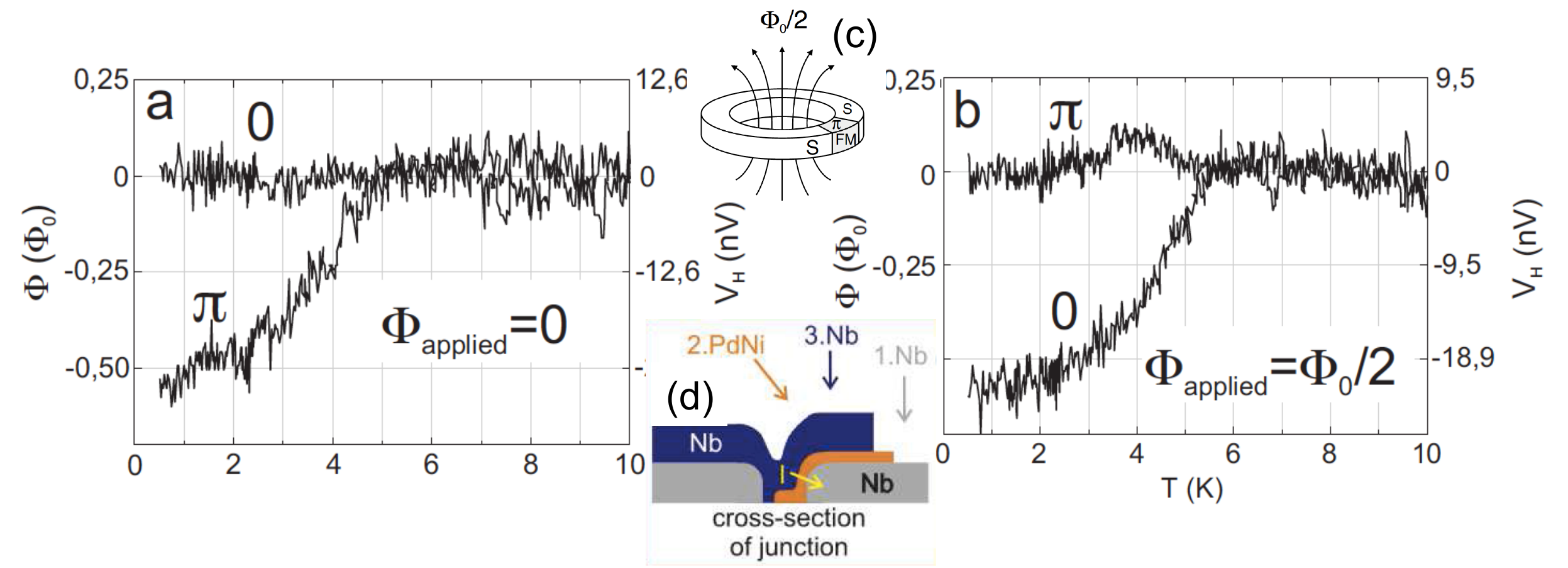}
\caption{
Temperature dependent magnetic
flux produced by a $\pi$- and a 0-loop when cooling (a) in zero field
and (b) in a magnetic field equal to half a flux quantum in the
loop. Below $T\approx $5 K the
loop develops a spontaneous
current when cooling down in zero field, while the magnetic
flux through the $0-$loop remains zero. When applying a field
equal to half a flux quantum $\Phi_0$ 
the roles of the
$\pi$- and the 0-loops are interchanged. 
At low temperatures the spontaneous flux saturates at a value close to $\Phi_0/2$
in each case.
(c) 
In zero external magnetic field, a spontaneous flux $\Phi=\Phi_0/2$ 
is required to maintain the condition of fluxoid quantization if a
$\pi$-junction is inserted into a superconducting loop 
(d)
Cross section of the planar Josephson junction.
From Bauer {\it et al.} \cite{Bauer04}.  
Copyright (2004) by the American Physical Society.
\label{Bauer04}
}
\end{figure}
The presence of spontaneous magnetic moments in superconducting (Nb) loops containing a ferromagnetic (PdNi) $\pi$ junction was experimentally demonstrated by Bauer {\it et al.} (see Figure \ref{Bauer04}) \cite{Bauer04}. The loops were prepared on top of a micro-Hall sensor. The authors observed asymmetric switching of the loop between different magnetization states when reversing the sweep direction of the magnetic field. The presence of a spontaneous current near zero applied field was studied as function of temperature, and the magnetic moment approached half a flux quantum at low temperatures.

An rf SQUID geometry where the macroscopic ground state
of a ferromagnetic (PdNi) Josephson junction shorted by a 0 weak link was experimentally investigated by Della Rocca {\it et al.}
and showed spontaneous half quantum vortices with random sign ($\pm \Phi_0/2$) \cite{DellaRocca05}; it was found that $0-\pi$ junctions behave as classical spins.

\subsection{$\phi$-Josephson junctions}

\begin{figure}
\includegraphics[width=0.58\linewidth]{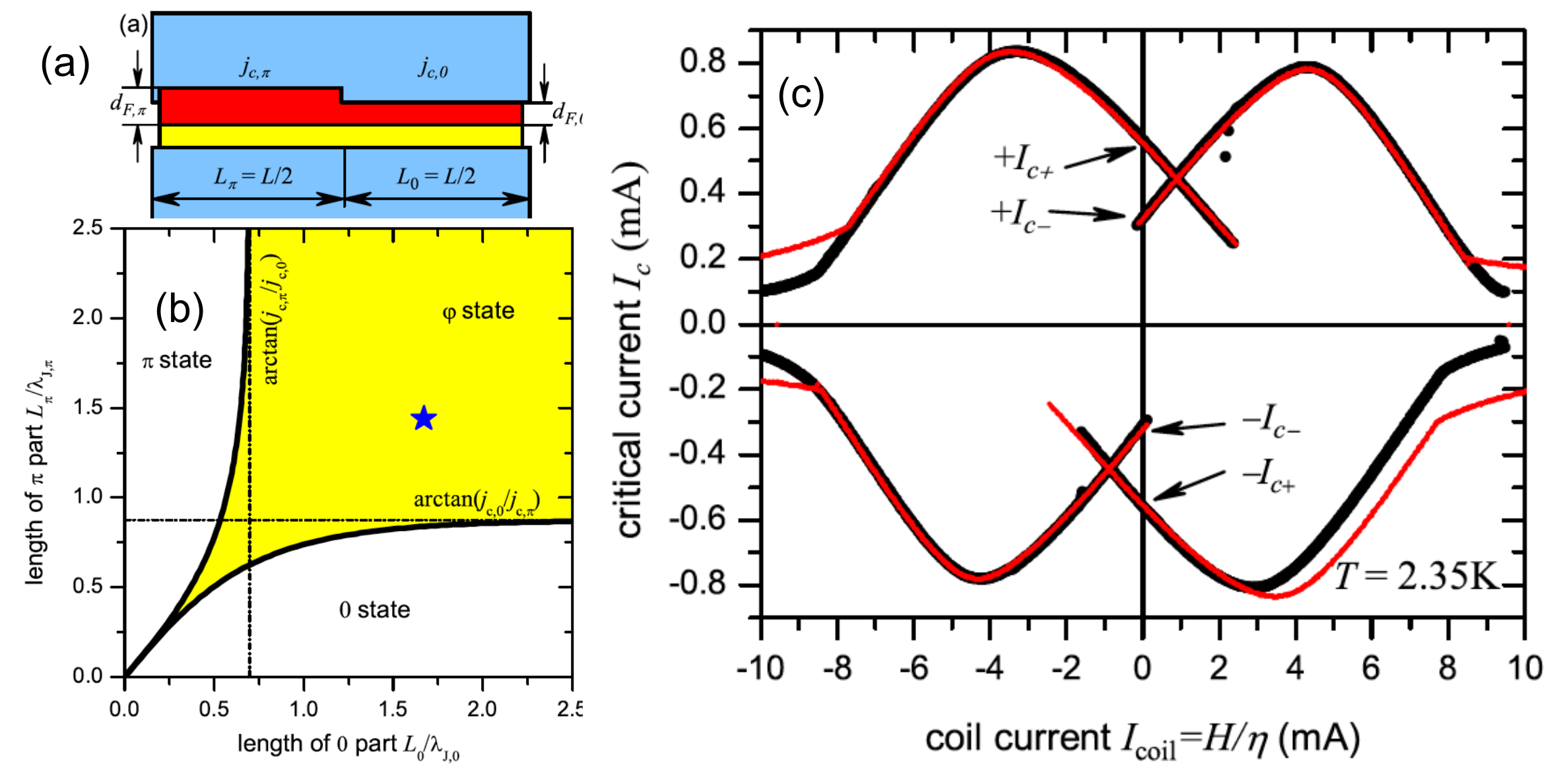}
\includegraphics[width=0.42\linewidth]{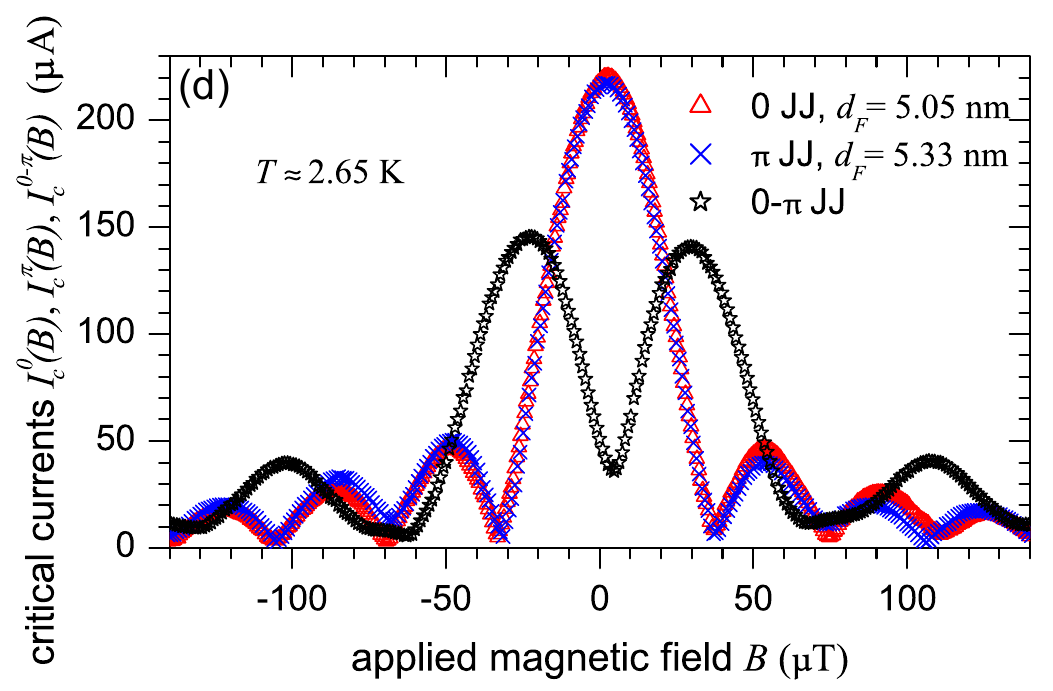}
\caption{
(a) Sketch (cross section) of the investigated SIFS 
0-$\pi$ Josephson junctions 
(Nb/Al-Al$_2$O$_3$/Ni$_{0.6}$Cu$_{0.4}$/Nb) 
with a step in the ferromagnet layer thickness.
(b) Domains of existence of 0-, $\pi$- and $\phi$-Josephson junctions.
The star shows the position of the investigated Josephson junction in (c).
(c)
Experimentally measured $I_{\rm c}(B)$
(black symbols) and theoretical predictions (red symbols).
(a)-(c) from Sickinger {\it et al.} \cite{Sickinger12}.
Copyright (2012) by the American Physical Society.
(d)
$I_c(B)$ of $0-$junction (red triangles), $\pi$-junction (blue triangles), and 0-$\pi$ junction (black spheres). 
From  Weides {\it et al.} \cite{Weides07}.
With kind permission from Springer Science and Business Media.
\label{Weides07}
}
\end{figure}

An interesting separate development concerns a geometry where a step in the thickness of the ferromagnet is present in a superconductor-ferromagnet bilayer, coupled to another superconductor via an insulating barrier. On one side of the step one has a 0 junction, on the other side a $\pi $ junction. A half flux quantum (semifluxon) is trapped in such a structure at the step \cite{Xu95,Goldobin02,Hilgenkamp03}, and a supercurrent circulates in the structure, similarly as in a $0-\pi$ SQUID \cite{Weides06,Weides07,Pfeiffer08,Kemmler10}. Such structures were studied in detail by Weides, Kohlstedt, Koelle, Kleiner, Goldobin and co-workers. In particular, there exists in such structures the possibility of a $\phi$-junction, which has neither 0 nor $\pi$ phase difference as a ground state. A method to realize a $\phi$ Josephson junction by combining alternating 0 and $\pi$ parts with intrinsically non-sinusoidal current-phase relation was suggested in Ref.~\cite{Pugach10}.
The realization of an SIFS $\phi $-junction has been experimentally achieved in 2012 (see Figure \ref{Weides07}) \cite{Sickinger12}.
Properties of zero-$\pi$ junctions which act as a Josephson junction with an equilibrium phase difference $0<\phi <\pi$ are discussed in Ref.~\cite{Lipman14}, where the current-phase relation is calculated numerically and in certain limiting cases analytically.

In Refs.~\cite{Bakurskiy13,Heim13a}, a planar Josephson junction with a ferromagnetic weak link located on top of a thin normal metal film is considered. It is shown that this Josephson junction is a promising candidate for the realization of a $\phi$-junction.

\section{Long-range triplet supercurrents}
\subsection{Length scales for superconducting correlations in ferromagnets}
As detailed in the sections \ref{secNorm} and \ref{secUsadel}, proximity induced superconducting correlations in ferromagnets with sufficiently strong exchange splitting $|\vJ|$ come in pairs of short-range and long-range amplitudes. The former ones are 
the singlet amplitude and the triplet pair amplitude with zero spin projection, $S_z=0$, on the magnetization axis of the ferromagnet. These short-range amplitudes decay in ballistic structures algebraically with a reduced magnitude of order $\Delta/|\vJ|$, and in diffusive structures exponentially on the length scale $\xi_J=\sqrt{\hbar D/|\vJ|}$. The other two components, called long-range amplitudes, decay 
on the thermal coherence length scale $\xi_T$, given for ballistic structures by
$\hbar v_f/(2\pi k_BT+2\alpha_{\rm s})$ and for diffusive structures by $\sqrt{\hbar D/ (2\pi k_BT+2\alpha_{\rm s})}$.
They are characterized by triplet amplitudes with a spin projection $S_z=\pm 1$ on the magnetization direction of the ferromagnet. Consequently, they are called equal-spin triplet pair correlations. Equal spin pairs do not suffer from having to populate spin-split pairs of Fermi surfaces. Both electrons of the pair are situated on the same spin component of the Fermi surface. For this reason, they behave like in a normal metal, with the coherence length determined by the Fermi surface properties of one spin component only (which in general differ for $S_z=+1$ and $S_z=-1$). The quest for such long-range pair amplitudes has a long history, and is one of the success stories of interaction between experiment and theory.

\subsection{Experimental ``pre-history''}
As in the 1990's the field of spintronics, i.e. 
functional nanometer-size devices based on spin-dependent transport 
phenomena, developed rapidly \cite{Zutic99}, a strong motivation to search for a 
long-range proximity effect in ferromagnets was established.
Such long-range amplitudes would lead to long-range supercurrents
in Josephson devices, and would ultimately lead to valuable
applications. The ultimate goal is to obtain completely spin-polarized
supercurrents, which would necessarily have to be triplet supercurrents.

It was for this reason that in the second half of the 1990's experimental
efforts intensified to study proximity effects in strongly spin-polarized
ferromagnets. 
Petrashov, Antonov, Maksimov, and Sha\u{i}kha\u{i}darov found in 1994 that the effect of superconducting islands, deposited on the surface of ferromagnetic Ni, can be still seen in the resistance of the Ni layer distances greater than 2 $\mu$m (exceeding by more than 30 times what the length $\hbar v_f/|\Jv|$ would suggest) away \cite{Petrashov94}.

Lawrence and Giordano studied in 1996 the resistance as a function of temperature and magnetic field in structures containing a narrow ($\sim 2\mu$m) ferromagnetic strip connecting two superconducting films (In/Ni/In and Pb/Ni/Pb) \cite{Lawrence96}. They observed a magnetoresistance dip as function of magnetic field much too large to be accounted for by weak localization effects. In addition they found implausibly long phase coherence lengths, inconsistent with the usual superconducting proximity effect.
In a later study of Sn/Ni/Sn structures, where the Ni was a narrow wire ($\sim 40$ nm wide), they measured an unexpectedly long proximity length of about 50 nm (theory predicted 4 nm for these structures) when the Sn/Ni interfaces are clean, however with an oxide layer at the interfaces they observed an unexplained re-entrant behavior \cite{Lawrence99}. They also noted that their proximity length  was of the order of the thermal length $\sqrt{\hbar D/k_{\rm B}T}$ which would appear in a normal metal.

Giroud and co-workers studied in 1998 Co/Al structures with a ferromagnetic Co wire in contact with superconducting Al \cite{Giroud98}.
They observed below the superconducting transition that the Co resistance exhibited a significant dependence on temperature and voltage, and the differential resistance showed that the decay length for the proximity effect was much larger than expected from the exchange field of Co.

Petrashov and co-workers found in 1999 a giant mutual proximity effect in Ni/Al structures with an proximity-induced conductance on the Ni side two orders of magnitude larger than predicted by theory \cite{Petrashov99}. 
Aumedato and Chandrasekhar, by performing multi-probe measurements on Ni/Al structures, re-evaluated these studies and came to the conclusion that superconducting correlations cannot extend into a ferromagnet over distances larger than the exchange length, and associated the large changes in resistance seen in the previous experiments with the superconductor/ferromagnet interface or the superconductor,
which had been measured in series or in parallel with the ferromagnet \cite{Aumentado01}.
The controversy lead to a vivid discussion about the existence of such long-range proximity components.

\subsection{Theory of long-range proximity amplitudes }
\subsubsection{Spiral magnetic inhomogeneities and domain walls}

The renewed interest created by the lack of understanding of long-range superconducting proximity effects in ferromagnets lead in the beginning of the 2000's to new theoretical developments.
Pivotal was a series of papers by Bergeret, Volkov, Efetov, and coworkers, who 
studied the effect of a spiral Bloch-type inhomogeneity close to an
S/F interface, where the magnetization vector rotates along the junction direction. They found that an equal-spin triplet pair amplitude
shows a long-range penetration into the ferromagnet \cite{Bergeret01,Bergeret05a,Konstandin05,Volkov06,Volkov08}.
The authors also noted that this, as an isotropic (``$s$-wave'') triplet pair component,
must be a realization of odd-frequency pairing. As impurities suppress all anisotropic pairing components, this odd-frequency amplitude is the only superconducting pairing amplitude present
in the ferromagnet.
Shortly after that, Kadigrobov, Shekhter and Jonson found a similar effect \cite{Kadigrobov01}.
A realization of these ideas motivated by experiment \cite{Robinson10} was studied in Ref.~\cite{Alidoust10}, where the focus of study was a setup of a Ho/Co/Ho trilayer sandwiched between conventional $s$-wave superconducting leads. Holmium is a conical ferromagnet with an intrinsic spiral structure.
In Ref.~\cite{Wu12} a bilayer consisting of an ordinary singlet superconductor and a magnet with a spiral magnetic structure of the holmium or erbium type was investigated by solving self-consistently Bogoliubov-de Gennes equations. A re-entrance behavior as function of temperature is observed for certain parameter ranges.

Another type of magnetic structure discussed in the literature is a domain structure of the N\'eel type, in which the magnetization vector lies parallel to the interface and rotates along a direction parallel to the interface. 
A setup with a chiral ferromagnet, exhibiting a homogeneous cycloidal spiral, placed between two conventional superconductors was studied in Ref. \cite{Champel08}. Depending on the spiral wave length, 0-$\pi$ transitions can be induced. In the case of a uniformly rotating spiral, however, it was shown that only short-range components exist \cite{Champel08}. In a more general case, with magnetic domains separated by N\'eel walls, long-range triplet components are present and arise at the domain walls, decaying inside the domains \cite{Volkov05,Fominov07}.

Refs.~\cite{Linder09a,Alidoust10a} investigate
diffusive SF and SFS structures with a domain wall in the F layer, using a Riccati representation of the non-linear Usadel equations \cite{Eschrig04} and including spin-mixing parameters for interfaces.
They study local density of states, induced magnetization, and spin-polarized Josephson currents.
It is found that the spin polarization of the spin current in SFS structures is determined by the magnetization profile and that the spin current shows discontinuities at the zero-$\pi$ transition points of the critical Josephson current. Similar as in Ref.~\cite{Grein09}, a non-zero spin-current even for zero superconducting phase difference is reported.
In Ref.~\cite{Buzdin11} the contribution of domain walls to the Josephson current through a ferromagnetic metal is examined for both ballistic and diffusive systems. It is found that in the clean limit domain walls enhance the Josephson current even for a collinear magnetic domain structure, whereas in the diffusive limit a non-collinear domain structure is necessary to enhance the effect.
In Ref.~\cite{Linder14} the influence of the location of a domain wall within a Josephson junction on the ground state properties of the junction was the topic of investigation. The authors considered both the diffusive and ballistic limits. They find that the location of the domain wall determines if the junction is in a zero-state or in a $\pi$-state and influences the transition temperature of the junction.

A different type of system with inhomogeneous magnetic structure was studied in Refs. \cite{Kalenkov11,Silaev12}, where a ferromagnetic vortex in a mesoscopic ferromagnetic disc is brought in contact with a singlet superconductor. Proximity-induced long-range triplet components are generated by the ferromagnetic vortex under certain circumstances, leading to zero- or $\pi$-junction behavior depending on the contact geometry.

\subsubsection{Multi-layer geometries}
Instead of considering inhomogeneous magnetization profiles within a ferromagnetic layer as sources of long-range triplet correlations, one can also obtain the same effect in a multilayer geometry with non-collinear arrangement of the magnetizations of the various layers. 
In Ref.~\cite{Golubov02} a diffusive SFIFS junction is investigated by self-consistently solving 
Usadel equations with appropriate boundary conditions. It is found that for antiparallel ferromagnet magnetizations the critical current is enhanced by the exchange energy, and for parallel magnetization the junction exhibits a transition to a $\pi$-state. A switching behavior between zero- and $\pi$-state can be observed when going from the antiparallel to the parallel alignment.
In Refs.~\cite{Volkov03,Bergeret03} a multilayered superconductor-ferromagnet structure with non-collinear alignment of the magnetizations of the ferromagnetic layers is considered, and the resulting long-range odd-frequency triplet condensate amplitude is calculated. It is found that this setup allows for a Josephson effect with possible zero- or $\pi$-junction behavior. It is also pointed out that the chirality of the magnetization profile matters for the direction of the Josephson current.
In Ref.~\cite{Lofwander05} a fully self-consistent calculation was performed for a ferromagnet-superconductor-ferromagnet trilayer setup with non-collinear magnetization, and the induced spin magnetization in the entire structure was determined. This work includes self-consistency of the order parameter in the superconducting layer at arbitrary temperatures, arbitrary interface transparency, and any relative orientation of the exchange fields in the two ferromagnets.
The long-range Josephson effect through a ferromagnetic trilayer was the topic of Ref.~\cite{Houzet07}, where conditions were derived under which the long-range triplet proximity effect dominates the short-range proximity effect. Both diffusive and clean limits were studied.

Halterman and Valls have investigated numerically ballistic S/F/S trilayer and S/F/S/F/S pentalayer Josephson junctions by self-consistently solving Bogoliubov-de Gennes equations, concentrating on the thermodynamic stability of zero- and $\pi$-junctions as function of material parameters and geometry \cite{Halterman04,Halterman06}. 

Zero-$\pi$ transitions in Josephson junctions with a diffusive pentalayer-structure of the form F/S/F/S/F with misaligned magnetizations in the three ferromagnetic layers were explored in Ref.~\cite{Lofwander07}. The transition temperatures for zero- and $\pi$-junctions as function of misalignment angle were determined numerically using an effective and fast procedure, which is described in detail in this publication.
In Ref.~\cite{Volkov10} a diffusive S/F$'$/FF/F$'$/S structure is discussed, with two different ferromagnetic materials F and F$'$. The middle FF double layer is considered to be parallel or antiparallel magnetized, whereas the F$'$ layers were allowed to be non-collinear. This setup is in particular adapted to the experiments by Khaire {\it et al.} \cite{Khaire10}. A similar study for clean or moderately diffusive materials was performed in Ref.~\cite{Trifunovic10}.

An SF$_1$F$_2$S junction in the ballistic case (including moderate disorder in the ferromagnets) was considered in Refs.~\cite{Trifunovic11a,Knezevic12}, with misaligned ferromagnetic layers F$_1$ and F$_2$. 
It was found that
the long-range spin-triplet correlations lead to a dominant
second harmonic in the Josephson current-phase relation of asymmetric junctions \cite{Trifunovic11}, an effect also found for diffusive structures in Ref.~\cite{Richard13}. In the latter reference this phenomenon is traced to the long-range coherent propagation of two triplet pairs of electrons, a process first pointed out and discussed in Ref.~\cite{Grein09}, and called there ``crossed pair transmission''.

Magnetic moment manipulation by triplet Josephson currents in Josephson junctions with multilayered ferromagnetic weak links is discussed in Ref.~\cite{Pugach12}. It is reported that by tuning the Josephson current, one may control a long-range induced
magnetic moment, which appears due to non-collinear magnetization of the layers. Alternatively, applying a voltage one can in such a geometry generate an oscillatory magnetic moment. 
The spin-switching behavior as function of rotation of the magnetization in one ferromagnetic layer in diffusive SFSFS and SFSFFS Josephson junctions was investigated in Ref.~\cite{Alidoust14}. 

The proximity effect in clean superconductor-ferromagnet structures caused by either the 
spatial or momentum dependence of the exchange field was discussed in Ref.~\cite{Melnikov12}. It was shown that in symmetric junctions the long-ranged proximity effect is present for any non-collinear ferromagnetic moment orientation and a dominant second harmonic appears for any asymmetric junction.
The Josephson coupling between two $s$-wave superconductors separated by a ferromagnetic trilayer with non-collinear magnetization was reconsidered in Ref.~\cite{Meng13}, where in particular the dependence on strength of the exchange field and thickness of the interface layers was studied, and the competition between long-range and short-range components was investigated.

The appearance of a spin supercurrent in a diffusive SF$_1$F$_2$S Josephson junction was subject of study in Ref.~\cite{Shomali11}. It was found that a spin current with spin polarization normal to the magnetization vectors flows between the ferromagnetic domains. This spin current is even as function of superconducting phase difference, and odd as function of misalignment angle between F$_1$ and F$_2$.

A diffusive S-(FNF)-S double proximity structure, in which an FNF trilayer with mutually misaligned ferromagnetic magnetization vectors bridges two superconducting banks, was investigated in Ref.~\cite{Karminskaya08}. In this case long-range equal-spin triplet correlations are generated by the misalignment of the two ferromagnetic layers in the bridge.

\subsubsection{Singlet-Triplet conversion at interfaces}

The mechanisms discussed in the previous subsections employ a quasiclassical approximation, which is not valid for length scales as small as the Fermi wavelength. In fact, for the mechanism to work it is crucial that short-range components can enter the ferromagnet over a sufficiently long distance to feel the magnetic inhomogeneity, which was assumed to vary on the scale much larger than the Fermi wavelength.
This poses the problem of what happens at interfaces between a superconductor and a ferromagnet with large exchange splitting, so that the length scale $\xi_J$ becomes comparable to the Fermi wave length.

An appropriate mechanism for structures with strong exchange splittings was proposed, and first exemplary studied for the extreme case of a half-metallic ferromagnet (where 
one spin band is insulating and one spin band metallic), in \cite{Eschrig03,Kopu04,Eschrig07,Eschrig09}. 
The mechanism is based on spin-dependent scattering phases under reflection and transmission, which in combination with a misalignment with respect to the 
ferromagnetic bulk magnetization of the interface magnetic moment 
in an atomically thin interface layer leads to the creation of long-range triplet pairs. 
It was shown subsequently, that this mechanism persists for the full
range from ballistic to diffusive systems \cite{Eschrig08}. Note that such a mechanism invokes the magnetic inhomogeneity directly in an {\it interface barrier}, rather than in a metallic region. It thus differs from the mechanism suggested by Bergeret {\it et al}. 
Indeed, the two mechanisms are complementary, with the first one
being important for ferromagnets where the exchange splitting is small 
compared to the Fermi energy, and the second one being important for
the case of comparable exchange splitting and Fermi energy, both being
much larger than all superconducting energy scales. The crossover between these two asymptotic cases cannot be described within either theory, as no useful expansion parameter is present in the crossover region.

The mechanism of long-range triplet pair creation at interfaces was generalized to Josephson structures involving strongly spin-polarized ferromagnets with two itinerant spin bands in Ref.~\cite{Grein09}.

\subsubsection{Supercurrents in strongly spin-polarized itinerant ferromagnets and half metals}
\label{geometric}

\begin{figure}
\begin{minipage}{0.55\linewidth}
\begin{overpic}[width=1.0\linewidth]{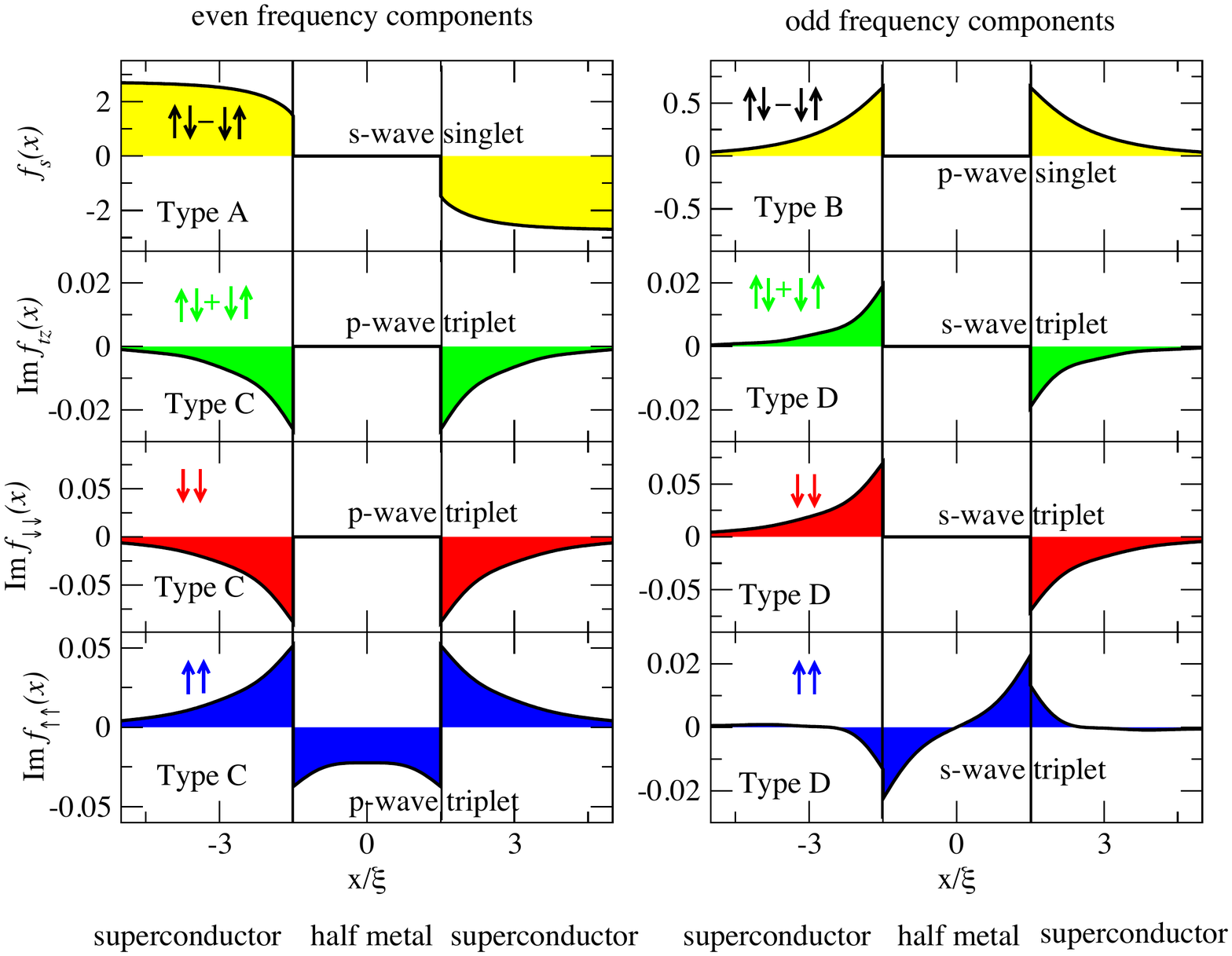}
\put(5,70){\makebox(0,2){(a)}}
\end{overpic}
\end{minipage}
\begin{minipage}{0.45\linewidth}
$\,$ \begin{overpic}[width=0.9\linewidth]{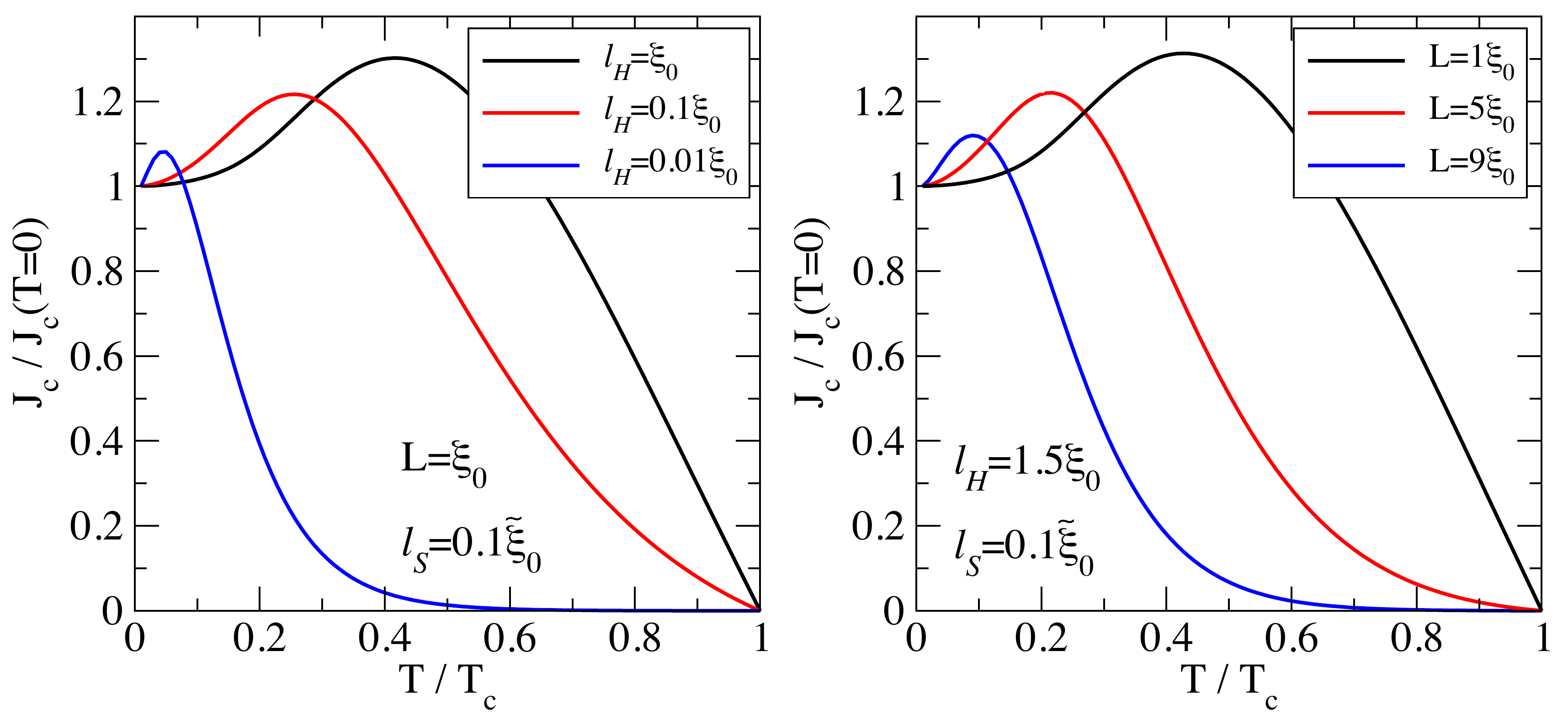}
\put(0,40){\makebox(0,2){(b)}}
\put(52,40){\makebox(0,2){(c)}}
\end{overpic}\\
\hspace*{0.2cm}\begin{overpic}[width=0.45\linewidth]{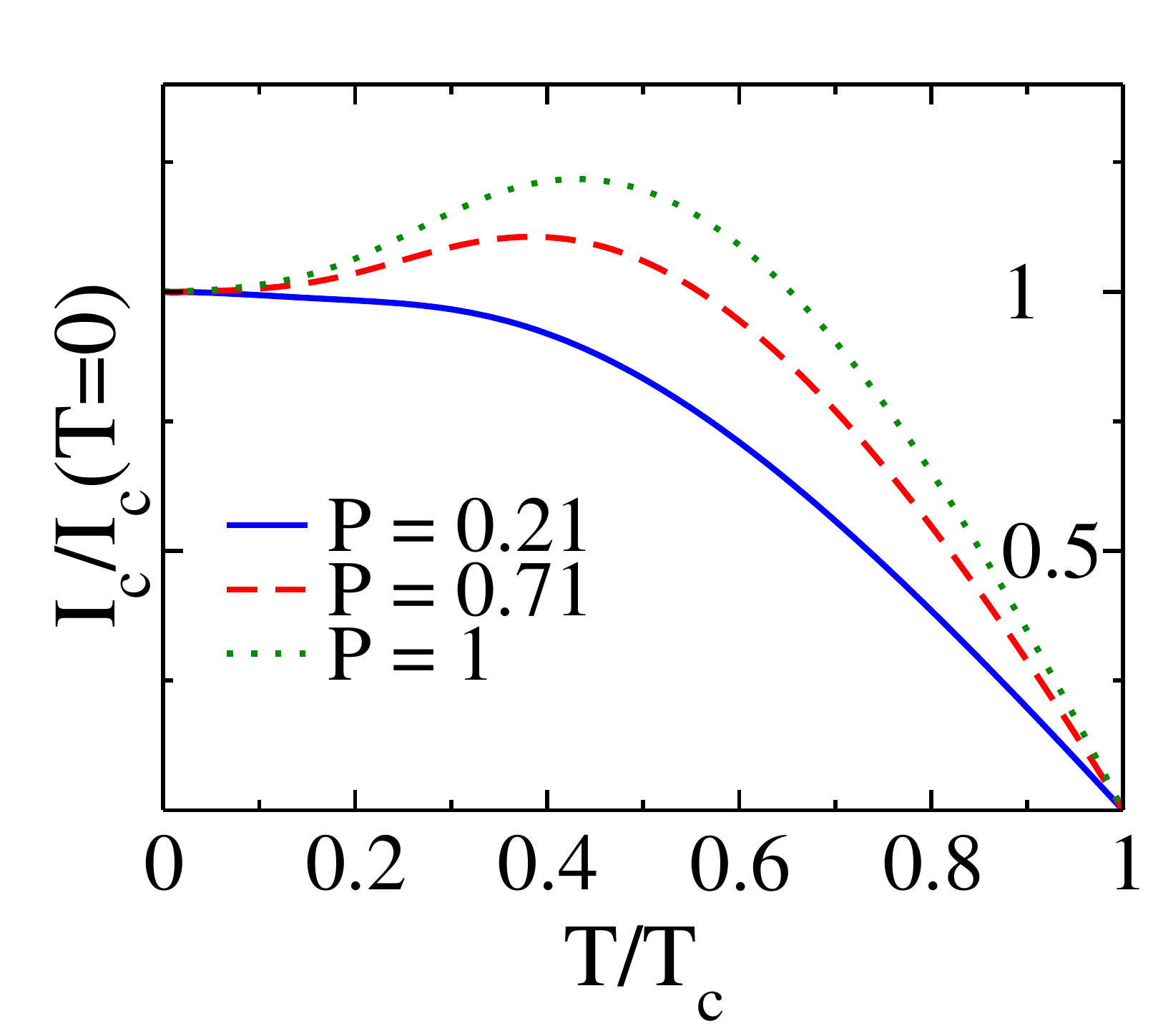}
\put(3,75){\makebox(0,2){(d)}}
\end{overpic}
\begin{overpic}[width=0.5\linewidth]{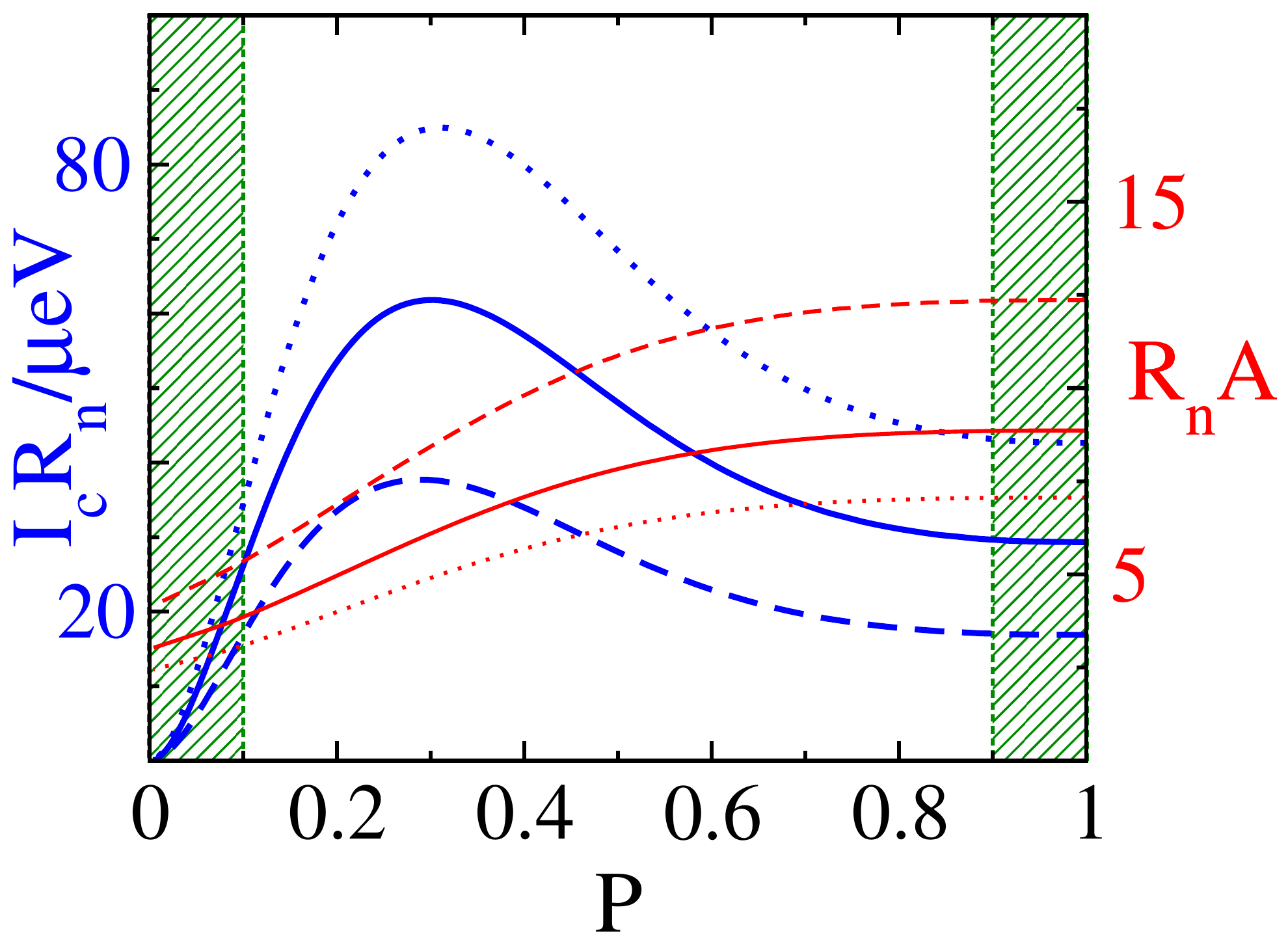}
\put(4,70){\makebox(0,2){(e)}}
\end{overpic}
\end{minipage}\\
\caption{
\label{Esch08}
(a) All symmetry components in a ballistic S/F/S $\pi$-Josephson junction with a half-metallic ferromagnet (H) as F-layer (self-consistent calculation for model barrier as in Fig.~\ref{Fig_alpha0} (b), with $V_{\rm S}=0.5$, $d=0.5/k_0$); barrier spin-polarized in $x$-direction, half-metal spin-polarized in $z$-direction; temperature $T=0.2T_c$;
(b)-(e) Non-monotonic temperature dependence of critical Josephson current density in an S/F/S junction, normalized to its value at zero temperature; 
(b) for half-metallic ferromagnet, dependence on the mean free path $\ell_H$ in H for a fixed junction length $L = \xi_c=\hbar v_H /2\pi T_c $, where $v_H$ is the Fermi velocity in H;
(c) as in (b), but showing the dependence on junction length $L$ for fixed $\ell_H=1.5\xi_c$;
(d) for a ferromagnet in the ballistic limit, with various spin polarizations $P=(p_{F\downarrow}-p_{F\uparrow})/(p_{F\downarrow}+p_{F\uparrow})$; $L=\hbar v_F /2\pi T_c $ with $v_F$ the Fermi velocity in F. 
(e) $I_cR_n$ product and normal-state resistance $R_nA$ [in units of $(e^2N_Sv_S)^{-1}$, where $N_S$ and $v_S$ are density of states per spin at the Fermi level and the Fermi velocity in the superconductor, respectively] as function of $P$ for $T=0.5T_c$, and various strengths of interface transmission.
In all plots the interface misalignment angle is $\alpha = \pi/2$.
(b) and (c) after \cite{Eschrig08}, (d) and (e) from \cite{Eschrig09}.
Copyright (2009) by the American Physical Society.
}
\end{figure}

We discuss first the main predictions of a long-range Josephson effect in a structure involving a long half-metallic ferromagnetic region, as discussed in Refs.~\cite{Eschrig03,Eschrig07,Eschrig08,Eschrig09,Eschrig15}. As shown in Fig.~\ref{Esch08}, which is for a fully self consistent calculation using a model barrier as in Fig. \ref{Fig_alpha0} (b), all possible symmetry components shown in Fig.~\ref{Symmetry} are indeed generated at the interface. This simply reflects the fact that both spin rotational symmetry and inversion symmetry are broken by the interface. The calculation assumes that the effective magnetic moment of the interfaces is perpendicular to the bulk magnetization of the half-metallic ferromagnet. As a consequence, according to Eq. (\ref{triplet_z}), long-range amplitudes are generated, linking the two singlet superconductors [see lower panels in Fig. \ref{Esch08} (a)]. Thus, we have an example of an {\it indirect} Josephson coupling \cite{Eschrig03}, via a long-range triplet component which is generated in the first place from the singlet component by microscopic processes in the interface region. Note also, that the thermodynamically stable configuration for identical interfaces is a $\pi$-junction
(for asymmetric junctions the appearence of a $\phi_0$-junction is predicted \cite{Eschrig07,Eschrig08}),
for which the system tries to minimize the odd-frequency triplet amplitudes of Type D in the structure. 
This is a general principle, as odd-frequency amplitudes {\it increase} the free energy density, instead of decreasing it as even-frequency singlet amplitudes do. Only if the $p$-wave components (and all higher orbital components) are suppressed by impurity scattering, can the odd-frequency triplet amplitude dominate.

A prominent feature in such a structure is a non-monotonous temperature dependence of the critical Josephson current density, with a pronounced maximum at a temperature which can be linked to the Thouless energy of the ferromagnetic layer (see Fig.~\ref{Esch08} (c) and Ref.~\cite{Eschrig08}). 
When the junction becomes effectively long compared with the diffusive-limit coherence length, 
$L \gg \xi_d =\sqrt{\xi_0 \ell_H/3}$, the current is dramatically suppressed and the peak is shifted to a lower temperature.
This maximum moves also to low temperatures in diffusive structures, as is seen in Fig.~\ref{Esch08} (b), which shows its dependence on the mean free path in the ferromagnet. 
If one replaces the half-metallic ferromagnet by a strongly spin-polarized ferromagnet with two itinerant spin bands, then the maximum remains only for sufficiently strong spin polarization, as seen in Fig.~\ref{Esch08} (d). 
If one projects the Fermi surfaces on the contact plane, then only the ``half-metallic'' momentum directions contribute to the maximum, which are inside the Fermi sea for one spin projection and outside for the other \cite{Grein09}. Furthermore, there is an optimal spin polarization which maximizes the $I_cR_n$-product of the Josephson junction, which for the particular model interface studied in Ref.~\cite{Grein09} is around $P\sim 0.3$. The green shadowed regions in Fig.~\ref{Esch08} (d) for small and large spin polarizations are regions beyond the validity of the theory, as other processes which are neglected become relevant.

Following the early theoretical work on triplet supercurrents in half-metallic ferromagnets \cite{Eschrig03,Kopu04}, which treats a fully developed triplet proximity effect self-consistently, numerous groups confirmed the existence of triplet supercurrents 
under various conditions in half-metallic ferromagnets.
In Ref.~\cite{Eschrig08} the work of Refs.~\cite{Eschrig03,Kopu04,Eschrig07} was generalized to include impurity disorder bridging between the two limits of ballistic and diffusive transport. In this work, the full Eilenberger equations with impurity self energy in self-consistent Born approximation
was solved. In Ref.~\cite{Braude07} a fully developed triplet proximity effect, just like in Ref.~\cite{Eschrig03} for the clean limit, was treated for the diffusive limit. 
Further studies in the diffusive limit were performed in Refs.~\cite{Asano07,Asano07a,Eschrig15}.
The role of magnon creation in the half-metallic ferromagnet was studied in \cite{Takahashi07}, with a resulting temperature dependence of the critical current reproducing the temperature dependence found in \cite{Eschrig03,Eschrig07}.
The case when the superconductor is unconventional was studied in \cite{Yokoyama07,Linder10,Enoksen12}.
In Refs. \cite{Galaktionov08,Eschrig09} the problem was advanced on the analytical level assuming a constant singlet order parameter in the superconductor.
Self-consistent solution of Bogoliubov-de Gennes equations in the clean limit \cite{Halterman09}
found results in agreement with Ref.~\cite{Eschrig03,Eschrig07}, including a strong influence of the junction behavior by subgap Andreev bound states.

Most of these theoretical studies deal with either the diffusive limit within the Usadel approximation or the clean limit. 
The full range of impurity scattering from clean to diffusive limit for a superconductor-half metal proximity structure was first covered in the theory of Ref. \cite{Eschrig08}. 
A treatment in the quantum limit was given by B\'eri {\it et al.} \cite{Beri09}.

\begin{figure}
\begin{overpic}[width=0.5\linewidth]{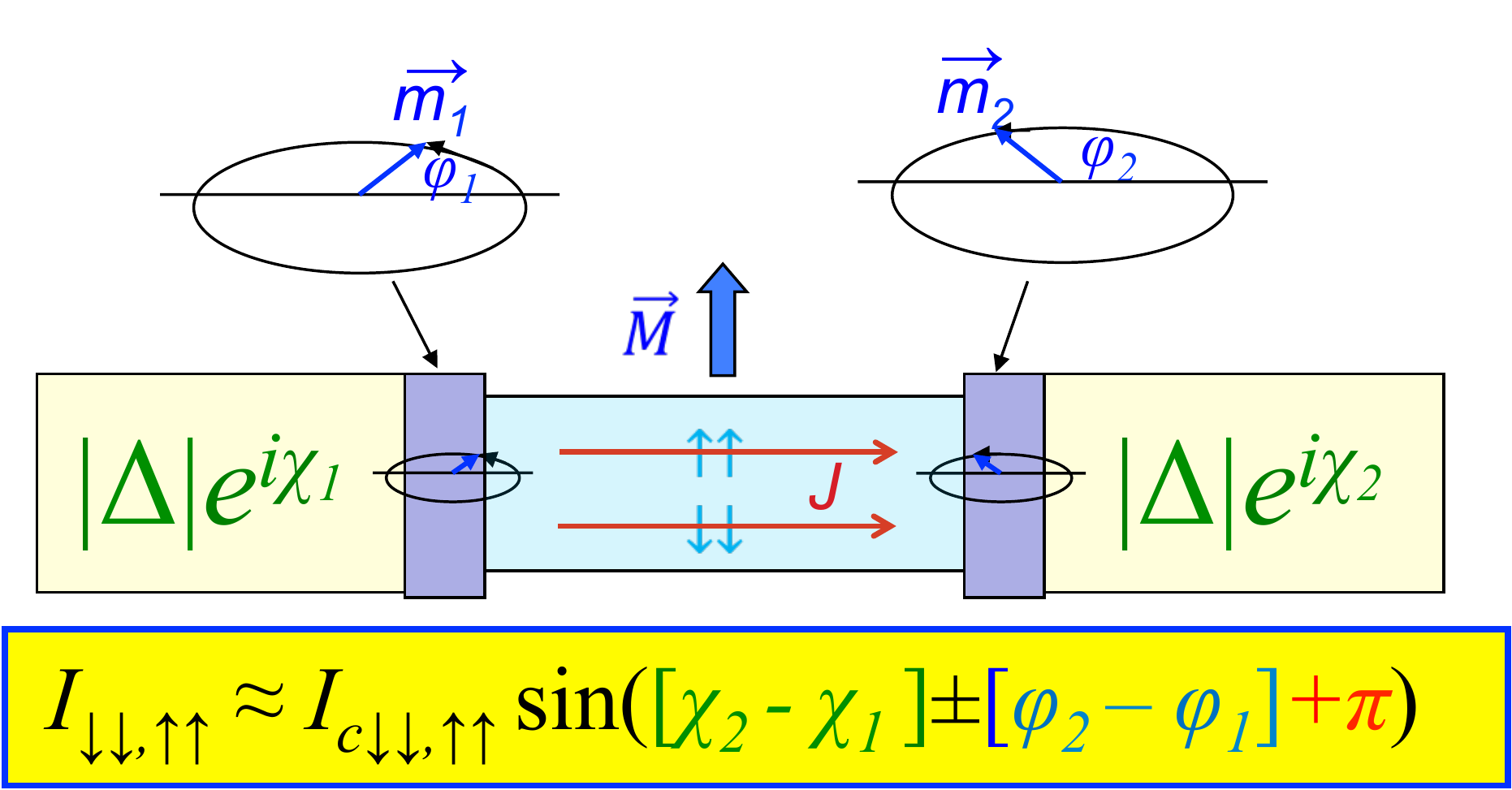}
\put(4,50){\makebox(0,2){(a)}}
\end{overpic}
\begin{overpic}[width=0.5\linewidth]{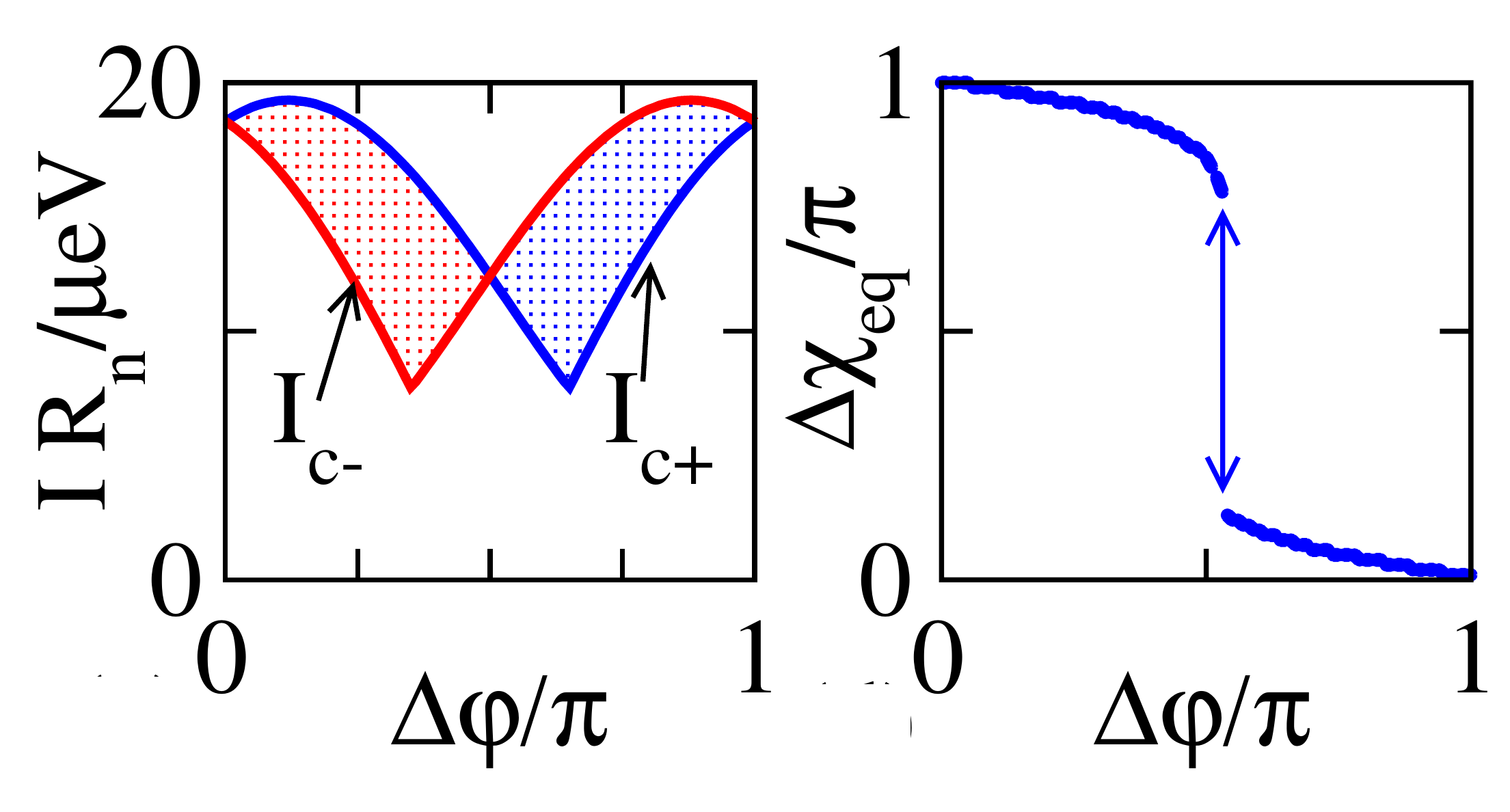}
\put(2,50){\makebox(0,2){(b)}}
\put(54,50){\makebox(0,2){(c)}}
\end{overpic}
\caption{
\label{Grein}
(a) S/F/S Josephson junction with a strongly spin-polarized ferromagnet with magnetization $\vec{M}$, having interface barriers with misaligned magnetic moments $\vec{m}_1$ and $\vec{m}_2$. Only equal-spin pair amplitudes are  present in each of the spin bands of the ferromagnet. The azimuthal angles of $\vec{m}_1$ and $\vec{m}_2$ with respect to the magnetization axis $\vec{M}$ are denoted $\phi_1$ and $\phi_2$.
(b) Critical current in positive ($I_{c+}$)/negative ($I_{c-}$) bias direction vs $\Delta \phi=\phi_2-\phi_1$. (c) The zero current equilibrium phase difference $\Delta \chi_{\rm eq}=(\chi_2-\chi_1)_{\rm eq}$ vs $\Delta \phi$ varies from
$\pi$ to $0$. Here, $T=0.2T_c$, $P=0.21$. The polar angle of $\vec{m}_1$ and $\vec{m}_2$ is $\alpha=\pi/2$. The length of the junction is $L=\hbar v_S/2\pi T_c$, with $v_S$ the Fermi velocity of the superconductor. (b) and (c) 
from \cite{Grein09}.
Copyright (2009) by the American Physical Society.
}
\end{figure}

In Ref.~\cite{Grein09} a number of effects were pointed out that are absent in the half-metallic case, and which are due to the phase coherent transport of Cooper pairs in the two itinerant spin bands of a ferromagnet. These are illustrated in Fig.~\ref{Grein}. As seen from Eq.~(\ref{triplet_z}), equal-spin triplet correlations in the ferromagnet acquire an additional phase from the configuration of the magnetic moments in the interface region, $\vec{m}_1$ and $\vec{m}_2$, with respect to the bulk magnetization $\vec{M}$. 
If the three magnetizations are {\it non-coplanar}, there is an important geometric phase involved, which is related to the projection of the interface magnetic moments of the two interfaces to the plane perpendicular to the bulk ferromagnetic magnetization\footnote{In the coplanar but non-collinear case there still can be an important geometric phase of $\pi$ introduced by the triple of magnetic vectors if $(\vec{m}_1\times \vec{M})\cdot (\vec{m_2}\times\vec{M})<0$.}. The relative angle between these two projected magnetic moments defines the angle $\Delta \phi=\phi_2-\phi_1$, which is a gauge invariant quantity independent of the choice of the global spin quantization axis. This angle directly enters the current-phase relation for each spin band, and precisely with opposite sign for the two spin projections. Thus, the first-order processes (single Cooper pair tunneling) lead to contributions of the form
\numparts
\begin{eqnarray}
\label{firstorder1}
I_{\uparrow\uparrow} &=& -I_{c\uparrow\uparrow} \sin(\Delta \chi + \Delta \phi) \\
\label{firstorder2}
I_{\downarrow\downarrow} &=& -I_{c\downarrow\downarrow} \sin(\Delta \chi - \Delta \phi) 
\end{eqnarray}
\endnumparts
with $\Delta \chi=\chi_2-\chi_1$ the difference of the superconducting phases of the singlet pair potential on either side of the Josephson junction. 
The equilibrium configuration for zero current amounts to\footnote{For zero phase difference a spontaneous current $I(\Delta\chi=0)=-(I_{c\uparrow\uparrow}-I_{c\downarrow\downarrow}) \sin (\Delta \phi)$ appears.}
\begin{eqnarray}
\Delta\chi_{\rm eq}=\pi-\arctan \left(\frac{I_{c\uparrow\uparrow}-I_{c\downarrow\downarrow}}{I_{c\uparrow\uparrow}+I_{c\downarrow\downarrow}} \tan \Delta \phi \right).
\end{eqnarray}
Thus, the system can act as a phase battery. For the half-metal case ($I_{c\downarrow\downarrow}=0$) this has been noted 
using Eilenberger equations in Refs. \cite{Eschrig07,Eschrig08} and using Usadel equations in Ref. \cite{Eschrig15}.
Going beyond the tunneling limit, we find a non-zero equilibrium phase difference between the superconducting leads, which shows a jump as function of $\Delta \phi$, indicating a first order transition 
(two competing minima of the free energy located at different values of phase difference $\Delta \chi$), 
which is illustrated in Fig.~\ref{Grein} (c). Note that the current-phase relation fulfills the symmetry $I(-\Delta \chi,\Delta \phi)=-I(\Delta \chi,-\Delta \phi)$ following from the behavior of the current density under time reversal. The equilibrium configuration carries a non-zero spin current density $I_s=2I_{\uparrow\uparrow}(\Delta \chi_{\rm eq},\Delta \phi) \ne 0$.
The presence of a Josephson current at zero superconducting phase difference $\Delta \chi$ was confirmed in calculations using Bogoliubov-de Gennes equations in a tri-layer geometry \cite{Liu10}.
A study of a superconductor/ferromagnetic-insulator/superconductor junction on the surface of a three-dimensional topological insulator revealed similar effects \cite{Tanaka09a}.

\begin{figure}
\begin{center}
\includegraphics[width=0.45\linewidth]{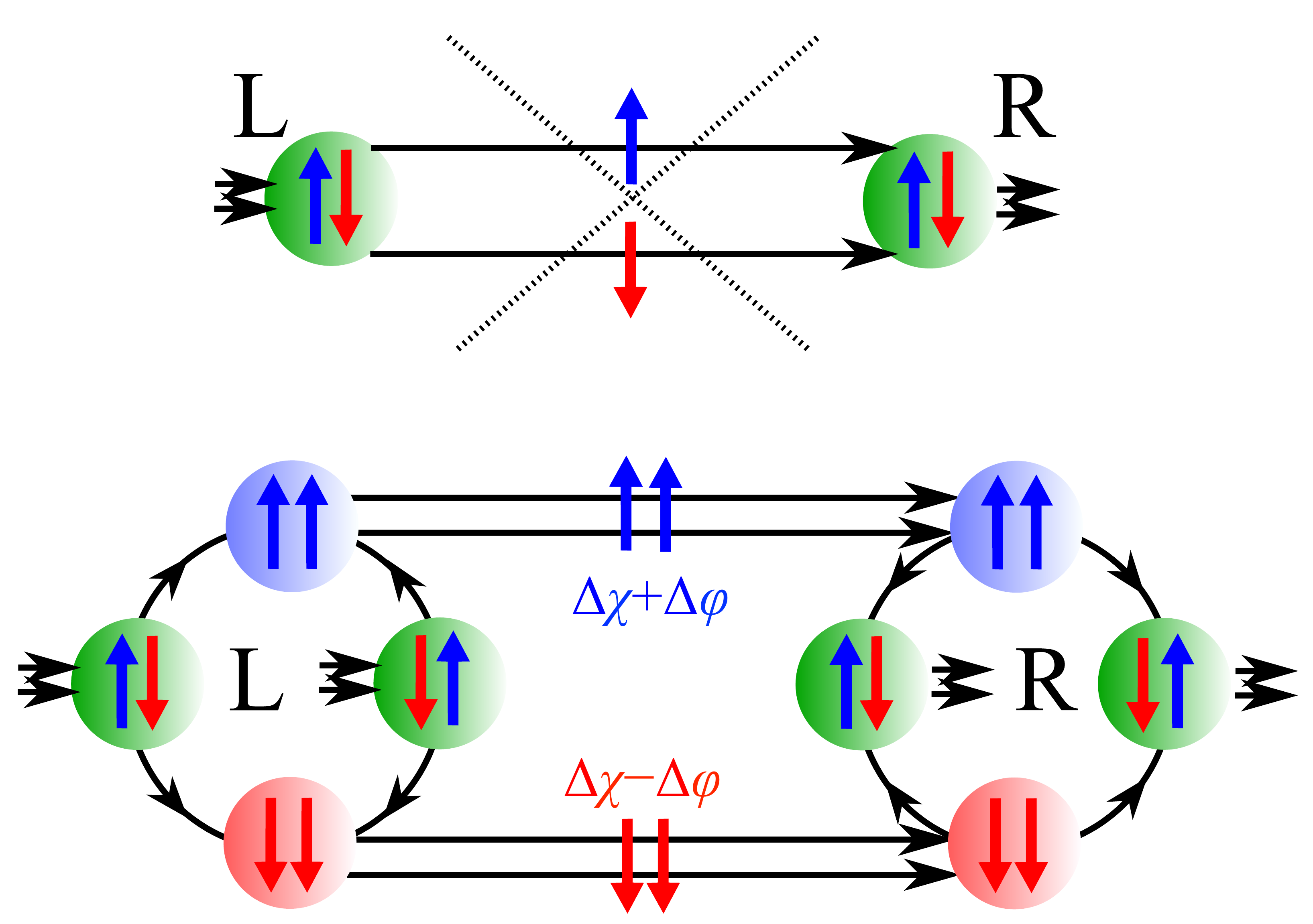}
\end{center}
\caption{
Schematic drawing of the crossed pair transmission process. 
Cooper pairs with opposite spin projection on the magnetization vector do not contribute to the spin supercurrent in strongly spin-polarized ferromagnets (top panel). However,
due to magnetic inhomogeneity across the interfaces 
singlet-triplet conversion processes generate out of two singlet pairs two equal-spin triplet pairs 
with opposite spin polarization. These
two equal-spin pairs are transmitted simultaneously in a crossed pair transmission process (bottom panel). 
Their phases correspond to $\Delta \chi \pm \Delta \phi$.
A geometric phase shift $2\Delta \phi$ appears between the two types of Cooper pairs.
Copyright (2009) by the American Physical Society \cite{Grein09}.
\label{CPT}
}
\end{figure}
The first-order processes described by Eqs.~(\ref{firstorder1})-(\ref{firstorder2}) compete with a second order process, where two (or an even number of) pairs are transmitted simultaneously. In analogy to the so-called crossed Andreev reflection process, we call this process {\it crossed pair transmission} (see figure \ref{CPT}).
For transmission of all pairs in equal direction,
this process involves phases which are multiples of $(\Delta \chi+\Delta \phi)+(\Delta\chi-\Delta\phi)=2\Delta \chi$, for which the dependence on $\Delta \phi$ drops out. Consequently, the corresponding contribution to the Josephson current, $I_{\rm cp}$, is independent of $\Delta \phi$ and $\pi$-periodic. More generally, a transmission process of $m$ spin-$\uparrow\uparrow$ pairs and $n$ spin-$\downarrow\downarrow$ pairs involves a phase $(m+n)\Delta \chi + (m-n)\Delta \phi + (m+n)\pi$. Thus, the corresponding Josephson charge current can be written as
\numparts
\begin{eqnarray}
I &=& 2e\;\frac{1}{2}\sum_{mn} (m+n)(-1)^{n+m}I_{mn} \sin [(m+n)\Delta \chi + (m-n)\Delta \phi]
\end{eqnarray}
whereas the spin current is
\begin{eqnarray}
I_s &=& \hbar\;\frac{1}{2}\sum_{mn} (m-n) (-1)^{n+m}I_{mn}\sin [(m+n)\Delta \chi + (m-n)\Delta \phi] .
\end{eqnarray}
\endnumparts
As $I_{-m,-n}=I_{mn}$, the factor $\frac{1}{2}$ can be omitted if we restrict the summation to $m\ge0$, and for $m=0$ to positive $n$.
Note that the spin-current is even present in the case $\Delta \chi=0$, provided that $\Delta \phi\ne0$. An example of such a case was discussed in \cite{Grein09}, where one of the superconductors was replaced by an insulator, and a pure spin supercurrent remains, giving rise to a {\it spin-Josephson effect}. In this case only terms with $m+n=0$ remain, with a pure spin supercurrents $I_s=\hbar \sum_m 2m I_{m,-m} \sin(2m\Delta \phi)$.
On the other hand, crossed pair transmission processes in equal direction have zero spin current and correspond to $m=n$, leading to a contribution to charge current of the form $2e\sum_m 2m I_{mm} \sin(2m\Delta \chi)$.
For illustrative purposes we consider the example of only first order terms and leading order crossed pair transmission terms, i.e. only terms associated with $I_{01}$,$ I_{10}$, $I_{11}$, and $I_{1,-1}$. Then
\numparts
\begin{eqnarray}
I/2e &=& -I_{10} \sin (\Delta \chi + \Delta \phi)-I_{01} \sin (\Delta \chi - \Delta \phi) +
2I_{11} \sin (2\Delta \chi)  \\
I_s/\hbar &=& -I_{10} \sin (\Delta \chi + \Delta \phi)+I_{01} \sin (\Delta \chi - \Delta \phi) +
2I_{1,-1} \sin (2\Delta \phi) .
\label{spinc}
\end{eqnarray}
\endnumparts
The last term in each of these equations corresponds to the crossed pair transmission process (in equal and opposite direction, correspondingly).
The critical current density for positive and negative current bias differs for this case, similar as shown in Fig.~\ref{Grein} (b). Furthermore, for sufficiently large $I_{11}$ multiple minima of the free energy as function of $\Delta \chi$ appear, leading to the characteristic jump at a certain value of $\Delta \phi$ as illustrated in Fig.~\ref{Grein} (c). If $\Delta \phi$ were continuously varied, a hysteresis would occur, typical for a first order phase transition. Finally, when one of the superconductors is replaced by an insulating material then $I=0$ and the spin-Josephson current is $I_s=2\hbar I_{1,-1} \sin(2\Delta \phi)$ \cite{Grein09}.

\subsection{Recent experimental observations}

\subsubsection{Triplet supercurrents in half-metallic ferromagnets}

Half-metallic ferromagnets are especially promising for applications, as they are fully spin polarized and thus give the largest spin filtering effect possible. The recent developments and applications for half-metallic ferromagnets have been reviewed e.g. in Refs. \cite{Pickett01,Felser07,Katsnelson08}.

First indications for an coupling of superconductors through a half-metallic ferromagnet came from experiments by Pe\~na {\it et al.} on ferromagnetic La$_{0.7}$Ca$_{0.3}$MnO$_3$, coupled with the high-temperature $d$-wave singlet superconductor YBa$_2$Cu$_3$O$_{7-\delta}$ in trilayers and superlattices \cite{Sefrioui03,Pena04}. These authors found a long-range proximity effect over a length scale of 100 nm.

In 2006, Keizer {\it et al.} reported a triplet supercurrent in the half metal CrO$_2$ 
\cite{Keizer06}. The authors used NbTiN as superconducting electrodes and CrO$_2$ grown on TiO$_2$.
After it had proved difficult to reproduce the effect, it was in 2010 when finally Anwar {\it et al.} in the group of Aarts 
reported that they had found long-range supercurrents through CrO$_2$ grown on Al$_2$O$_3$ (sapphire)  \cite{Anwar10}.
The current was observed over a distance of 700 nm between two superconducting amorphous Mo$_{70}$Ge$_{30}$ electrodes. 
The effect was interpreted in terms of odd-frequency pairing correlations (although really the experiment only indicates equal-spin pairing correlations and does not provide experimental insight about the orbital and frequency symmetry of the pairs).
However, the group was not able to find the long-range supercurrent in CrO$_2$ grown on TiO$_2$, the material used by Keizer {\it et al.} 
They noted that their CrO$_2$ films showed a uniaxial anisotropy in contrast to the biaxial anisotropy present in the Keizer {\it et al.} samples \cite{Anwar10,Anwar11}. 

There had been criticism that based on previous point contact spectroscopy experiments, CrO$_2$ might not be fully spin-polarized.
These data were summarized by L\"owfander {\it et al.} in 2010, and it was shown that all point contact data on CrO$_2$ are indeed consistent with full spin polarization, if the theory is extended to take into account realistic interfaces \cite{Lofwander10}.

Anwar {\it et al.} have in a subsequent study extended their work \cite{Anwar12} and find that if a Ni/Cu sandwich between the CrO$_2$ film and the Mo$_{70}$Ge$_{30}$ electrodes is used, a long-range supercurrent is observed also for TiO$_2$ substrate over a distance of almost a micrometer. The critical current density is 100 times larger in these Ni/Cu/CrO$_2$ junctions on TiO$_2$ substrate than if sapphire is used as a substrate, and is comparable to that observed by Keizer {\it et al.}
This proves that in addition to the substrate on which CrO$_2$ is grown, the interface characteristics between CrO$_2$ and the superconductor play a decisive role, and spin-mixing due to misaligned spins as predicted before by theory \cite{Eschrig03} is the crucial ingredient for explaining the singlet-triplet conversion in these structures.

Further evidence for triplet supercurrents came from a report by Sprungman {\it et al.},
who found a Josephson effect with ferromagnetic Cu$_2$MnAl Heusler barriers \cite{Sprungmann10}. 
It was observed that the critical current density versus temperature shows a pronounced peak around 4.5 K (electrodes were Nb), decreasing towards lower and higher temperatures. Such a maximum was predicted in \cite{Eschrig03,Eschrig08,Eschrig09}. However, there is some intrinsic gradient of the degree of L2$_1$-type Heusler structure ordering inside the Heusler layers, with a low degree if order at the interfaces and higher degree of order in the interior \cite{Sprungmann10}. For that reason, it is too early to decide about the origin of this maximum in $I_{\rm c}(T)$ in these experiments.

Another half-metallic ferromagnet is La$_{0.7}$Ca$_{0.3}$Mn$_3$O, which has been studied in the context of YBa$_2$Cu$_3$O$_7$/La$_x$Ca$_{1-x}$MnO$_3$ multilayers
by Pe\~na {\it et al.} in 2005. Kalcheim {\it et al.} \cite{Kalcheim12} performed scanning tunneling spectroscopy experiments on La$_{0.7}$Ca$_{0.3}$Mn$_3$O film epitaxially grown on superconducting Pr$_{1.85}$Ce$_{0.15}$CuO$_4$, and find long-range penetration of superconductivity into half-metallic La$_{0.7}$Ca$_{0.3}$Mn$_3$O. Visani {\it et al.} \cite{Visani12} find quasiparticle and electron interference effects in the conductance across a La$_{0.7}$Ca$_{0.3}$Mn$_3$O/YBa$_2$Cu$_3$O$_7$ interface that demonstrate long-range propagation of superconducting correlations across the half metal. The effect is interpreted in terms of equal-spin Andreev reflections (equal-spin Andreev reflection was introduced in Ref. \cite{Grein10} under the term ``spin-flip Andreev reflection'', or SAR). 
The peculiarities of Andreev reflection at a half-metal interface with a singlet superconductor are connected with the spin-active interfaces that unavoidably will be involved, and have been clarified in Appendix C of Ref. \cite{Eschrig09}, and in Refs. \cite{Grein10,Lofwander10,Kupferschmidt11,Wilken12}. 

\subsubsection{Multilayer converter}
\begin{figure}
\includegraphics[width=1.0\linewidth]{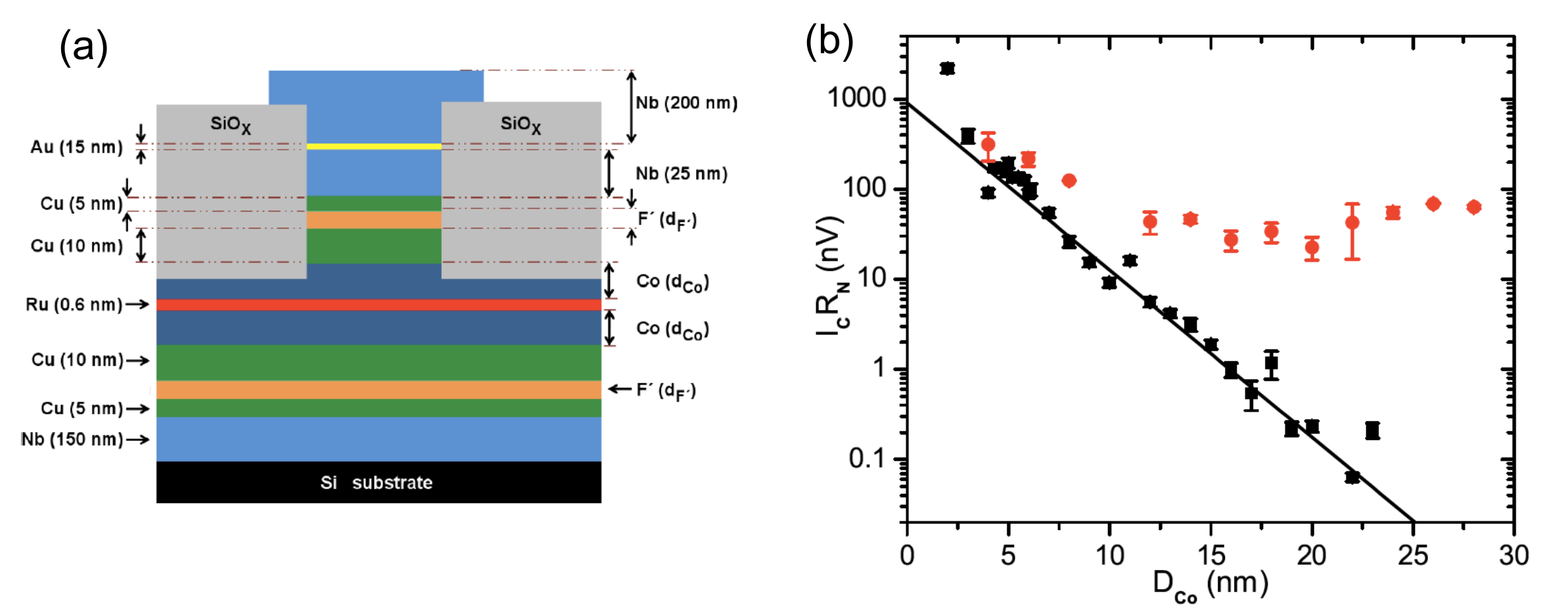}
\caption{
(a) Schematic diagram of the Josephson
junctions used in the work of Khaire {\it et al.}, shown in cross-section.
From Klose {\it et al.} \cite{Klose12}.
Copyright (2012) by the American Physical Society.
(b)
Product of critical current times normal state resistance,
$I_{\rm c} R_{\rm N}$, as a function of total Co thickness,
$D_{\rm Co}= 2 d_{\rm Co}$. Red circles represent junctions with 
F'$ = $PdNi and $d_{\rm  PdNi} = $4 nm, whereas black squares represent
junctions with no F' layer (taken from Ref. \cite{Khasawneh09}). As
$D_{\rm Co}$ increases above 12 nm, $I_{\rm c} R_{\rm N}$
hardly drops in samples with
PdNi, but drops very rapidly in samples without. (The solid
line is a fit of the data without PdNi to a decaying exponential, from Ref. \cite{Khasawneh09}.) 
From Khaire {\it et al.} \cite{Khaire10}.
Copyright (2010) by the American Physical Society.
\label{Khaire10}
}
\end{figure}

Experiments utilizing a multilayer geometry, in some sense manufacturing an inhomogeneous magnetization profile by using different materials for the various layers, led to a successful generation of triplet supercurrents in a thick Co layer by the group of Birge in 2010 \cite{Khaire10}.
In these experiments two layers of ferromagnetic alloys were used to generate triplet amplitudes, which then are passed as long-range supercurrent through a strongly spin-polarized ferromagnet.
The crucial role of the misalignment of the magnetizations was demonstrated. With this
breakthrough a simple and reliable way was found to produce long-range triplet supercurrents. 
The geometry is shown in Figure \ref{Khaire10} together with the results. It makes use of the previously acquired knowledge of the group that a Co/Ru/Co trilayer instead of a single Co layer will give much better control over the junction behavior [see Figure \ref{Khasawneh09}(b)] \cite{Khasawneh09}.
When the extra outer layers are present, a long-range effect persists, whereas if they are absent, there are only short-range supercurrents present \cite{Khaire10,Khasawneh11}. It was later shown experimentally that the triplet supercurrent is enhanced up to 20 times after the samples are
subject to a large in-plane magnetizing field aligning the two Co layers perpendicular to the magnetizations of the two thin outer layers \cite{Klose12}, exactly as predicted e.g. in \cite{Eschrig03,Eschrig07,Houzet07}.
In Ref.~\cite{Gingrich12} a triplet supercurrent in S/F'/F/F'/S Josephson junctions was studied, where S was superconducting Nb, F' a thin Ni layer with in-plane magnetization, and F a Ni/[Co/Ni]$_n$ multilayer with out-of-plane magnetization. It was found that the supercurrent decays very slowly with F-layer thickness and is much larger when the F' layer are present than if not. This confirmed that the spin-triplet supercurrent is maximized by the orthogonality of the magnetizations in the F and F' layers.

Theoretical treatments of the geometry used by Khaire {\it et al.} have appeared in Refs. \cite{Volkov10,Trifunovic10,Alidoust10}.
It was subsequently demonstrated experimentally that
the critical current scales linearly with area in magnetized junctions, confirming the homogeneity of the superconducting phase difference across the junction area \cite{Wang12}.

\subsubsection{Long-range effects in ferromagnetic nanowires}

There have been also reports for a long-range proximity effect in long ferromagnetic nanowires.
An early overview over studies of ferromagnetic nanowires with superconducting electrodes was given by Petrashov {\it et al.} \cite{Petrashov00}.
The controlled growth of nanowires has improved over the last decade, allowing to create crystalline nanowires of several hundreds nanometers length. Wang and coworkers used a single-crystalline ferromagnetic Co nanowire and reported zero resistance in wires up to 600 nm length \cite{Wang10}.
A theory for a long-range singlet proximity effect in ferromagnetic nanowires was given by Konschelle {\it et al.} \cite{Konschelle10}. Takei {\it et al.} developed a theory of a ferromagnetic nanowire-superconductor proximity structure based on Rashba spin-orbit coupling in the barrier that induces $p$-wave superconductivity in the ferromagnet \cite{So12}. 

Almog {\it et al.} \cite{Almog11} report measurements of the dynamical conductance of a double In/Co/In device, where two Co wires are connected to superconducting In electrodes, running parallel less than a coherence length apart from each other. They find a spin polarization of the superconducting order parameters at the interfaces which depends on the relative spin polarization of the two wires.
Colci {\it et al.} \cite{Colci12} study a similar structure with two superconducting electrodes being bridged by two parallel ferromagnetic wires forming an SFFS junction. They concentrate on the phenomenon of crossed Andreev reflection. At low temperatures and excitation energies below the superconducting gap, they find that the resistance corresponding to antiparallel alignment of the magnetization of the ferromagnetic wires is higher than that of parallel alignment. Spin-dependent interface scattering was found to be important in order to understand these findings \cite{Sun13}.

\subsubsection{Spiral magnetic order}
A realization of the ideas by Bergeret, Volkov, and Efetov was 
reported in Ref. \cite{Petrashov06},
by employing the conical ferromagnet holmium. A long-range
proximity effect was observed. In this work,
superconducting phase-periodic conductance oscillations in ferromagnetic Ho wires in contact with conventional superconductors were measured.  The distance between the interfaces was much larger than the singlet superconducting penetration depth. 

\begin{figure}
\includegraphics[width=1.0\linewidth]{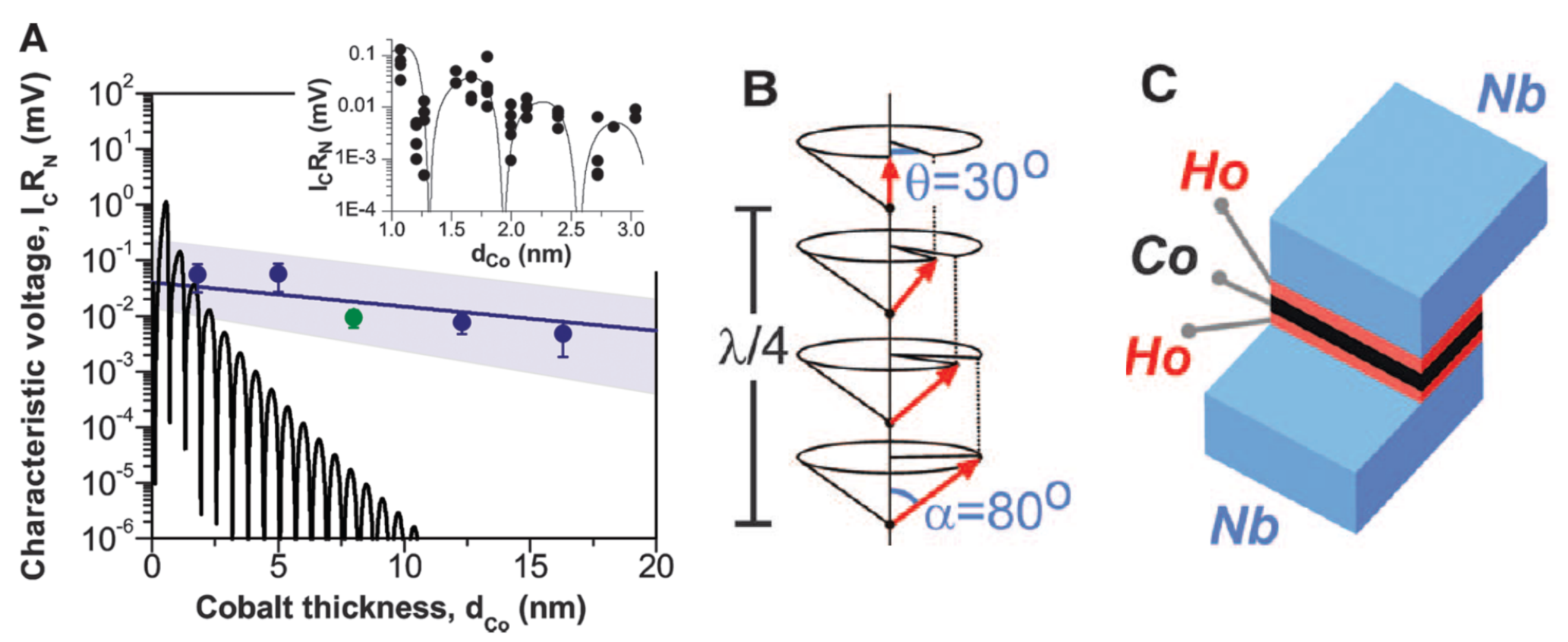}
\caption{
(A)
Slow decay at 4.2 K in the characteristic voltage $I_cR_N$ of
Nb/Ho(4.5 nm)/Co( $d_{\rm  Co}$)/Ho(4.5 nm)/Nb
junctions (blue circles) and a Nb/Ho(10 nm)/Co($ d_{\rm Co}$)/Ho(10 nm)/Nb junction (green circle) versus Co barrier
thickness ($ d_{\rm Co}$). 
Inset: Comparative data (black circles) from \cite{Robinson09}
showing the behavior of Nb/Rh/Co/Rh/Nb
junctions. The oscillating curves in the inset and main panel are theoretical fits to the experimental data in the inset, as described in \cite{Robinson09}.  
(B) The conical magnetic configuration of idealized
Ho below its Curie temperature (20 K), showing an antiferromagnetic spiral rotating in-plane $\theta = 30^{\rm o}$ per atomic plane and pitched
$\alpha = 80^{\rm o}$ out-of-plane. 
The moments (arrows) rotate about the surface of a cone with the spiral wavelength,
$\lambda $, corresponding to a Ho thickness of ~3.4 nm. 
(C)~Device layout consisting of two
superconducting Nb electrodes coupled via a Ho/Co/Ho trilayer.
From Robinson {\it et al.} \cite{Robinson10}. Reprinted with permission from AAAS.
\label{Robinson10}
}
\end{figure}

A triplet supercurrent has been found by Robinson, Witt, and Blamire
with a setup using a cobalt layer sandwiched between two holmium layers,
constituting another breakthrough in 2010 \cite{Robinson10}.
The main results of this work are shown in Fig.~\ref{Robinson10}. Without the Ho layers 
between the Nb and Co (or with the Ho replaced by Rh) 
the characteristic voltage $I_cR_N$ shows fast oscillations and a short-range decay over six orders of magnitude on a scale of 10 nm. If additional Ho layers are inserted between Nb and Co, the characteristic voltage decays very slowly, by about an order of magnitude on a scale of 50 nm. The effect is spectacular especially for Co thicknesses above 10 nm.
Theoretical treatments of this geometry have been given in Refs. \cite{Halasz09,Halasz11,Fritsch14}.

Further reports concentrate on creating artificial
non-collinear magnetic structures that can be brought in contact with 
superconductors.  In Ref.~\cite{Gu10} it is reported that
a new structure using an exchange-spring magnet was fabricated that can be tuned from a collinear to a non-collinear state by a rotating external magnetic field in a controllable way. With this setup the authors found an increase of superconductivity in the structure (transition temperature and conductance) when going to a non-collinear state.

Robinson {\it et al.}
studied Nb/Fe/Cr/Fe/Nb junctions where the thickness of the Cr layer determines the relative alignment of the Fe layers, and find a substantial enhancement of the critical Josephson current when a non-parallel configuration was realized  \cite{Robinson10a}.

Witt {\it et al.} find that in structures with a conical magnetic spacer layer of Ho the critical Josephson current decays exponentially with layer thickness, corresponding to a coherence length of 4.34 nm in Ho, and without showing any oscillatory behavior \cite{Witt12}. Comparing this with the mean free path of 0.28-0.87 nm, they conclude that their Ho structures are in the dirty limit.

A realization of field-tunable in-plane Bloch domain walls in the rare-earth magnet gadolinium if grown between non-collinearly aligned ferromagnets was suggested in Ref.~\cite{Robinson12}. It was found that supercurrents flow through magnetic Ni/Gd/Ni nanopillars, the magnitude of which strongly depends on the domain wall state in Gd. The authors explain this result in terms of the inter-conversion of triplet and singlet pairs, the efficiency of which depends on the magnetic helicity of the structure.

For a recent review by Blamire and Robinson see \cite{Robinson14}.

\section{Modern developments}
In the following I selectively give examples for exciting modern developments of the field, without claiming to cover the entire spectrum of activities. 

\subsection{Spin-valve devices and spin-filter junctions}
Although currently the research still concentrates on studying fundamental questions, applications can be imagined in various ways. The most obvious application would 
be the development of a spin-valve device. Re-entrant phenomena are predicted 
not only as function of thickness of the adjacent ferromagnetic layers, but
also as function of other parameters, e.g. the degree of magnetic inhomogeneity, 
as in ferromagnets with spiral order. A high degree of experimental control has been achieved in producing large batches of samples with gradually varying thicknesses, allowing for a tailoring of the Fulde-Ferrell-Larkin-Ovchinnikov state \cite{Eschrig12A}. A controllable Josephson spin-valve has been realised recently \cite{Iovan14}.

Superconducting spin-filter tunnel junctions have been studied recently experimentally \cite{Senapati11} and theoretically \cite{Bergeret12}.
In Ref.~\cite{Senapati11}, GdN barriers are used as ferromagnetic insulator barriers, and it is shown that the field and temperature dependence of the critical Josephson current is strongly modified by the ferromagnetic insulator. It is found that the strong suppression of Cooper pair tunneling by the spin filtering of the barrier can be modified by magnetic inhomogeneity in the barrier.
In theoretical work \cite{Bergeret12} it was also demonstrated that the differential conductance may exhibit peaks at different values of the voltage depending on the polarization of the spin filter, and the relative angle between the exchange fields and the magnetization of the barrier.

\subsection{Flux Qubits and semifluxons}
Superconducting electronics is undergoing a revival at present, with pressing need for cryogenic 
memory for flux quantum logic applications as well as new superconducting Qubit applications. 
Metal spintronics is already widely employed in computer hard disc technology, and 
superconducting spintronics may lead to energy efficient memory and logic for 
supercomputing applications. 
Superconducting digital single-flux-quantum circuits using superconductor-ferromagnet-superconductor sandwich technology to insert $\pi$-Josephson junctions into a circuit is promising due to its high operation speed and low energy consumption \cite{Feofanov10}. Such systems also solve the problem of high element densities on-chip for operating and holding magnetic flux quanta. A combination of magnetic Josephson junctions and conventional Josephson junctions can be used to form addressable memory cells, energy-efficient memory periphery circuits and programmable logic elements \cite{Ryazanov12,Vernik13}. 

Another interesting development is semifluxon physics in zero-$\pi$ Josephson junctions.
Fluxons are traveling, solitonic waves of magnetic flux in long Josephson junctions, created by external magnetic fields. In a zero-$\pi$-junction a Josephson vortex of fractional magnetic flux is pinned at the zero-$\pi$-boundary. 
A memory cell based on a $\phi$-Josephson junction has been suggested \cite{Goldobin13}, where writing is done by applying a magnetic field and reading by applying a bias current. 
Storage at low temperatures is passive, without any bias or magnetic field applied.

A high-frequency cryogenic generator operating at about 200 GHz, and based on flipping a semifluxon in a Josephson junction has recently been demonstrated in Ref.~\cite{Paramonov14}. A $\pi$ junction is artificially created by current injection using a technique by Ustinov \cite{Ustinov02}, giving rise to a semifluxon \cite{Goldobin04}. A conversion efficiency of $\sim 10$\% of dc input power to ac output power was achieved, including the losses on the way from the generator to the on-chip detector.
This type of Josephson oscillator is comparable with those based on flux-flow, with advantages like small size and insensitivity to injection current.

Realization of a 0-$\pi$ junction in an atomic bosonic quantum gas has been proposed in Refs.~\cite{Kulic07,Walser08}. In Ref.~\cite{Walser08} it is suggested that
two-state atoms in a double-well trap are coupled and an all-optical 0-$\pi$ Josephson junction is created by the phase of a complex-valued Rabi frequency, exhibiting modes similar to semifluxons. It is suggested that pairs of semifluxons can be created by starting from a flat-phase state in long, optical 0-$\pi$-0 Josephson junctions formed with internal electronic states of atomic Bose-Einstein condensates \cite{Grupp13}.

\subsection{Non-equilibrium quasiparticle distribution}

A particular exciting subject is the possibility to utilize non-equilibrium quasiparticle distribution in conjunction with quantum coherence in so-called Andreev interferometer geometries. These devices were introduced by Petrashov, Antonov, Delsing, and Claeson in 1994 \cite{Petrashov94a}. A superconducting wire (e.g. aluminum) is attached at two points to a normal mesoscopic conductor (e.g silver or antimony) to form a loop. The conductance of the wire then oscillates as function of the magnetic flux through the loop. The oscillations are attributed to phase transfer from the superconducting condensate to normal electrons via Andreev reflections at the N-S interfaces. 
The so-called $\pi$-SQUID employs an idea of Volkov \cite{Volkov95}, where by applying a control voltage directly to a normal metal the electron distribution function in the normal metal is modified.
This non-equilibrium distribution spreads to the superconductor and influences the transport in a nonlocal way. This allows for a direct control of the supercurrent through the device, including switching between zero-junction and $\pi$-junction behavior. Such a device was realized experimentally 
by Morpurgo, Baselmans, van Wees, and Klapwijk in 1998-99 \cite{Morpurgo98,Baselmans99}.
Corresponding effects are intricately connected with non-locality. 
Cadden-Zimansky, Wei, and Chandrasekhar \cite{Cadden09} 
combined an Andreev interferometer arrangement with injection of non-equilibrium excitations into a normal wire, giving rise to a  current that adds to the supercurrent in the Andreev interferometer and leads to a voltage at the contacts. Due to current conservation each current influences the other, and together they allow for tuning of the contact voltage by changing the flux through the Andreev interferometer that controls the supercurrent \cite{Eschrig09NV}. Coherent voltage oscillations
as functions of external flux are the result.

The combination of non-equilibrium quasiparticle distribution with the Josephson effect seems particularly exciting, and is largely unexplored so far.
Recent studies by Bobkova and Bobkov \cite{Bobkov11,Bobkova12} address the problem of spin-dependent non-equilibrium quasiparticle distribution. It is found that the interplay between the spin-dependent quasiparticle distribution and the triplet superconducting correlations induced by the proximity effect between the superconducting leads and ferromagnetic elements of the interlayer leads to the appearance of an additional contribution to the Josephson current. The interplay between short-range and long-range proximity effect is elucidated, and a long-range penetration of opposite-spin Cooper pairs under non-equilibrium conditions is proposed. An increase of the critical Josephson current by few orders of magnitude as a result of the non-equilibrium population in the ferromagnetic layer is suggested. 
Long-range spin and charge accumulation in mesoscopic superconductors with Zeeman splitting was recently also studied by
Silaev {\it et al.} \cite{Silaev14}.

A quantum interference transistor involving textured ferromagnets was suggested in Ref.~\cite{Alidoust13}. It was shown that such a device acts as an ultra-sensitive magnetometer and allows for singlet-triplet switching by tuning a bias voltage.
A combination of two spin injectors with an SFS Josephson junction was discussed in Refs.~\cite{Malshukov12,Meng14}.

\subsection{Dynamical effects}
Dynamical effects in superconductor-ferromagnet structures have moved in the focus of interest recently. In particular, it is clear that magnetization dynamics will be linked closely with Josephson dynamics, leading to potentially new effects. The geometric phases discussed in section \ref{geometric}, when time dependent, are expected to lead to spin accumulation effects via an imbalance between the equal-spin pair electrochemical potentials, very similar to the voltage linked to a dynamical superconducting phase in the ac Josephson effect \cite{Grein09}. 

The Josephson current in a diffusive superconductor-ferromagnet-superconductor junction with precessing bulk magnetization was calculated in Ref.~\cite{Houzet08}. It was found that when the junction is phase biased, a dc Josephson current without ac component can still flow under this non-equilibrium condition. Long-range triplet amplitudes are induced by the precessing magnetization.

In Refs. \cite{Holmqvist11, Holmqvist12} non-equilibrium effects in a Josephson junction with two $s$-wave singlet superconducting leads coupled via a precessing spin in a quantum dot was examined. 
An external magnetic field leads to a Larmor precession of the spin, rendering the magnetically active interface time-dependent. The authors analyze the non-equilibrium population of Andreev sidebands and dynamical spin currents. It is found that the supercurrent is enhanced and the critical Josephson current density shows a non-monotonous behavior as function of temperature, accompanied by a corresponding change in spin-transfer torques acting on the precessing spin.

Supercurrent-induced magnetization dynamics in superconductor-ferromagnet Josephson junctions
was explored in Refs.~\cite{Braude08,Konschelle09,Cai10,Linder11,Kulagina14,Linder15}. It is found that the spin supercurrent can induce magnetization switching that is controlled by the superconducting phase difference. 
The authors in Refs.~\cite{Linder11,Kulagina14} confirm the finding of Ref.~\cite{Grein09}, that the effect of chiral spin symmetry breaking of the structure leads to additional geometric phases that allow for the stabilization of
a $\phi$-junction. 

In Ref.~\cite{Mai11} the dynamics of 
superconductor-ferromagnet-insulator-ferromagnet-superconductor (SFIFS) junctions
with a thin ferromagnetic layers investigated. The coupled dynamics of the magnetization and the Josephson phase leads to Josephson plasma waves coupled to oscillations of the magnetization, affecting the form of the current-voltage characteristics in weak magnetic fields.

\subsection{Spin-orbit coupling and topological materials}
\begin{figure}[b]
\includegraphics[width=1.0\linewidth]{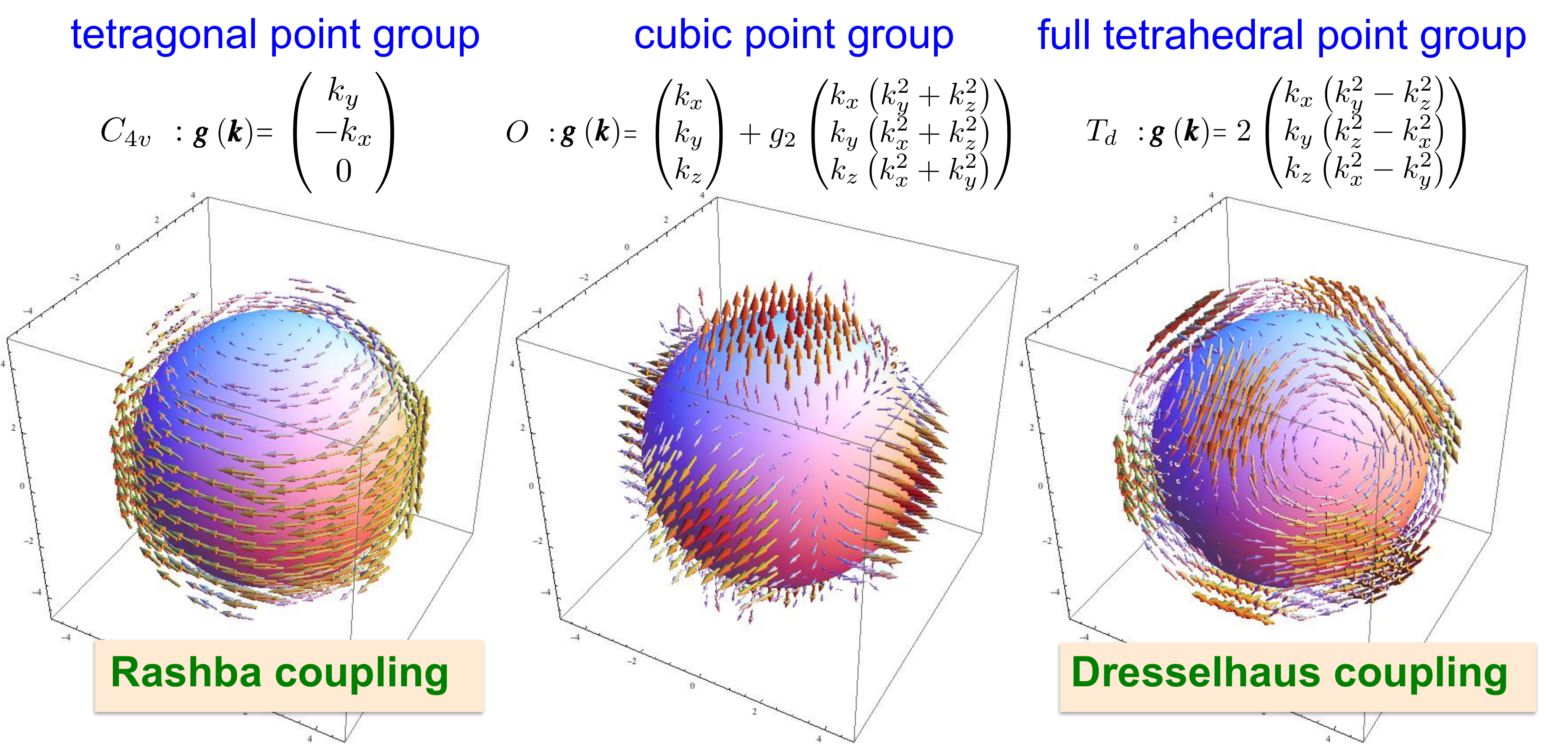}
\caption{
Various types of spin-orbit coupling in crystals, relevant for topological superconductivity. The crystal point group sets restrictions to the allowed spin-orbit vector fields. The lowest order expansion terms in crystal momentum are shown.
\label{SO}
}
\end{figure}
Spin-orbit effects play an important role whenever the inversion symmetry is broken, either in the bulk material (when it is lacking a center of inversion), or at interfaces and surfaces \cite{Samokhin09}. A prominent example for spin-orbit effects at surfaces is the Rashba-Bychkov spin-orbit coupling \cite{Rashba60}, and in the bulk an example is the Dresselhaus coupling due to bulk-inversion asymmetry \cite{Dresselhaus55}. 

Whereas in materials with a center of inversion all band-diagonal matrix elements of the spin-orbit coupling vanish, this is not the case in non-centrosymmetric materials.
Spin-orbit coupling in non-centrosymmetric materials is strongly enhanced due to band-diagonal contributions, leading to a splitting of the Fermi surface into spin-orbit bands, sometimes also called `helicity bands' (although a well defined helicity is only in special cases present).
The kinetic part of the Hamiltonian in a one-band model has the form
\begin{equation}
{\cal H}_{\rm kin} = \sum_{\vec{k}} \sum_{\sigma \sigma'\in \{\uparrow,\downarrow\}}
\left[ \varepsilon(\vec{k}) + \vec{g}(\vec{k}) \cdot \bsp \right]_{\sigma\sigma'} a_{\vec{k}\sigma}^\dagger a_{\vec{k}\sigma'}
\end{equation}
with the spin-orbit vector being odd in $\vec{k}$: $\vec{g}(-\vec{k})=-\vec{g}(\vec{k})$.
In figure \ref{SO} various spin-orbit vector fields compatible with the point group of the crystal are visualized on a hypothetical spherical Fermi surface. All three cases are relevant for materials: $C_{4v}$ for CePt$_3$Si \cite{Bauer04a}, CeRhSi$_3$ \cite{Kimura05} and CeIrSi$_3$ \cite{Sugitani06}, $T_d$ for Y$_2$C$_3$ \cite{Amano04}, and $O$ for Li$_2$(Pd$_{1-x}$Pt$_x$)$_3$B \cite{Togano04}.
The kinetic part of the Hamiltonian can be diagonalized, leading to the above-mentioned helicity bands. The spin-orbit interaction locks the orientation of the quasiparticle spin with respect to its momentum in each band.

One interesting aspect of a non-centrosymmetric ferromagnetic Josephson junction is the modification of the Josephson relation from an odd function in the superconducting phase difference to a current-phase relation where this symmetry is broken.
For example, near the critical temperature the current-phase relation $I=I_c\sin (\Delta \chi )$ can be modified to
\begin{equation}
I=I_c\sin(\Delta \chi-\phi_0),
\label{AN}
\end{equation}
where $\Delta \chi $ is the superconducting phase difference, and $\phi_0$ is a phase shift proportional to the magnetic moment perpendicular to the gradient of the asymmetric
spin-orbit potential \cite{Buzdin08}. The possibility of a $\phi_0$-junction has been 
already taken into account by Josephson \cite{Josephson62}, and was later considered in Josephson junctions involving unconventional superconductors \cite{Geshkenbein86}.
Examples in Josephson junctions with half-metals and strongly spin-polarized ferromagnets we discussed in Section \ref{geometric}.
A similar effect has also been predicted for Josephson junctions with a spin-polarized quantum point contact in a two-dimensional electron gas with spin-orbit coupling in an external magnetic field \cite{Reynoso08}.
The spin-dynamics with such a $\phi_0$ Josephson junction has been discussed in Ref.~\cite{Konschelle09}. 
The competition between a Zeeman interaction and a Rashba spin-orbit interaction has been in the center of attention since a while. 
In connection with one-dimensional quantum wires, an anomalous $\phi_0$ shift was predicted in Refs.~\cite{Krive04,Yokoyama13}. 
In a model for a mesoscopic multilevel quantum dot the conditions for an anomalous Josephson current Eq. (\ref{AN}) were found to be 
a finite spin-orbit coupling, a suitably oriented Zeeman field, and the dot being a chiral conductor \cite{Zazunov09,Reynoso12}. In Refs. \cite{Liu10,Malshukov10} an anomalous Josephson current was predicted in junctions coupled with a two-dimensional electron gas exhibiting coexistence of spin-orbit coupling and Zeeman field. 

A recent development concerns a gauge-covariant approach to establish the transport
equations, treating the charge and spin degrees of freedom
on equal footing. In this approach both the electromagnetic and
spin interactions are described in terms of U(1) Maxwell
and SU(2) Yang-Mills equations, respectively \cite{Frolich92,Berche13}. The starting point is
a quasiclassical expansion of the microscopic Gor'kov equations
in a gauge-covariant manner \cite{Konschelle14,Bergeret14}.
The idea is that one can re-write a Hamiltonian of the type
\begin{equation}
{H}_{\rm kin}= \frac{p_x^2+p_y^2}{2m} - \mu + \vec{h}\cdot \bsp + v_{\rm s.o.}(\hat\vec{z}\times \vec{p} ) \cdot \bsp
\end{equation}
as
\begin{equation}
{H}_{\rm kin}= \frac{(p_x-mv_{\rm s.o.}\sigma_y)^2}{2m} 
+ \frac{(p_y+mv_{\rm s.o.}\sigma_x)^2}{2m}
- \mu - \frac{mv_{\rm s.o.}^2}{2}+ \vec{h}\cdot \bsp ,
\end{equation}
which motivates to introduce
$A_x= -mv_{\rm s.o.}\sigma_y/\hbar $ and $A_y= mv_{\rm s.o.}\sigma_x/\hbar $ as two non-abelian gauge potentials. 
In this approach, the covariant derivative
\begin{equation}
\mathfrak{D} G= \partial_{\bf R} G + i [\vec{A},G]
\end{equation}
with the gauge potential $\vec{A}$ is associated with the gauge field
\begin{equation}
\vec{F}_{\mu\nu}= \partial_\mu \vec{A}_\nu - \partial_\nu \vec{A}_\mu + i [\vec{A}_\mu,\vec{A}_\nu]
\end{equation}
which is antisymmetric in its indices, and has only one component, $\vec{F}_{xy}=-2(mv_{\rm s.o.})^2\sigma_z/\hbar^2$. 
This leads in the transport equation to the modification
\begin{equation}
\label{modification}
i\hbar \vec{v}_f \cdot \nabla_{\vec{R}} \hat g \to
i\hbar \vec{v}_f \cdot \nabla_{\vec{R}} \hat g + imv_{\rm s.o.}^2 \partial_{\phi} \left\{ \sigma_z,\hat g \right\}
\end{equation}
with $\partial_\phi$ the derivative along the Fermi surface (which is assumed a circle for simplicity). Using this modified equation, 
the role of spin-orbit coupling as source of long-range triplet proximity effect in superconductor-ferromagnet structures has been explored in Ref.~\cite{Bergeret14}. Furthermore, the connection to $\phi_0$ Josephson junction behavior has been elucidated \cite{Konschelle14a,Bergeret15}. It should be, however, cautened that the modification in Eq.~(\ref{modification}) does not contain all terms of the same order as the additional term.
A quasiclassical theory of disordered Rashba superconductors has also been proposed in Ref. \cite{Houzet15}.

One way to manipulate spin in spintronics devices is the so-called spin Hanle effect \cite{Hanle23,Zutic04,Awschalom07}, which describes the coherent rotation of a spin in an external magnetic field. A recent theoretical study of this effect is given in Ref.~\cite{Silaev15}.
It is demonstrated that superconductivity can strongly influence the coherent spin rotation, depending on the type of spin relaxation mechanism being dominated either by spin-orbit coupling or spin-flip scattering at impurities.

Another manifestation of coherence in systems with spin-orbit coupling is the Aharonov-Casher effect \cite{Aharonov84}, leading to a phase on the Josephson current through a semiconducting ring attached to superconducting leads \cite{Liu09}. This effect is the charge-spin dual effect to the Aharonov-Bohm effect \cite{Aharonov59}, which describes the relative phase shift between two charged particle paths enclosing a magnetic flux. 
Both effects are manifestations of geometric phases acquired by the quantum mechanic wave function under adiabatic changes \cite{Berry84,Aharonov87}. The particle's spin acquires such a geometric phase in systems with spin-orbit interaction \cite{Meir89,Mathur92,Aronov93,Qian94,Nitta99,Loss99,Frustaglia04}.
The Aharonov-Casher effect was, e.g., observed experimentally in ring structures of HgTe/HgCdTe quantum wells \cite{Konig06}. 
In superconducting Josephson rings, or closed Josephson junction arrays, such phases 
lead to oscillations of the Josephson current due to the Aharonov-Casher phase
\cite{Matveev02,Liu09,Pop12,Bell15}. The effect allows for control of the Josephson current through the control of the Aharonov-Casher phase by the gate voltage.
Thus, this effect is a promising candidate for realizing new types of controllable devices in superconducting spintronics based on geometric phases.

Recently Mironov {\it et al.}
\cite{Mironov14} studied double path interference during Cooper pair transport through a single nanowire with two conductive channels. It is found that multi-period magnetic oscillations appear due to quantum mechanical interference between channels affected by spin-orbit coupling and Zeeman coupling. The model is relevant to recent observations of interference phenomena in Bi nanowires \cite{Kasumov14}.

A Josephson junction containing a spacer with strong spin-orbit interaction was considered in Ref.~\cite{Hoffman12}. A nonlinear dynamical coupling between magnetic moment and charge current was found, and magnetic torque and charge pumping was investigated in such a system. 
The intricate coupling between spin and charge currents in systems with strong spin-orbit coupling is previously known 
from non-centrosymmetric superconductors, where spin-polarized Andreev states play a prominent role \cite{Vorontsov08}. 

This connects to the current hot topic of topological materials, in particular topological insulators and superconductors, both systems with strong spin-orbit coupling. 
As an example for the extremely rich plethora of effects involving topological materials 
we mention here the possibility to observe chiral Majorana modes in one-dimensional channels built at a superconductor/ferromagnetic-insulator/superconductor junction on top of a topological insulator \cite{Tanaka09a}, where a $\phi_0$ junction is predicted.
The effect is interpreted as tunneling process between two Majorana edge channels at the two interfaces between the superconductor and the ferromagnetic insulator.
Majorana fermions have the property of being their own antiparticle, i.e. their field operators fulfill $\gamma^{\,}_{\alpha }=\gamma^\dagger_{\alpha }$ \cite{Majorana37,Wilczek09}.
A non-trivial superconducting phase is also obtained in proximity junctions involving
semiconductors with Rashba spin-orbit coupling and a time-reversal symmetry breaking Zeeman term in the Hamiltonian \cite{Sato09,Sau10}. 

The combined effect of spin-orbit interaction, magnetic field, and Coulomb charging for a multilevel quantum dot tunnel contacted by two superconductors was analyzed in Ref.~\cite{Brunetti13}. Majorana bound states in a double dot variant of this system are predicted to leave a clear signature in the $2\pi$-periodic current-phase relation.
Majoranas in spin-orbit coupled ferromagnetic Josephson junctions were investigated in Ref.~\cite{Schaffer11}, were it was shown that
two delocalized Majorana fermions with no excitation gap appear in a $\pi$-junction.
Josephson currents through Majorana bound states in topological insulators have been studied in Refs.~\cite{Kitaev01,Law11,Jiang11,Ioselevich11,Heck11,Jose12,Pikulin12,Zazunov12,Dominguez12,Houzet13}.
Such Josephson junctions carry 4$\pi$-periodic bound states \cite{Kitaev01}. 
It has been shown that under certain conditions this periodicity manifests itself by an even-odd effect in Shapiro steps \cite{Jiang11,Dominguez12}. In addition, a peak in the current noise spectrum at half the Josephson frequency has been predicted \cite{Houzet13}.

The diverse spectrum of effects and phenomena in
hybrid systems between singlet superconductors and topological insulators 
(see e.g. Refs.~\cite{Fu08,Qi09,Chung11,Alicea12,Grein12,Tanaka12,Beenakker13,Xu14}) are promising examples of how the field can be brought forward, playing a prominent role in various modern developments at the forefront of international research.

\section*{Acknowledgments}
I acknowledge the hospitality and inspiring atmosphere of the Aspen Center of Physics,
financial support from the Lars Onsager Award committee at NTNU,
as well as stimulating discussions within the Hubbard Theory Consortium.
This work is supported by the Engineering and Physical Science
Research Council (EPSRC Grant No. EP/J010618/1).

\end{document}